\documentclass{aa}  
\usepackage{graphicx}
\usepackage{natbib}
\bibpunct{(}{)}{;}{a}{}{,}
\usepackage{amssymb}
\usepackage{txfonts}
\usepackage{ifpdf}
\ifpdf%
\usepackage{pdflscape}
\else
\usepackage{lscape}
\fi
\begin{document}
   \title{Modeling water emission from low-mass protostellar envelopes}

  \subtitle{}

   \author{T.A. van Kempen
          \inst{1}
	  \and
	  S.D. Doty\inst{2}
          \and
          E.F. van Dishoeck\inst{1},\inst{3}
          \and
          M.R. Hogerheijde\inst{1}
	  \and
	  J.K. J{\o}rgensen\inst{4}
}

   \offprints{T. A. van Kempen: kempen@strw.leidenuniv.nl}

   \institute{$^1$Leiden Observatory, Leiden University, P.O. Box 9513,
              2300 RA Leiden, The Netherlands\\
	      $^2$Department of Physics and Astronomy, Denison
              University, Olin 109 Granville, OH  43023, USA\\
	     $^3$Max-Planck Institut f\"ur Extraterrestrische 
             Physik (MPE), Giessenbachstr.\ 1, 85748 Garching, Germany \\
	      $^4$Argelander-Institut f{\"u}r Astronomie, Universit\"at Bonn, 
       Auf dem H{\"u}gel 71, 53121, Bonn, Germany \\
              \email{kempen@strw.leidenuniv.nl}
                     }

   \date{}
 \titlerunning{Water emission from low-mass protostars}

 
  \abstract 
   {Within low-mass star formation, water vapor plays a key role
   in the chemistry and energy balance of the circumstellar material.
The Herschel Space Observatory will open up the possibility to observe
 water lines originating from a wide range of excitation energies.
}
{Our aim is to simulate the emission of rotational water lines 
from envelopes characteristic of embedded low-mass
  protostars. A large number of parameters that influence the water
  line emission are explored: luminosity, density, 
density slope and water abundances.}
{Both dust and water emission are modelled using full radiative
  transfer in spherical symmetry.  The temperature profile
  is calculated for a given density profile. The H$_2$O level
  populations and emission profiles are in turn computed with a
  non-LTE line code. The
  results are analyzed to determine the diagnostic value of different
lines, and are compared with existing observations. }
{Lines can be categorized in: (i) optically thick lines, including
ground-state lines,
mostly sensitive to the cold outer part; (ii) highly excited
($E_u>200-250$ K) optically thin lines sensitive to the abundance in
the hot inner part; and (iii) lines which vary from optically thick to
thin depending on the abundances.
Dust influences the emission of water significantly by
becoming optically thick at the higher frequencies, 
and by pumping optically thin lines.}
   {A good physical model of a source, including a correct treatment of
dust, is a prerequisite to infer the water abundance structure and possible
jumps at the evaporation temperature from
observations.  
The inner warm ($T>$100 K) envelope can be probed by
highly-excited lines,
   while a combination of excited and spectrally resolved ground state
   lines probes the outer envelope. Observations of H$_2^{18}$O lines,
although weak, provide even stronger constraints on abundances. }

   \maketitle
%

\section{Introduction}

Low-mass ($M<$3 M$_{\odot}$) young stellar objects (YSOs) form
through gravitational collapse of cloud cores. Most of their mass is
accreted over periods of less than 10$^6$ years
\citep{Andre00,Myers00}.  These early stages of low-mass star
formation, the so-called Class 0 and Class I stages, are characterized
by the presence of centrally condensed envelopes which veil the
central protostars and disks \citep{Lada87,Andre93}. The envelopes also contain
most of the mass ($M\sim$0.1-1 M$_{\odot}$) of the total system
until the emergence of the central star and disk
\citep{Adams87,Shirley00,Jorgensen02}.  Water plays a pivotal role in
these early evolutionary stages.  First, the high dipole moment of
H$_2$O makes its rotational lines excellent coolants of warm gas in
the inner region of protostellar envelopes, allowing material to
continue collapsing at higher temperatures
\citep{Goldsmith78,Neufeld93,Ceccarelli96}.  Second, water is one of
the dominant reservoirs of the non-refractory oxygen budget in dense
clouds \citep{vanDishoeck93}. The partitioning of gaseous oxygen
between H$_2$O, OH and O, as well as the ratio between H$_2$O gas and
ice, affect the chemistry of all other oxygen-containing molecules
including the complex organics \citep{Charnley92}.

The evaporation point of H$_2$O ice in interstellar space is
around 100 K \citep{Fraser01}, so only the small warm inner
regions of protostellar envelopes are likely to contain fractional
abundances of water vapor as high as $10^{-4}$ with respect to H$_2$,
perhaps even exceeding that of CO
\citep[e.g.,][]{Ceccarelli96,vanDishoeck98,Boonman03}. In colder
regions, water is the main ingredient of icy grain mantles as shown by
infrared spectroscopy
\citep[e.g.,][]{whittet83,smith89,boogert02,pontoppidan04}, with
orders of magnitudes lower gas-phase water abundances.  This abundance
jump makes water an excellent tracer of warm, dense gas and an unique
probe into the inner region close to the forming star.

Observations of interstellar gas-phase water from the ground are
limited by the large amounts of water vapor in the Earth's atmosphere.
Water maser lines at radio wavelengths long-ward of 3 mm (100 GHz), such as at
22 GHz, are readily seen, even from low-mass embedded YSOs
\citep[e.g.,][]{Furuya03,Park07}, but are constrained to small hot-spots
and often variable in nature \citep{Felli07}. They provide little
information about the general physical and chemical evolution during
star formation \citep{Elitzur89}. One of the few exceptions of a
non-masing water line observable from the ground is the 183 GHz
$J=3_{13}-2_{02}$ transition, which can be used to trace wide-spread
water emission under exceptional conditions
\citep[e.g.,][]{Cernicharo90,Cernicharo96,Doty00}.

Observations of deuterated water can also be done from the ground, in
particular through the HDO $1_{01}-0_{00}$ 464 GHz, $3_{12}-2_{21}$
225 GHz and $2_{11}-2_{12}$ 241 GHz lines
\citep[e.g.,][]{Turner75,Jacq90,Helmich96,Stark04,Parise05} and the
D$_2$O $1_{10}-1_{01}$ 316 GHz line \citep{Butner07}, but the large
uncertainties in the D/H ratio owing to fractionation make such
observations a poor tracer of water itself.  The H$_2^{18}$O
isotopologue can be detected as well, but only through its high
excitation temperature ($E_{\rm{up}}$ = 193 K) 3$_{13}$-2$_{20}$
transition at 203 GHz \citep[e.g.,][]{Jacq88,Gensheimer96,vanderTak06}.

To observe water emission lines systematically for a wide range of
energy states, one has to turn to space-based observatories. Gaseous
water has been detected in a number of star forming regions using the
Infrared Space Observatory (ISO), the Sub-millimeter Wave Astronomy
Satellite (SWAS) and ODIN \citep[e.g.,][]{Helmich96,Harwit98,Nisini99,
Ceccarelli99,Snell00,Bergin02,Maret02,Nisini02,Boonman03,Bergin05,Ristorcelli05,Cernicharo05}. However,
due to large beams, low spectral resolution or a combination of both,
such observations were unable to accurately resolve the water
abundance profile and trace the origin of the water emission,
especially for low-mass YSOs.  Two instruments on-board the Herschel
Space Observatory\footnote{see http://herschel.esac.esa.int/} can be used
for future observations of rotational water lines at unprecedented
sensitivity, resolution and spectral coverage
\citep{Ceccarelli01}. The \textit{Heterodyne Instrument for the Far
Infrared} (HIFI) \citep{Degraauw01} and \textit{Photodetector Array
Camera and Spectrometer} (PACS) \citep{Poglitsch01} provide high
spectral (up to 125 kHz) and spatial (20$''$ at 1 THz) resolution,
which will allow a thorough analysis of water emission through a much
wider range of observable transitions. In addition, the high
sensitivity of the instrument will allow observations of various lines
at low abundance, such as the optically thin isotopologues of water, in particular
H$_2^{18}$O. However, to infer abundances from such observations,
detailed radiative transfer calculations are needed.

Several modelling efforts have been carried out to analyze the
observed water lines, either based on models of shocks at high and low
velocities \citep[e.g.,][]{Kaufman96,Giannini01,Nisini02} or on
thermal emission from the envelope itself
\citep[e.g.,][]{Maret02,Boonman03}. 
As pointed out most recently by \citet{vanderTak06}, the high water
abundance combined with its high dipole moment and its susceptibility
to far-infrared emission from dust results in high optical depths and
make this molecule particularly difficult to model.  As a
result, no systematic parameter studies have yet been carried out.  We
present here a parameter study of the water emission from low-mass
protostellar envelopes using a full radiative transfer method
including the effects of dust and applied to realistic physical models
of low-mass YSOs. Specific predictions for Herschel are made by
convolving the computed emission with the Herschel observing
beams. \citet{Poelman07} present similar models for high-mass
protostars. 

This paper is organized as follows.  $\S$2 presents the physical and
molecular parameters necessary to accurately model water. $\S 3$ takes
an in-depth look at a single Class 0 source, L~483, and explores the
effects of possible water abundance profiles. $\S 4$ presents the
results from the range of parameters that influence water
emission. $\S 5$ investigates the influence of dust on water line
fluxes and profiles, which is often ignored. $\S 6$ gives an analysis
of a selected sample of observed water lines from ISO, SWAS and ODIN
to test the models, and to infer implications for observations with
future instruments such as HIFI and PACS. Final conclusions and
recommendations can be found in $\S 7$.

\section{Model}

 The modelling has been carried out in two stages. First, the dust
 radiative transfer is calculated for a given power-law density
 profile ($\S 2.1$). This produces dust temperature profiles for given
 density structures. Subsequently water level populations and line
 emission profiles are modelled using the line radiative transfer code
 RATRAN ($\S 2.2$) \citep{Hogerheijde00}, with the assumption that the gas
 temperature is equal to that of the dust.  For an overview of the
 steps, see \citet{vanDishoeck03} (Figure 2). Our method refers to
 the `Empirical method' presented there. See Table \ref{table:results}
 for an overview of the parameters.

\subsection{Physical structure of the envelope model}
\subsubsection{Dust radiative transfer}
Our protostellar envelope models are taken to be spherically
symmetric, with the protostar located at the center. It has been shown
that this assumption is valid for the analysis of both the continuum
as well as line emission down to scales of a few hundred AU
\citep[e.g.,][]{Shirley00,Jorgensen02, Schoeier02, Jorgensen05}.  The
density $n$ at radius $r$ within the protostellar envelope is
described by:
\begin{equation}
 n(r)=n_0(r/r_0)^{-p},
\end{equation}
with the power law exponent $p$ and density $n_0$ at a reference
 radius $r_0$ as parameters. Within the context of protostellar
 collapse theories, both parameters depend on the age of the system
 and on the original core conditions, such as its initial mass.

Both the gas and dust temperatures inside the envelope are often
also described by a power-law profile.
However, this approximation fails at small
radii where the dust becomes optically thick
\citep[e.g.,][]{Doty94,Jorgensen02,Young05}.  The actual temperature structure
depends on a combination of the density structure and the central source
luminosity, and a dust radiative transfer calculation is needed to
determine it.  
The radiation is produced by the energy released through
accretion of material onto the protostar.

The continuum radiative transfer through the envelope was calculated
using the spherically symmetric dust radiative transfer code DUSTY
\citep{Ivezic97}. DUSTY is a scale-free 1-D code, and uses the ratio
$Y$ of the outer over inner radius, $R_{\rm{out}}$/$R_{\rm{in}}$, the
dust opacity $\tau$ at 100 $\mu$m, the temperature at $R_{\rm{in}}$
and the power law index of the density profile $p$ as its
variables. Results are then scaled by using the luminosity and
distance of specific sources.  The inner edge of our envelope
$R_{\rm{in}}$ was set at the radius where the temperature of the
envelope reaches 250 K.  Material closer to the star is assumed to be
located in a disk, which is not taken into account. In addition, we
assume that the disk has no influence on the temperature and density
distribution of the envelope.  The exact inner radius in AU depends on
the luminosity, which differs for each model. $R_{\rm{in}}$ is
typically of the order of a few tens of AU, much smaller than the
Herschel beams (see Table \ref{tab:mass}).  In all our models $Y$ is
taken to be 200. Values of $R_{\rm{out}}$ are a few thousand to ten
thousand AU, larger than the Herschel beam (20$''$ at 1113 GHz) at all
but the lowest frequencies, but generally smaller than the values
given in \citet{Jorgensen02}.  The adopted dust opacity's are the
often-used OH tables, column 5 of \citet{Ossenkopf94}
throughout. Figure \ref{Fig:structure} shows a typical temperature and
density profile.

\subsubsection{Gas radiative transfer}
The gas temperature is taken to be equal to the dust temperature for
the entire spatial extent of our model, and the gas:dust mass ratio is
taken to be 100:1. Warm dense gas can efficiently cool through the
emission of excited molecular and atomic lines such as CO, H$_2$O, OH
and [O I] \citep{Giannini01}. Water can also heat the gas through
far-infrared pumping followed by collisional de-excitation
\citep[e.g.,][]{Ceccarelli96}.  As a result, dust and gas temperatures
can differ slightly, especially in the 100-200~K range
\citep[e.g.,][]{Ceccarelli00,Maret02}. However, at densities of $10^7$
cm$^{-3}$ or higher the gas and dust in the warm inner region couple
efficiently. At low densities ($<$10$^5$ cm $^{-3}$) and
  temperatures ($<$50 K) any differences between gas and dust
temperature depend on the source parameters (e.g., luminosity,
  total envelope mass) and the external interstellar radiation field
  (ISRF). \citet{Doty97} show in their Figure 12 that these
  differences are generally small, of order 10 K or less for low-mass
  YSOs, and typically only exist at radii larger than 2000 AU.
\citet{Boonman03} show that for high-mass YSOs, the water line
emission for a full radiative transfer calculation of the gas
temperature differs negligible from that calculated using the
$T_{\rm{dust}}=T_{\rm{gas}}$ assumption.

Our choice of $R_{\rm{out}}$=200 $R_{\rm{in}}$ assumes that water
vapor at larger radii does not contribute to the total water
line emission due to the low densities ($<10^5$ cm$^{-3}$) and
temperatures ($<$15 K) in these regions. This assumption
  mostly affects the transitions leading to the ground states. Any
  such material is able to absorb water emission for optically thick
  lines, and lead to weak emission lines in observations taken with a
  beam that is much larger than our adopted source sizes. However,
  such extended envelope material is almost indistinguishable from any
  larger scale cloud complexes not directly related to the source.
  Such contributions from large-scale clouds need to be taken into
  account on a source-by-source basis, but are beyond the scope of
  this work.  The protostellar envelopes are assumed to be shielded
from strong external radiation fields by surrounding molecular cloud
material. If directly exposed to outside radiation, the temperature in
the outer envelopes is affected by the ISRF, but material is not
heated by more than 40 K for the ISRF found within the Orion
star-forming region \citep{Jorgensen06}. Low-mass star-forming regions
are expected to have significantly lower ISRF. External radiation is
not considered for these models.

\begin{table}[!htp]
\caption{Parameters for low-mass protostellar envelopes }
\label{table:results}  
\begin{center}
\begin{tabular}{l l}     
\hline \hline       
\multicolumn{2}{c}{Variables} \\ \hline
$n_{\rm{H}_2}$(1000 AU) (10$^6$ cm$^{-3}$)  & 0.4, 1, 5\\
$p$ & 1.5, 2.0\\
Luminosity (L$_{\odot}$)& 2, 7, 25\\
$X_0$ (water) & $10^{-4}$, $10^{-6}$\\
$X_{\rm{d}}$ (water) & $10^{-6}$, $10^{-7}$, $10^{-8}$\\ \hline
\multicolumn{2}{c}{Fixed Parameters} \\ \hline
Distance & 150 pc \\
Stellar temperature & 5000 K \\
Gas temperature & = $T_{\rm{dust}}$\\
$R_{\rm{in}}$ (inner radius) & 250 K \\
$R_{\rm{out}}$ (outer radius) & 200 $R_{\rm{in}}$\\ 
$r_0$ & 1000 AU \\
Water freeze-out & 100 K \\
$o:p$ ratio H$_2$O$^{a}$ & 3:1 \\
$o:p$ ratio H$_2$ & 1:1\\
Dust properties & OH 5$^{b}$\\
Velocity dispersion $\Delta V$ & 1.67 km s$^{-1}$ \\ \hline
\end{tabular} 
\end{center}
$^{a}$For the L483 models, ortho- and para-water were modelled independently. See \S 3 
$^{b}$See Ossenkopf \& Henning (1994) column 5
\end{table} 

\begin{table}[!htp]
\caption{The masses and inner radii  of the model envelopes.}
\label{tab:mass}  
\begin{center}
\begin{tabular}{c c c c c} 
\hline \hline       
$L_{\rm{bol}}$ & $n_{\rm{H}_2}$($r_0$)  & $M_{\rm{env}}$ & $R_{\rm{in}}$ & $R_{\rm{100 K}}$\\
L$_{\odot}$ & 10$^6$ cm$^{-3}$ & M$_{\odot}$ & AU & AU\\ \hline
\multicolumn{5}{c}{$p=1.5$} \\ \hline
2 & 0.4  & 0.06 & 8.22& 23\\
2 & 1   & 0.15 & 8.22 & 23\\
2 & 5   & 0.76 & 8.22 &23\\
7 & 0.4  & 0.15 & 15.3& 42\\
7 & 1   & 0.39 & 15.3 &42\\
7 & 5   & 1.93 & 15.3&42\\
25 & 0.4 & 0.40 & 29 & 80 \\
25 & 1  & 1.01 & 29  &80 \\
25 & 5  & 5.04 & 29 &80\\ \hline
\multicolumn{5}{c}{$p=2$}\\\hline
2 & 0.4& 0.07  &8.22 &21\\
2 & 1  & 0.18 & 8.22&21\\
2 & 5  & 0.90 & 8.22&21\\
7 & 0.4 & 0.13 & 15.3&38\\
7 & 1  & 0.33 & 15.3&38\\
7 & 5  & 1.65 & 15.3&38\\
25 &0.4 & 0.24 & 29&72\\
25 & 1 & 0.63 & 29&72\\
25 & 5 & 3.15 & 29&72\\
\hline
\end{tabular} 
\end{center}
\end{table} 

\subsubsection{Grid parameters}

To limit the model complexity, the large range of parameters was
reduced to four variables (see Table \ref{table:results}).  The
density at 1000 AU, $n_0(\rm{H}_2$), and the exponent of the density
power law $p$ are two of these.  \citet{Jorgensen02} observed many
low-mass YSOs and inferred densities at 1000 AU between $1 \times
10^5$ to $5 \times 10^6$ cm$^{-3}$. Accordingly, three values,
$0.4\times10^6, 1\times10^6$ and $5\times10^6$ cm$^{-3}$ were chosen
to sample this range. Similarly, $p$ was chosen to be either 1.5 or
2.0.
The stellar variables are reduced to a single variable, the
luminosity. Models using luminosities of 2, 7 and 25 L$_{\odot}$
are explored to represent the large range of luminosities observed in
low-mass protostars. Although the surface temperature of the protostar
cannot be derived directly, it is assumed to be around 5000 K for
low-mass young stars, regardless of the luminosity or age. Test models
show that due to the reprocessing of stellar photons into IR/thermal
photons very close to the star, the stellar effective temperature has
little influence on the modelling.  The final variable is the
gas-phase water abundance, to be discussed below. Table
\ref{table:results} summarizes all the adopted parameters. The
resulting masses and inner and outer radii for each envelope model can
be found in Table \ref{tab:mass}.

\begin{figure}
\centering

\caption{ FIGURE TOO LARGE FOR ASTRO-PH. CAN  BE DOWNLOADED FROM WEBSITE http://www.strw.leidenuniv.nl/~kempen/water.php. Example of the physical structure of a model envelope with
the density (solid), temperature (dashed) and abundance (dotted)
displayed versus radius. The adopted parameters are
$n_0(\rm{H}_2)$=10$^6$ cm$^{-3}$, $p$=1.5 and $L=2$
L$_{\rm{\odot}}$. }
\label{Fig:structure}
\end{figure}

\subsection{H$_2$O line modelling}
Because of the gradient in temperature, the water abundance $X$ with
respect to H$_2$ is expected to vary within YSO envelopes.  Previous
studies of low- and high-mass YSOs used a variety of H$_2$O
observations with ISO-LWS, ISO-SWS and/or SWAS to constrain the water
abundance profile in protostellar envelopes to have a 'jump' in
abundance at a characteristic temperature around 100~K
\citep[e.g.,][]{Maret02,Boonman03,vanderTak06}. Such `jumps' have also
been confirmed by ground-based observations of other molecules such as
CH$_3$OH and H$_2$CO
\citep[e.g.,][]{vanderTak00,Ceccarelli00,Schoeier02,Jorgensen04,Maret04}.
The abundance of water in the inner region $X_0$ is high for two
reasons. First, most water exists in the gas phase
above 100 K, the temperature at which water ice evaporates from the
grains.  Second, gas-phase reactions of atomic oxygen with molecular
hydrogen drive all oxygen into water at high temperatures ($T$$>$230
K) \citep{Charnley01}.
However, in regions where the 
temperature drops below 100 K, water is mostly frozen out onto the
grains and only trace amounts of gaseous water are present
with an abundance $X_{\rm{d}}$. 

Accordingly, 'jump' abundances are assumed for all models with $X_0$,
the abundance of the warm inner region, taken to be 10$^{-4}$ or
10$^{-6}$ , and $X_{\rm{d}}$, the abundance in the cold outer region,
taken to be 10$^{-6}$, 10$^{-7}$ or 10$^{-8}$. These values cover the
range of inferred abundances from observations to date for embedded
protostars. Although beam averaged H$_2$O abundances below
  10$^{-8}$ have been observed for B 68 and Oph-D \citep{Bergin02},
  such clouds are starless. At the extremely low temperatures of $<$15
  K virtually all heavy elements are frozen out. Abundances as low as
  10$^{-11}$ were tested by \citet{Boonman03}, but models with outer
  abundances of 10$^{-8}$ were preferred in fitting the combined SWAS
  and ISO-LWS data. Note that one combination of abundances is a
constant abundance at 10$^{-6}$, included to investigate the
differences between `jump' and constant abundance profiles. A finer
grid of abundances within this range is explored for the L 483 model.
 
Figure \ref{Fig:structure} includes the assumed abundance structure.
Our models do not consider a so-called `drop' abundance profile
\citep[see][]{Schoeier04}, in which the abundance in the outermost
part is higher due to longer timescales for freeze-out at lower
densities and due to photodesorption of ices \citep{Bergin05}. Such an
abundance profile would result in more self-absorption in the
ground-state lines, equivalent to absorption caused by cold
fore-ground material not associated with the protostellar system.

The effects of micro-turbulence within the envelope are
represented by a velocity dispersion FWHM, $\Delta V$=1.67 km
s$^{-1}$ ($b=$1 km s$^{-1}$, with $b$ the Doppler parameter),
consistent with the typical line widths observed for optically thin
rotational emission in low-mass YSO envelopes
\citep{Jorgensen02,Jorgensen04a,Jorgensen05}. For a more thorough
  discussion of the effects of the choice of this value,
  see $\S$ 5.2. An infall velocity profile was included, based on the
velocity profiles for collapsing clouds as found by
\citet{Shu77}. This velocity profile was scaled to a velocity of 4 km
s$^{-1}$ at the inner radius $R_{\rm in}$.
At radii of a few hundred AU, the micro-turbulent velocity becomes
dominant over the infall velocity, which is the scale probed mostly by
single-dish observations \citep[e.g.,][]{Jorgensen04a}.  H$_2$ is
assumed to be in a 1:1 ortho to para ratio. Water itself is in 3:1
ortho to para ratio.  Collisional rate coefficients for both ortho-
and para-H$_2$O with ortho- and para-H$_2$ were obtained from the
LAMDA database \citep{Schoeier05} based on calculations by
\citet{Green93} for temperatures in the range from 20 to 2000
K. New rate coefficients have recently been published
  \citep{Phillips96,Dubernet02,Grosjean03,Faure07}\footnote{The rate
    coefficients from \citet{Faure07} only became available most of
    the models had been run}, but those new calculations often do not
  include a sufficiently large range of temperatures nor the higher
  excited energy levels, necessary for the analysis of YSOs. Test
  models using different published rate coefficients show that
  absolute line fluxes may differ up to 30$\%$ \citep{Poelman07}. Even
  larger differences are found at low temperatures ($T<$ 20 K) and
  densities ($n(\rm{H}_2)<10^4$ cm$^{-3}$) for the ground state line
  intensities and profiles
  \citet[e.g.,]{Poelman05,Dubernet06}. However, the general trends
  found in this paper remain valid for different rate coefficients.
. For a general discussion of uncertainties in molecular data, see
\citet{Schoeier05}.

\begin{table}[!htp]
\caption{Key water lines observable with Herschel HIFI and PACS.}
\label{table:lines}
\begin{center}
\begin{tabular}{c c c c c}
\hline \hline
Transition  & Freq. & Wavelength & $E_{\rm{up}}$& Beam$^a$ \\ 
 & (GHz) & ($\mu$m) & (K) & ($''$)   \\ \hline
\multicolumn{5}{c}{HIFI} \\ \hline
\multicolumn{5}{c}{Ortho-H$_2$O transition} \\ \hline
1$_{10}$-1$_{01}$ & 556.93607  & 538.66147 &61.0 & 39  \\
2$_{12}$-1$_{01}$ & 1669.90496 & 179.65094 &114.4 & 13\\
2$_{21}$-2$_{12}$ & 1661.00802 & 180.61322 &194.1 & 13\\
3$_{12}$-2$_{21}$ & 1153.12682 & 260.16219 &249.4 & 19\\
3$_{12}$-3$_{03}$ & 1097.36505 & 273.38213 &249.4 & 21\\
3$_{21}$-3$_{12}$ & 1162.91187 & 257.97312 &305.3 & 19\\  \hline
\multicolumn{5}{c}{Para-H$_2$O transition} \\ \hline
1$_{11}$-0$_{00}$ & 1113.34306 & 269.45872 &53.4& 21 \\
2$_{02}$-1$_{11}$ & 987.92670  & 303.66625 &100.8 &21 \\
2$_{11}$-2$_{02}$ & 752.03323  & 398.91854 &136.9 &30\\
2$_{20}$-2$_{11}$ & 1228.78902 & 244.14281 &195.9 &19\\ 
3$_{31}$-4$_{04}$ & 1893.68681 & 158.42113 &410.4 &13\\
4$_{22}$-3$_{31}$ & 916.17164  & 327.44956 &454.3 &25\\
4$_{22}$-4$_{13}$ & 1207.63890 & 248.41863 &454.3 &19\\
5$_{24}$-4$_{31}$ & 970.31524  & 309.17787 &598.8&21\\ \hline 
\multicolumn{5}{c}{PACS} \\ \hline
\multicolumn{5}{c}{Ortho-H$_2$O transition} \\ \hline
2$_{12}$-1$_{01}$ & 1669.90496 & 179.65094 & 114.4 & 13\\
2$_{21}$-1$_{10}$ & 2773.97691 & 108.14798 & 194.1 & 8\\
2$_{21}$-2$_{12}$ & 1661.00802 & 180.61322 & 194.1 & 13\\
3$_{03}$-2$_{12}$ & 1716.76979 & 174.74678 & 196.8 & 12\\
4$_{14}$-3$_{03}$ & 2640.47434 & 113.61595 & 323.5 & 8\\\hline
\multicolumn{5}{c}{Para-H$_2$O transition} \\ \hline
2$_{20}$-1$_{11}$ & 2968.74895 & 101.05267 & 195.9& 7 \\
3$_{13}$-2$_{02}$ & 2164.13230 & 138.62371 & 204.7& 10\\
4$_{04}$-3$_{13}$ & 2391.57290 & 125.44046 & 319.5& 9\\ \hline
\end{tabular}
\end{center}
$^a$ taken from http://herschel.esac.esa.int/home.shtml
\end{table}

The excitation and radiative transfer of the rotational water lines
were calculated using the spherically symmetric version of the RATRAN
code \citep{Hogerheijde00}. First, level populations were computed for
all H$_2$O levels with energies up to 2000 K for the model envelope
using full Monte Carlo radiative transfer. Individual iterations use a
convergence criterion of $10^{-6}$ on the relative error of the
fractional population for the solution of the statistical
equilibrium. Iterations continue with increasing amounts of photons,
until the largest relative difference between fractional populations
over all cells and all levels of three subsequent iterations is less
than 1/6. In practice this means that the vast majority of cells and
levels are converged to much smaller relative errors.  Far infrared
radiation from the dust and CMB are included in the H$_2$O excitation
and line formation. This inclusion makes it possible to investigate
both infrared pumping, self-absorption of a line and absorption
against the dust continuum.  Even though many levels are very sparsely
populated, tests show that it is necessary to include the higher
levels to obtain an accurate population distribution throughout the
model.

The populations are then used to reconstruct detailed line profiles
through ray-tracing, which can be convolved with any beam or area. It
can also be used to make velocity maps of the source. The adopted spectral
resolution is 0.05 km s$^{-1}$.  RATRAN was bench-marked
against many codes \citep{vanZadelhoff02}. Furthermore, it was
part of a recent benchmark test\footnote{see
http://www.sron.rug.nl/$\sim$vdtak/H2O/} for line modelling of
H$_2$O \citep{vanderTak05}.

Although level populations were calculated for all levels up to
$E_{\rm{up}}$=2000 K, only lines with an upper energy level below 500
K are selected for the presentation of results (see Table
\ref{table:lines}). Both H$_2$O and H$_2^{18}$O were modelled with a
focus on lines that fall into the Herschel wavelength range. The
$^{16}$O:$^{18}$O isotope ratio is assumed to be 550:1, the ratio in
the local ISM \citep{Wilson94}.
All generic models are placed at 150 pc, the typical distance of most
nearby low-mass star forming regions such as Ophiuchus, Taurus and
Chameleon. Results are convolved with the appropriate Herschel beams
and given in main beam temperatures.  For reference, the typical
sensitivities expected with HIFI (5$\sigma$, 0.5 hr, 0.5 km s$^{-1}$ resolution)
are 60 mK for 557 GHz, 180 mK at 1130 GHz and 600 mK at 1669 GHz. For
PACS, limiting line fluxes (5$\sigma$, 0.5 hr, point source) range
from 6$\times$10$^{-18}$ W m$^{-2}$ to 35 $\times$10$^{-18}$ W m$^{-2}$
\footnote{Calculated with Hspot, see
http://herschel.esac.esa.int/}. To convert from main beam temperatures
$T_{\rm MB}$ to fluxes $S_\nu$ in Jy
(10$^{-26}$ W m$^{-2}$ Hz$^{-1}$) use the following formula.
\begin{equation}
S_{\nu} = 2.65 \times 10^{8} \times T_{\rm MB} ({\rm K}) \times
\theta_0^2(') \times \lambda^{-2} (\mu {\rm m})
\end{equation}
This assumes a source size equivalent to the beam size.For
  sources smaller than the beam, a correction needs to be taken into
  account.  The Herschel beam $\theta_0$ ranges from 40$"$
(0.66$'$) at 557 GHz (539 $\mu$m) to 7$"$ (0.12$'$)
at 3000 GHz (100 $\mu$m) (see Table \ref{table:lines}).

\section{Example : L~483}
\subsection{Physical structure}

As an illustrative model, the isolated Class 0 source L~483,
\citep{Motte01} 
located at 200 pc, was modelled to investigate the sensitivity of the
water line intensities to different abundances.  This source has been
studied extensively by \citet{Jorgensen02} and \citet{Jorgensen04} and
its physical and chemical structure is well constrained through
continuum and line data from (sub)millimeter single-dish telescopes
and interferometers.  The well-determined envelope and source
parameters ($L_{\rm{bol}}=9$ L$_{\odot}$ and
$n_0(\rm{H}_2)$=1$\times10^6$ cm$^{-3}$) are roughly in the middle of
our parameter range, making L~483 an excellent test-case to model water
emission. Only the inferred density slope, $p$=0.9, is somewhat
smaller than that of other sources, but this does not influence the
model trends.  For CO, a jump abundance at an evaporation temperature
of $T\approx$30 K was found to explain the observed molecular
lines. Any compact disk is constrained to a mass of 0.012
M$_{\odot}$ using interferometer observations 
with the Sub-Millimeter Array \citep{Jorgensen07}, negligible
compared to the total estimated envelope mass of 4.4 M$_{\odot}$
out to the 10 K radius of 10,000 AU.  In addition, this source has
been observed with ISO-LWS
\citep{Giannini01}.  The best-fit parameters from the dust radiative
transfer ($Y$=1400, $p$=0.9, $\tau_{100}$=0.3), which fit both
the SED and the spatial extent of the SCUBA continuum data,
determine the physical structure of the envelope. $R_{\rm{in}}$ (at
$T$=250 K) is 9.9 AU for this rather shallow density profile 
and $R_{\rm{out}}$ is $\sim$10,000 AU.  The radius where water freezes
out onto the grains ($T$=100 K) is located at 35 AU.  This L~483 model
was used to explore a wider variety of water abundances than the
generic models. Specifically values of $1\times10^{-4},
5\times10^{-5}, 1\times10^{-5}, 5\times10^{-6}$ and $1\times10^{-6}$
were used for $X_0$ and $1\times10^{-6}, 5\times10^{-7},
1\times10^{-7}, 5\times10^{-8}$ and $1\times10^{-8}$ for $X_d$.  All
other parameters were kept fixed. Ortho- and para-water were modelled
separately, and water abundances are given for ortho- or para-water
individually instead of the combined total water abundance with the used 3:1 ortho to para ratio. This
allows one to probe the ortho to para ratio independently.
Thus, $X_0$ and $X_d$ refer to either ortho- or para-H$_2$O, depending on the line in question.

\begin{table*}[!htp]
\caption{Integrated intensities, $\int T_{\rm{MB}}dV$ [K km
s$^{-1}$], for H$_2$O and H$_2^{18}$O lines for L~483 models with
various abundances. Negative values indicate strong (self-) absorbed
lines. 
}
\label{table:h2o1}

\begin{center}
\begin{tabular}{c c c c c c c c c c}
\hline \hline
 Transition & \multicolumn{3}{c}{X$_0=10^{-4}$}&\multicolumn{3}{c}{X$_0=10^{-5}$}&\multicolumn{3}{c}{X$_0=10^{-6}$} \\\cline{3-3} \cline{6-6} \cline{9-9}
 X$_d =$ & 10$^{-6}$ & 10$^{-7}$ & 10$^{-8}$ & 10$^{-6}$ & 10$^{-7}$ & 10$^{-8}$ & 10$^{-6}$ & 10$^{-7}$ & 10$^{-8}$ \\  \hline
\multicolumn{9}{c}{Ortho-H$_2$O transitions}   \\ \hline
1$_{10}$-1$_{01}$ & 2.7 & 2.4 & 0.83 & 2.8  & 1.0   & 8.8(-2)& 9.5 &3.8 &0.57 \\
2$_{12}$-1$_{01}$ & -0.2&-0.13& 0.27 & -0.35& -0.38 &-0.33   &3.0  &0.23 &-0.05 \\
2$_{21}$-2$_{12}$ & 1.7 &0.27 & 0.20 &0.17  & 0.32  &0.11    &1.7  &0.37 &0.12 \\
3$_{12}$-2$_{21}$ & 0.8 &0.17 & 0.10 & 0.47 &0.12   &7.8(-2) &0.95 &0.20 &7.1(-2) \\
3$_{12}$-3$_{03}$ & 0.8 &0.29 & 0.13 & 0.43 &0.22   &6.0(-2) & 0.94&0.39 &7.4(-2) \\
3$_{21}$-3$_{12}$ & 1.3 &0.35 & 0.11 & 0.73 &0.26   &5.4(-2) &1.5  &0.46 &8.2(-2) \\ \hline
\multicolumn{9}{c}{Para-H$_2$O transitions } \\ \hline
1$_{11}$-0$_{00}$ & -6.5(-3)& 1.23   & -0.17  & -0.12   & 5.2(-2) & 0.29    &3.12    &1.4     &-0.13 \\
2$_{02}$-1$_{11}$ & 1.1 & 0.59       & 0.26 &  1.4 &    0.6       & 0.45    &0.6     &1.4     & 0.26 \\
2$_{11}$-2$_{02}$ & 0.85   & 0.73    &0.16    & 0.99    &0.6      & 0.34    &0.96    &1.1     & 0.15 \\
2$_{20}$-2$_{11}$ & 0.38   &  0.61   &0.18    & 0.45    &0.50     & 0.31    &0.38    &0.83    & 0.13 \\
3$_{31}$-4$_{04}$ &  4.4(-2)& 6.3(-2) & 6.3(-2)& 1.9(-2) & 1.9(-2) & 2.7(-2) & 3.7(-3)& 1.9(-3)& 1.3(-3) \\
4$_{22}$-3$_{31}$ &  5.0(-2)& 6.8(-2) & 3.7(-2)& 5.2(-2) & 5.6(-2) & 5.1(-2) & 1.8(-2)& 1.5(-2)& 9.7(-3)\\
4$_{22}$-4$_{13}$ &  0.21   & 8.6(-2) & 5.3(-2)& 0.25    & 8.3(-2) & 8.0(-2) & 0.24   & 0.10   & 5.8(-3)\\
5$_{24}$-4$_{31}$ &  6.0(-2)& 8.8(-2) & 5.5(-2)& 3.2(-2) & 4.3(-2) & 5.5(-2) & 4.0(-3)& 5.9(-3)& 3.3(-3)\\ \hline
\multicolumn{9}{c}{Ortho-H$_2^{18}$O transitions}  \\ \hline
1$_{10}$-1$_{01}$  & -3.3(-2)& -5.6(-3)&2.3(-2) &-4.0(-2) & -2.0(-2)&5.2(-3) & -1.9(-2)& -2.4(-2)& -3.9(-4) \\
2$_{12}$-1$_{01}$  & -0.13   & -3.3(-2)& 1.2(-2)& -0.13   & -4.5(-2)&-6.7(-4)& -0.12   & -4.9(-2)& -5.2(-3) \\
2$_{21}$-2$_{12}$  & 4.1(-3) & 4.2(-3) & 3.7(-3)& 1.8(-3) & 9.6(-4) & 9.3(-4)& 1.0(-3) & 1.3(-4)& 1.1(-4)\\
3$_{12}$-2$_{21}$  & 5.6(-2) & 6.0(-2) & 5.9(-2)& 3.6(-3) & 3.0(-3) & 3.0(-3)& 5.9(-4) & 1.39(-4)&1.2(-4) \\
3$_{12}$-3$_{03}$  & 2.5(-2) & 2.6(-2) & 2.5(-2)& 8.5(-3) & 6.9(-3) & 6.8(-3)& 2.0(-3) & 4.6(-4)& 3.8(-4)\\
3$_{21}$-3$_{12}$  & 1.7(-2) & 1.9(-2) & 1.8(-2)& 4.5(-3) & 3.6(-3) & 3.5(-3)& 1.0(-3) & 2.1(-4)& 1.7(-4)\\ \hline
\multicolumn{9}{c}{Para-H$_2^{18}$O transitions} \\ \hline
1$_{11}$-0$_{00}$ &-0.21   & -7.7(-2)&8.7(-3)& -0.16  & -8.8(-2)& -4.7(-3)&-0.21   & -9.3(-2)&-1.2(-2) \\
2$_{02}$-1$_{11}$ & 5.3(-2)& 4.0(-2)& 3.8(-2)& 4.4(-2)& 2.0(-2)& 1.8(-2)  &2.5(-2) & 4.7(-3)& 3.1(-3)\\
2$_{11}$-2$_{02}$ & 2.5(-2)& 1.9(-2)& 1.9(-2)& 2.0(-2)& 7.9(-3)& 7.6(-3)  &9.6(-3) & 1.5(-3)& 1.1(-3)\\
2$_{20}$-2$_{11}$ & 1.7(-2)& 1.4(-2)& 1.4(-2)& 1.6(-2)& 6.5(-3)& 6.2(-3)  &7.2(-3) & 1.0(-3)& 6.9(-4)\\
3$_{31}$-4$_{04}$ & 9.2(-5)& 1.2(-4)& 1.2(-4)& 5.7(-6)& 4.6(-6)& 4.6(-6)  & 5.1(-7)& 2.5(-7)& 2.3(-7)\\
4$_{22}$-3$_{31}$ & 6.3(-4)& 8.4(-4)& 8.5(-4)& 2.8(-5)& 2.3(-5)& 2.3(-5)  & 9.1(-7)& 6.7(-7)& 6.4(-7)\\
4$_{22}$-4$_{13}$ & 2.5(-2)& 3.0(-2)& 3.0(-2)& 1.5(-3)& 1.2(-3)& 1.2(-3)  & 4.8(-5)& 3.5(-5)& 3.4(-5)\\
5$_{24}$-4$_{31}$ & 7.1(-5)& 1.1(-4)& 1.1(-4)& 1.6(-6)& 1.3(-6)& 1.3(-6)  & 4.6(-8)& 4.2(-8)& 4.0(-8)\\ \hline
\end{tabular}
\end{center}
In this and subsequent tables, the notation $A(B)$ indicates $A\times 10^{B}$.
\end{table*}

\begin{figure*}
\begin{center}
\end{center}
\caption{ FIGURE TOO LARGE FOR ASTRO-PH. CAN  BE DOWNLOADED FROM WEBSITE http://www.strw.leidenuniv.nl/~kempen/water.php.Contour plots of the integrated water line intensity for the L~483
models, plotted as functions of $X_{\rm{d}}$ and $X_0$. 
{\it left} : the o-H$_2$O 1$_{10}$-1$_{01}$ line (557 GHz) , showing a strong dependence on $X_{\rm{d}}$; 
{\it middle} : the p-H$_2^{18}$O 2$_{02}$-1$_{11}$ line (995 GHz), showing a clear cut between the dependency on $X_{\rm{d}}$ or $X_0$, depending on the abundance profile;
{\it right} : the p-H$_2$O 5$_{24}$-4$_{31}$ line (970 GHz), showing a strong dependence on $X_0$.}

\label{Fig:intensityexample}
\end{figure*}

\subsection{Results}
In Table \ref{table:h2o1} the integrated emission of both H$_2^{16}$O
and H$_2^{18}$O lines, observable with Herschel-HIFI (see Table
\ref{table:lines}), are presented for the reference grid of L~483,
whereas Table \ref{table:h2opeak} presents the peak
temperatures. Detailed contour plots of all HIFI lines as functions of
$X_0$ and $X_d$ are presented in Fig. \ref{Fig:intensityexample} the online appendix. Many of the trends discussed below are similar to those found for AFGL 2591, as discussed in \citet{Poelman07}
Overall, lines are found to belong to three categories. First, all
excited H$_2^{18}$O and several H$_2$O highly excited
($E_{\rm{up}}>$200 K) lines are completely optically thin.  They have
Gaussian profiles with widths depending on the turbulent width. Due to
the freeze-out of water onto grains below 100 K, these optically thin
lines probe the inner warm dense region. For the model where water has
a constant abundance of 10$^{-6}$, emission from the warm dense inner
region still dominates and is optically thin (either due to the low
population of the highly excited states or the low abundance of
H$_2^{18}$O). A good example is the H$_2$O 5$_{24}$-4$_{31}$ 970 GHz
line shown in Fig. \ref{Fig:intensityexample} ({\it right}).

\begin{figure*}[!hpt]
\centering
\includegraphics[width=120pt]{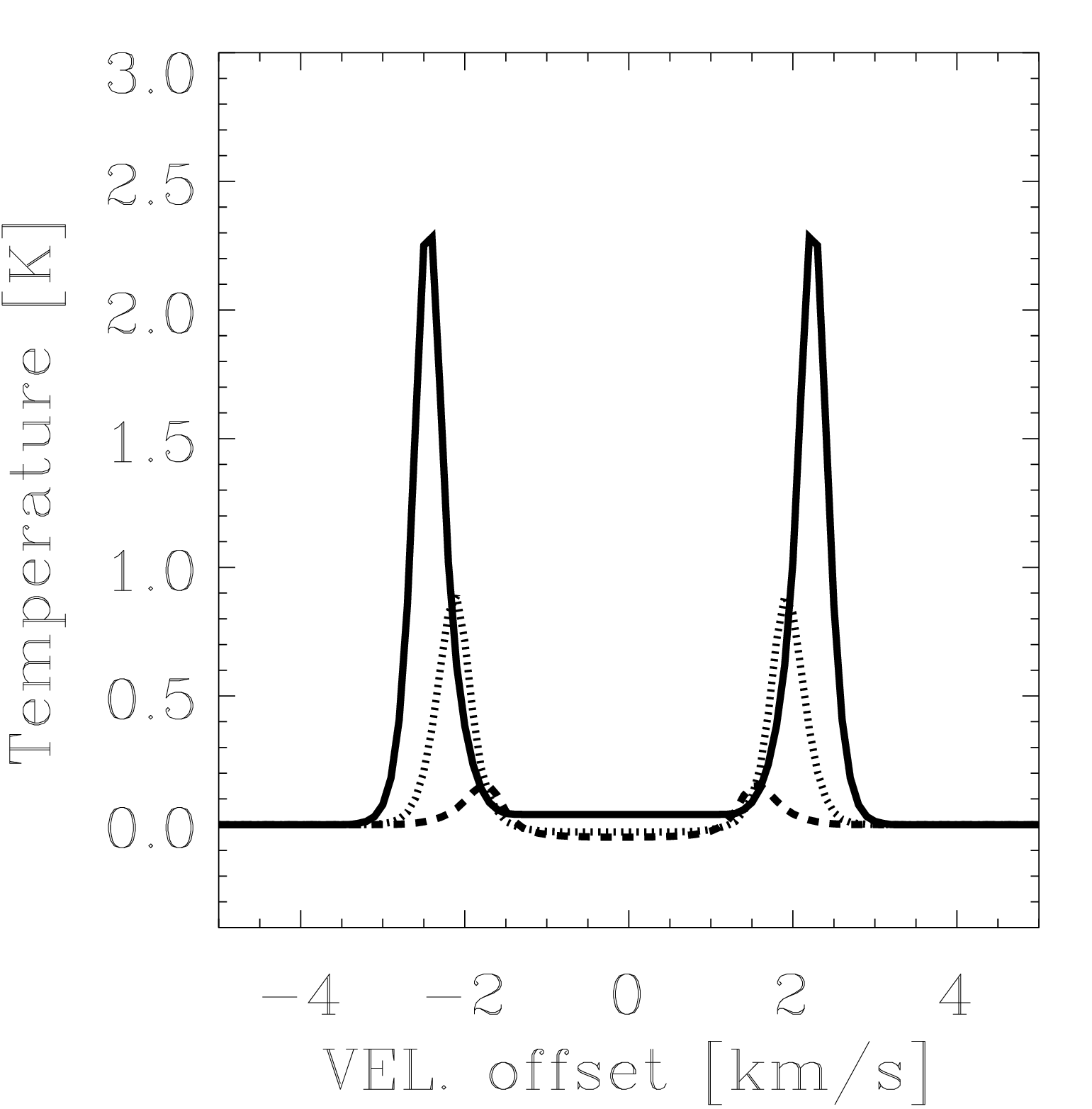}
\includegraphics[width=120pt]{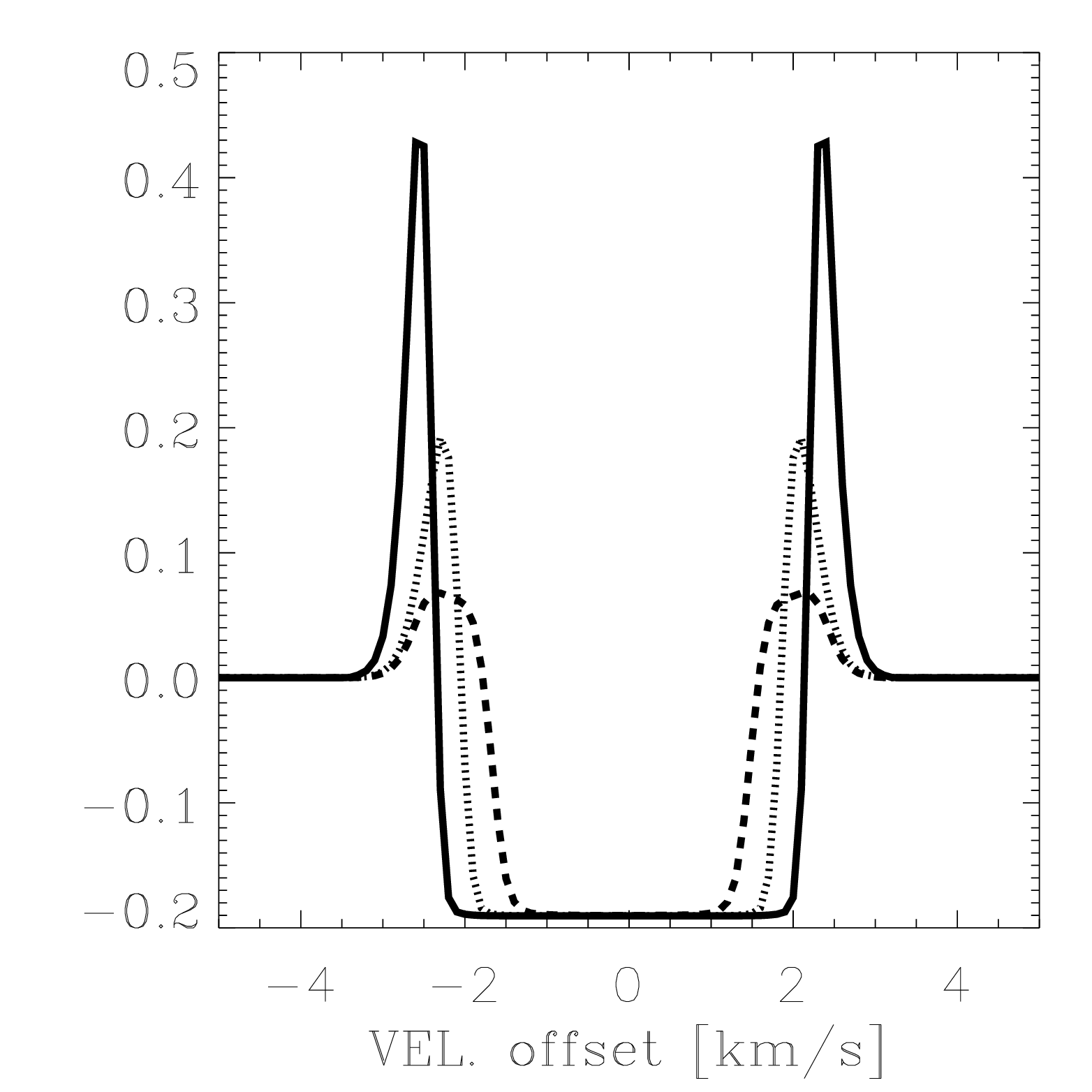}
\includegraphics[width=120pt]{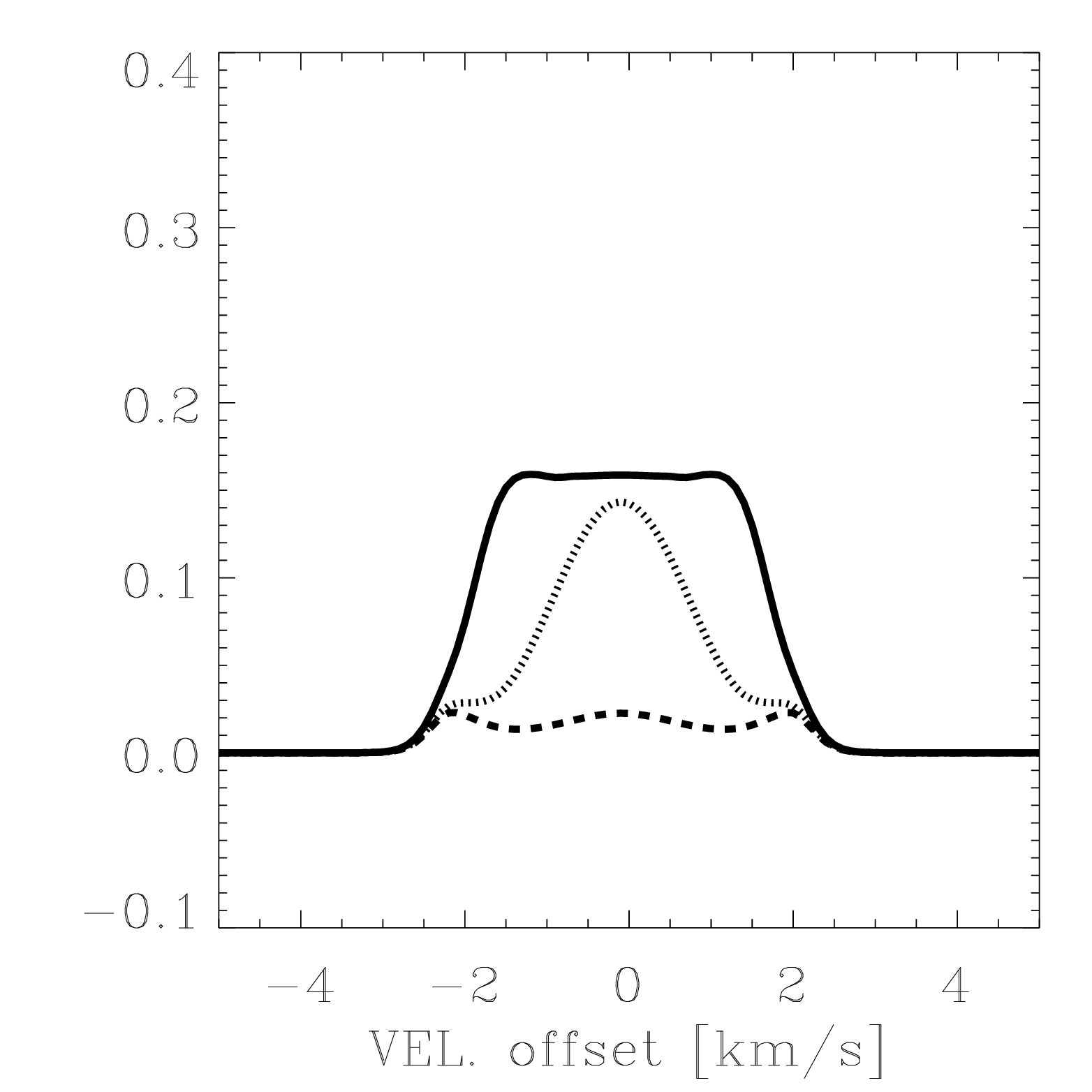}\\
\caption{H$_2$O spectra for an abundance $X_0$ of 10$^{-5}$ for the 1$_{10}$-1$_{01}$ (557 GHz, \textit{left}), 2$_{12}$-1$_{01}$ (1669 GHz \textit{middle}) and 3$_{12}$-3$_{03}$ (1097 GHz \textit{right}) lines. Lines are shown for $X_{\rm{d}}$ of 10$^{-6}$ (solid), 10$^{-7}$ (dot) and  10$^{-8}$ (dash). }
\label{Fig:examspec_L483}
\end{figure*}

\begin{figure*}[!hpt]
\centering
\includegraphics[width=120pt]{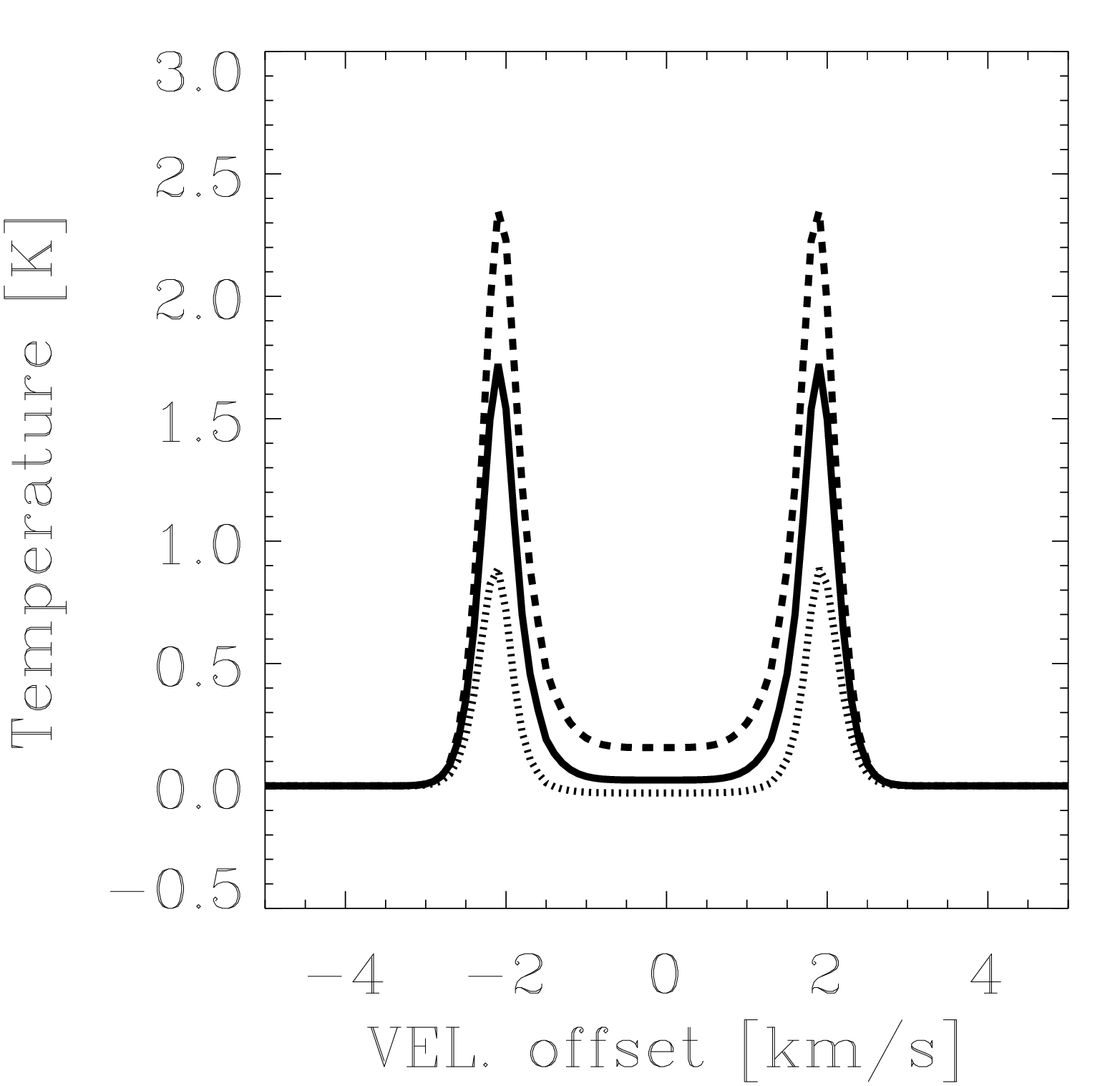}
\includegraphics[width=120pt]{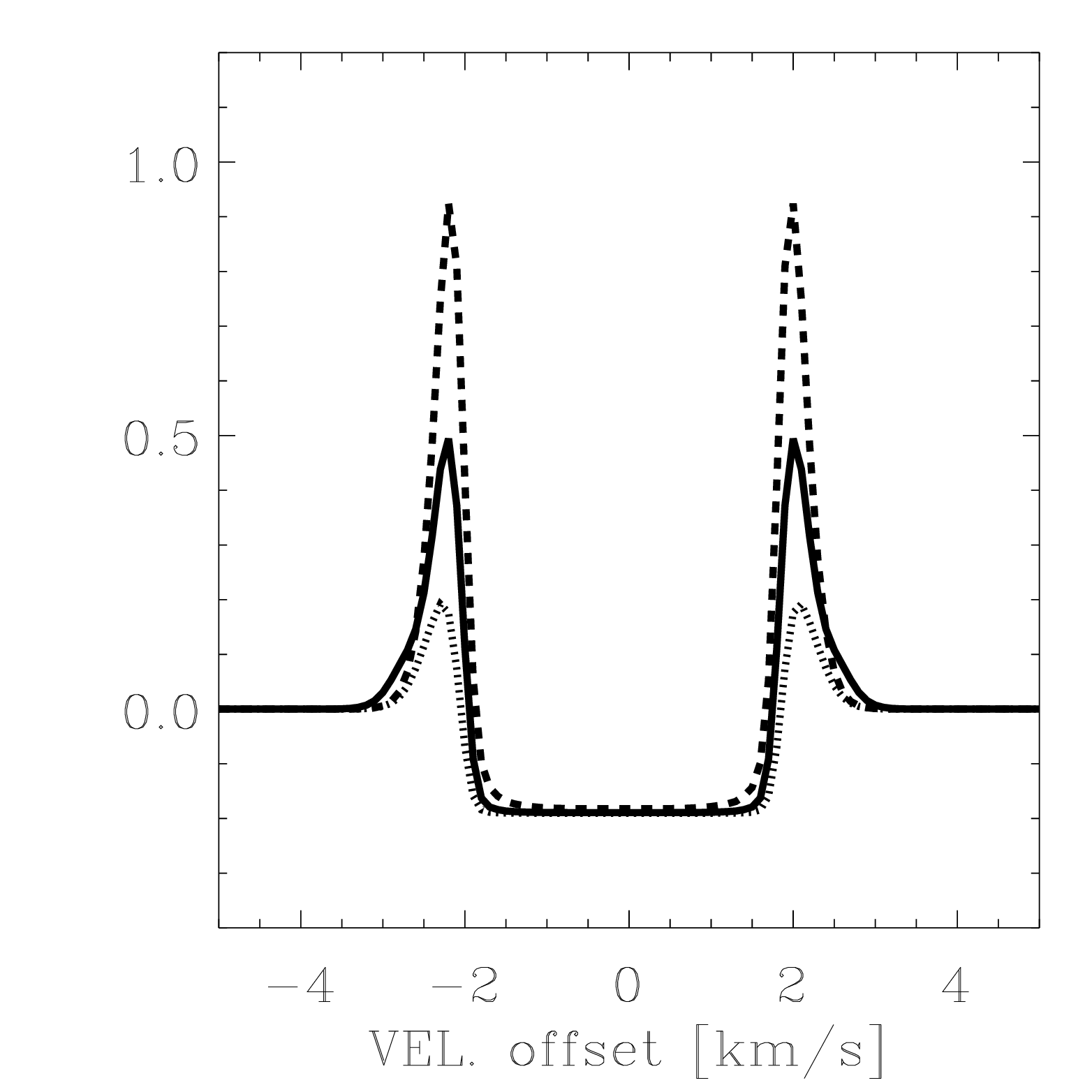}
\includegraphics[width=120pt]{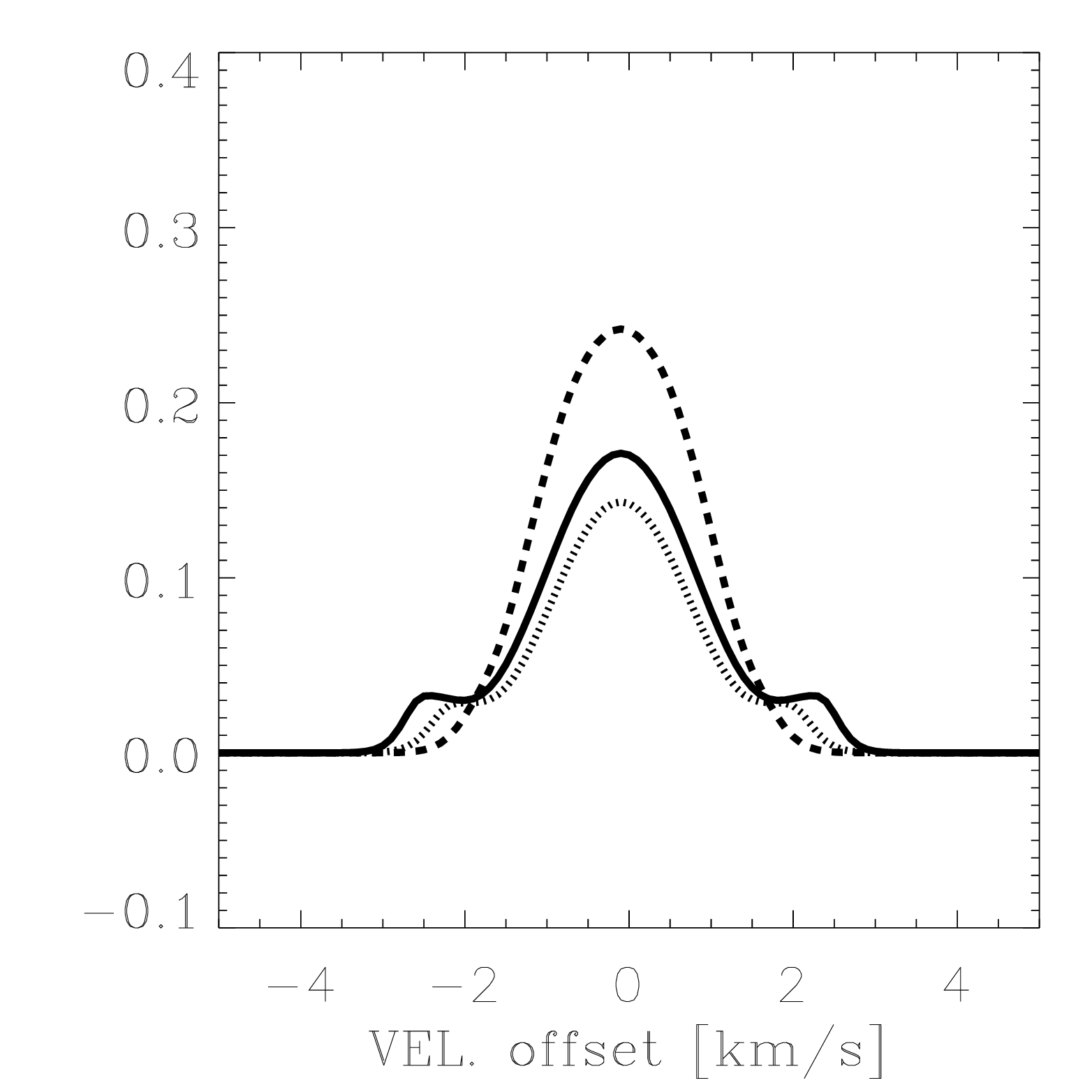}
\caption{H$_2$O spectra for an abundance $X_{\rm{d}}$ of 10$^{-7}$ for the 1$_{10}$-1$_{01}$ (557 GHz \textit{left}), 2$_{12}$-1$_{01}$ (1669 GHz \textit{middle}) and 3$_{12}$-3$_{03}$ (1097 GHz \textit{right}) lines. Lines are shown for $X_0$ of 10$^{-4}$ (solid), 10$^{-5}$ (dot) and  10$^{-6}$ (dash). }
\label{Fig:examspec2_L483}
\end{figure*}

The ground state lines of both H$_2^{18}$O and H$_2$O, together with
most H$_2$O excited lines with energies below 200~K, provide a strong
contrast with the optically thin lines. These lines only probe the
outer region. Even an abundance of water as low as 10$^{-8}$ in the
cold outer envelope produces a high enough column for these lines to
become optically thick. The extremely high optical depth ($\tau>25$)
and the absorptions into the dust further prevent the integrated
intensity from probing the entire envelope. Thus, these lines trace
the outer abundance $X_{\rm{d}}$, and self-absorption and absorption
of the dust continuum are commonly found in their line profiles.
Note, however, that most optically thick lines have wings that are
either optically thin or less optically thick.
These wings may be able to probe the inner region up to the
boundary layer around $T=100$ K. 
See the 1$_{10}$-1$_{01}$ line in Fig. \ref{Fig:intensityexample}
({\it left}), Table \ref{table:h2opeak} and the spectra in
Fig. \ref{Fig:examspec_L483} and \ref{Fig:examspec2_L483}
(discussed in \S 3.4).

Finally, some lines are optically thick for a high $X_0$-$X_{\rm{d}}$
combination, but are optically thin at low abundances. It is not known
a-priori if these lines originate within the cold outer envelope or
the warm dense inner regions. Examples are several of the highly
excited H$_2$O lines, but such dependencies are also seen for lowly
excited H$_2^{18}$O lines. Fig. \ref{Fig:intensityexample} ({\it
middle}) shows such an example with the p-H$_2^{18}$O 2$_{02}$-1$_{11}$ line. It can be seen that, depending on the
total column of water, this line either traces $X_{\rm{d}}$ at high
columns, or $X_0$ at lower columns.  For the H$_2^{18}$O lines a clear
dividing line between dependency on $X_0$ or $X_d$ lies at
$X_d(\rm{H}_2^{18}\rm{O})<10^{-9.5}$ (see online appendix). For
H$_2^{16}$O such a division is less clear.  Note also that the dust
can become optically thick, especially at high frequencies, so the
higher frequency lines will not be able to probe the inner region at
all. For the L~483 models, this only takes place at
frequencies higher than 1500 GHz, but see $\S$5 for further
discussion.  In the following, individual H$_2$O and H$_2^{18}$O lines
are discussed in more detail.


\subsection{Integrated emission}

\subsubsection{H$_2$O}

The main isotopologue of water, H$_2^{16}$O, is optically thick in
most transitions considered here. 
All lines with $J_{\rm{up}}$ $=$ 1 and 2 are optically thick and trace
the outer cold envelope.  Most other lines trace either the warm inner
or cold outer envelope.  The integrated intensities of these 'hybrid'
lines depend sensitively on the precise abundance parameters of the L
483 reference model. In particular, the radius where the line turns
optically thick is comparable to that where the temperature reaches
100 K for this model. These are the $2_{21}-2_{12}$, $3_{12}-2_{21}$,
$3_{12}-3_{03}$, $3_{21}-3_{12}$ and $4_{22}-4_{13}$ lines. At low
abundances of $X_{\rm{d}}$, they depend more on $X_0$. If the
abundance in the outer region is larger than 5$\times 10^{-8}$, they
depend on $X_{\rm{d}}$ instead. The integrated intensities range from
several hundreds of mK km s$^{-1}$ in absorption to a few K km
s$^{-1}$ in emission, with peak brightness from 100 mK to 1 K, readily
detectable with HIFI. The group of highly excited para-lines have
peaks up to a few tens of mK, requiring much longer integration times.

Dependencies of the line strengths on abundances are not linear. An
exception is if the line is completely optically thin over the entire
profile. For most lines this is not the case and parts of the line
profile will be optically thin and parts optically thick, causing
complicated dependencies.
For instance, the total integrated intensity of the 557 GHz
  line depends sensitively on the inner and outer
  abundances. Abundance combinations of 10$^{-4}$/10$^{-6}$ produce
  lines with less total emission than 10$^{-6}$/10$^{-6}$. The
  10$^{-6}$/10$^{-6}$ combination contains less water emission
  originating from the warm inner regions, but more dust continuum is able to escape the inner regions in the 10$^{-6}$/10$^{-6}$. This
  influences the population distribution in the outer region,
  resulting in less central absorption. Since absorption is a major
  feature within this line profile, the total integrated line
  intensity effectively goes up.

\subsubsection{H$_2^{18}$O}

Because of the low abundance of the H$_2^{18}$O isotopologue, most of
its lines are optically thin. Only the three transitions connecting to
the ground levels of ortho- and para-H$_2^{18}$O, $1_{11}-0_{00}$,
$1_{10}-1_{01}$ and $2_{12}-1_{01}$, have optical depths higher than 5
for all abundances. However, excited lines with $J_{\rm{up}}$=2 are
optically thick in their line centers for $X_0>10^{-5}$.
The integrated intensities of these H$_2^{18}$O lines range from a few
hundred mK km s$^{-1}$ in absorption to a few tens of mK km s$^{-1}$ in
emission, whereas peak temperatures range from a few mK to a few tens
of mK, thus requiring long integration times.  Due to the low
abundance, the dependence on either $X_0$ or $X_{\rm{d}}$ is much more
evident than for H$_2^{16}$O (see Fig.\ref{Fig:intensityexample} for
example). The optically thick ground-state lines accurately trace
$X_{\rm{d}}$, while optically thin excited lines depend only on
$X_0$. The few optically thick excited lines trace $X_0$ as well,
since the $\tau=1$ surface lies within the 100 K radius.
Higher excitation H$_2^{18}$O lines are not very strong, in the range
of a few micro-K to a few milli-K, too weak to be detectable with HIFI
in reasonable integration times.

\subsection{Line profiles}

In Figures \ref{Fig:examspec_L483} and \ref{Fig:examspec2_L483}, a
number of characteristic line profiles are presented.  The continuum
has been subtracted so that the profiles can be readily compared on
the same scale.

The total profile contains valuable information about the envelope
structure and velocity profile, none of which can be determined using
the integrated line intensity alone. As seen in the examples, there is
a large variety in model profiles depending on a number of
parameters. First, the excitation energy of the upper state, the total
water column density and the critical density of the line in question
will determine if an emission line is optically thick, thin or a
combination of both. Most lines show a combination, with optically
thin wings around an optically thick line center. Second, the width
depends on the systematic and turbulent velocity. Third, water in the
colder outer envelope will absorb photons and scatter the radiation,
causing an absorption feature. Finally, the optical depth of dust
itself blocks line emission at frequencies above 1500 GHz originating
in the inner region. In the following these profile shapes are
discussed in more detail. \\

\begin{table*}[!htp]
\caption{Peak temperatures [K] of H$_2$O lines for L~483 models of
various abundances. Negative values indicate primarily
absorption into the dust continuum. 
}
\begin{center}
\label{table:h2opeak}
\begin{tabular}{c c c c c c c c c c}
\hline \hline
 Transition & \multicolumn{3}{c}{X$_0=10^{-4}$}&\multicolumn{3}{c}{X$_0=10^{-5}$}&\multicolumn{3}{c}{X$_0=10^{-6}$} \\\cline{3-3} \cline{6-6} \cline{9-9}
 X$_d =$ & 10$^{-6}$ & 10$^{-7}$ & 10$^{-8}$ & 10$^{-6}$ & 10$^{-7}$ & 10$^{-8}$ & 10$^{-6}$ & 10$^{-7}$ & 10$^{-8}$ \\  \hline
\multicolumn{9}{c}{Ortho-H$_2$O transitions}   \\ \hline
1$_{10}$-1$_{01}$ & 2.8 & 2.1 & 0.65 & 2.8 & 1.1 & 0.2 & 5.4 & 2.8 & 0.5 \\
2$_{12}$-1$_{01}$ & 0.45 & 0.35 & 0.25 & 0.3 & -0.13 & -0.13 & 1.9 & -0.13 & 0.25 \\
2$_{21}$-2$_{12}$ & 1.1 & 0.1 & 0.05 & 0.2 & 0.1 & 3.8(-2) & 1.6 & 0.21 & 4.5(-2) \\
3$_{12}$-2$_{21}$ & 0.28 & 6.5(-2) & 2.4(-2) & 0.18 & 4.5(-2) & 2.1(-2) & 0.36 & 0.1 & 2.4(-2) \\
3$_{12}$-3$_{03}$ & 0.27 & 0.12 & 4.1(-2) & 0.12 & 0.1 & 1.5(-2) & 0.4 & 0.17 & 2.8(-2) \\
3$_{21}$-3$_{12}$ & 0.5 & 0.18 & 4.1(-2) & 0.33 & 0.13 & 1.5(-2) & 0.55 & 0.26 & 3.2(-2) \\ \hline
\multicolumn{9}{c}{Para-H$_2$O transitions } \\ \hline
1$_{11}$-0$_{00}$ & 1.5 & 1.3 & -9(-2) & 0.45 & 0.45 & 0.45 & 1.6 & 1.25 & 0.13 \\
2$_{02}$-1$_{11}$ & 1.6 & 0.55 & 8(-2) & 1.1 & 0.4 & 0.18 & 0.9 & 1.0 & 8(-2) \\
2$_{11}$-2$_{02}$  &0.8 & 0.22 & 7(-2) & 0.38 & 0.18 & 0.16 & 0.45 & 0.38 & 7.5(-2)\\
2$_{20}$-2$_{11}$ & 0.4 & 0.28 & 7.5(-2) & 0.12 & 0.24 & 0.17 & 0.1 & 0.35 & 7(-2) \\
3$_{31}$-4$_{04}$ & 2.8(-2) & 2.2(-2) & 2.2(-2) & 1(-2) & 1(-2) & 1.9(-2) & 2.5(-3) & 1.5(-3)&8(-4) \\
4$_{22}$-3$_{31}$ & 2.8(-2) & 2(-2) & 1(-2) & 2.3(-2) & 2.4(-2) & 1.5(-2) & 1.2(-2) & 1(-2) & 6(-3) \\
4$_{22}$-4$_{13}$ & 0.16 & 2.4(-2) & 1.4(-2) & 0.12 & 2.7(-2) & 1.7(-2) & 0.12 & 4.5(-2) & 2.2(-2) \\
5$_{24}$-4$_{31}$ & 2.2(-2) & 2.6(-2) & 1.6(-2) & 1.8(-2) & 1.9(-2) & 2.4(-2) & 2.5(-3) & 3.5(-3) & 2(-3) \\ \hline
\end{tabular}
\end{center}

\end{table*}

The profiles of optically thin lines are Gaussian and are good tracers
of the velocity profile. 
Optically thin lines always trace the inner
warm region, where the higher excitation states are populated. The
ratio of the line wing over the peak temperature can vary, depending
on the excitation energy of the line. Higher excitation lines are
broader, while low excitation lines have a narrower peak with very
weak line wings.  An example in Figure \ref{Fig:examspec2_L483} is the
3$_{12}$-3$_{03}$ 1097 GHz line, which is completely optically thin
for $X_{\rm{0}}$=10$^{-6}$. The line center becomes optically thick at
$X_{\rm{0}}$=10$^{-5}$ and thus decreases, but the optically thin line
wings are noticeably higher in emission. At $X_{\rm{0}}$=10$^{-4}$, the
optically thick line center is brighter than at 10$^{-5}$ due to the
high abundance, but still weaker than the optically thin emission at
$X_{\rm{0}}$=10$^{-6}$.

The profiles of optically thick lines are much more
complicated.  
Their wings can either be optically
thick or thin, depending on the abundances and frequency of the
line. This gives rise to three different profiles.

\begin{itemize}
\item Lines with a Gaussian line center, but with enhanced line
wings. A characteristic `bump' can be seen at velocities where the
line profile changes from optically thick to thin when going to more
extreme velocities. The line center is optically thick, but is heavily
pumped by the line radiation coming from water in the warmer regions
further inwards in the envelope.  This behaviour is often called
`effectively optically thin'. 
Such lines trace the inner abundance $X_0$, unless optically thick
dust veils the inner region.  An example is seen by the
$3_{12}-3_{03}$ 1097 GHz line with an $X_{\rm{0}}$ of 10$^{-4}$ in
Figure \ref{Fig:examspec2_L483}.

\item 'Flat top' lines, where the center of the line is optically
thick, but not self-absorbed. The center of the line is dominated by
emission from a small region with a specific temperature. Line wings
are optically thin. Material at larger radii is unable to cause
self-absorption and there is not enough material deeper within the
envelope to make such a line `effectively optically thin'.  Optically
thick dust can prevent warmer regions from influencing the line at
higher frequencies. An example is given by the $3_{12}-3_{03}$ 1097 GHz
line with $X_d=10^{-6}$ in Figure \ref{Fig:examspec_L483}.

\item Deeply (self)-absorbed lines with most of the emission in the
line wings. Absorption into the dust occurs, and is stronger for high
frequency lines, where the dust continuum is higher.
All three ground state lines show such profiles. Examples are the
1$_{10}$-1$_{01}$ and 2$_{12}$-1$_{01}$ lines in
Fig. \ref{Fig:examspec_L483} and \ref{Fig:examspec2_L483}. Both the
depth and the width of the absorption vary, as do the line wing
strengths:

(i) Absorption depth: If dust is optically thick (e.g., at 1669 GHz),
the depth is independent of both $X_0$ and $X_{\rm{d}}$ as long as
$\tau>>1$, and depends only on the radius where the dust turns
optically thick, setting the continuum level. In Figure
\ref{Fig:examspec_L483} and \ref{Fig:examspec2_L483} ({\it middle})
the 1669 GHz continuum-subtracted line is always at -0.2 K. If dust is
optically thin (e.g., the 557 GHz 1$_{10}$-1$_{01}$ line), the depth
v of the absorption depends on both $X_d$ and $X_0$ and is due to
self-absorption of the water in the cold outer region against the warm
emission.  Indeed, it is found that water at around 70--90 K acts as a
warm background emitter and water at 20 K as the cold absorber. \\

(ii) Absorption width: again there is a difference between high and
low optical depths of dust. For 1669 GHz the width of the absorption
depends on $X_{\rm{d}}$, but not on $X_0$. For 557 GHz, there is no
dependence on abundance. The width of the absorption is the same for
varying $X_0$ (the temperature at which the water emits changes
little), and even though the profiles appear different for different
$X_{\rm{d}}$, this can be attributed to the changing absorption depth,
not width. \\

(iii) Line wing strength: the strength of the line wing, which often
contains the bulk of the line emission, is determined by the optical
depth of the dust and gas. Most often, the peaks seen in the 
wings of the ground-state lines are still optically thick. The wings
trace both $X_{\rm{d}}$ (see 1$_{10}$-1$_{01}$ and 2$_{12}$-1$_{01}$
lines in Fig. \ref{Fig:examspec_L483}, left, middle), which still
contain large columns of the gas, as well as $X_0$ (1$_{10}$-1$_{01}$
and 2$_{12}$-1$_{01}$ Fig. \ref{Fig:examspec2_L483}, left, middle),
although in the latter case the relation is not straight forward.
Line pumping from the warm gas further inwards is the main origin of
the dependency from $X_0$=$10^{-4}$ to $10^{-5}$. However, if
$X_{\rm{d}}$ and $X_0$ are low enough the line wings turn optically
thin and are significantly enhanced (Fig \ref{Fig:examspec_L483},
dashed lines) because higher temperatures are probed. Such behaviour
is seen for a $X_{\rm{d}}$/$X_0$ combination of 10$^{-7}$/10$^{-6}$
where the line wings are brighter than for $X_{\rm{d}}$/$X_0$
combinations of 10$^{-7}$/10$^{-4}$ and 10$^{-7}$/10$^{-5}$. 
\end{itemize}

\subsection{HIFI diagnostic lines}
Although the beams for HIFI and the accompanying dilution
  factors are large compared to the physical size of the warm gas (see
  Table \ref{table:lines}), the inner warm region is detectable
  through the higher excited states. The outer cold envelope is
  completely transparent for such lines.  Considering the expected
sensitivities of HIFI, a combination of different lines can constrain
the water abundance profile. The H$_2^{18}$O 2$_{02}$-1$_{11}$ 994 GHz
line in HIFI Band 4 is mostly optically thin and traces $X_0$, except
at H$_2^{16}$O $X_{\rm{d}}$ abundances higher than
5$\times10^{-6}$. However, the line is weak (a few tens of mK) and
needs at least a few hours of integration to be detected. A possible
alternative is the high excitation H$_2^{16}$O 3$_{12}$-3$_{03}$ line
in HIFI Band 4, which can be detected in roughly an hour of
integration. At high $X_0$ abundances it traces $X_0$, but for $X_0$
lower than 10$^{-5}$, emission coming from the colder outer part
dominates, thus tracing $X_{\rm{d}}$. This emission originates in the
region with temperatures ranging from 70--100 K. Once the inner
abundance $X_0$ is well-constrained, ground-state lines can in turn be
used to constrain the abundance in the outer region. The bright wings
of the 557 GHz line (HIFI Band 1) are detectable within a few minutes,
but the entire line profile is needed for proper analysis (see \S
3.4). The dependence on $X_{\rm{d}}$ can be traced most accurately
using the line profiles of the 1$_{11}$-0$_{00}$ 1113 GHz para-H$_2$O
and 1101 GHz para-H$_2^{18}$O lines in Band 4, but resolving these
profiles requires much longer integration times than the 557 GHz line.
The 1667 GHz 2$_{12}$-1$_{01}$ ortho-H$_2$O line in Band 6 is intrinsically
brighter, but will take almost an hour of integration due to the low
sensitivity of HIFI at this frequency. In addition, dust can become
optically thick at this high frequency so that the line profile only
traces the structure in the outer envelope.

The optically thick H$_2^{16}$O 2$_{02}$-1$_{11}$ 988 GHz line, with
its optically thin H$_2^{18}$O counterpart are the best candidates to
determine the optical depth of water, but the H$_2^{18}$O lines
requires significant integration time (see above). An alternative is
provided by comparison between of the 548 GHz H$_2^{18}$O and 557 GHz
H$_2^{16}$O ground-state lines, but H$_2^{18}$O is still optically
thick so that detailed modeling is required.  The 1$_{11}$-0$_{00}$
1113 GHz, 2$_{02}$-1$_{11}$ 988 GHz and 2$_{11}$-2$_{02}$ 752 GHz
para-H$_2$O lines are well-suited to probe the physical structure of
the envelope: due to their different excitation energies and optical
depths, these lines probe different environments within the
protostellar envelope and their line ratios are not affect by the
uncertain ortho/para ratio.

\subsection{Comparison with ISO-LWS data}

L~483 was observed with ISO-LWS \citep{Giannini01}. The para
2$_{20}$-1$_{11}$ line at 2968 GHz (101 $\mu$m) was detected with a
flux of 8.6$\pm 1.0$ $\times$ 10$^{-20}$ W cm$^{-2}$ and the ortho
2$_{12}$-1$_{01}$ line at 1670 GHz (179.5 $\mu$m) was found to have an
upper limit of 4.3 $\times$ 10$^{-20}$ W cm$^{-2}$. The low LWS
resolving power of $\lambda/\Delta\lambda = 200$ ($\sim$ 1500 km
s$^{-1}$) was insufficient to determine the effect of any
self-absorption, or to disentangle outflow components from quiescent
envelope material. The best fitting LVG model for the high-$J$ CO
lines gave a component of gas with $T$=850--1800 K; it is expected
that such hot gas contributes significantly to the water emission
\citep{Giannini01}.

The above models, corrected for the ISO beam of 80$''$, are compared
to the observed intensity. Our models show a strong absorption into
the dust for both lines with optically thin line wings. Integrated
intensities range from -7 $\times$ 10$^{-22}$ up to 17 $\times$
10$^{-20}$ W cm$^{-2}$ for the 2$_{20}$-1$_{11}$ line and from 0.9
$\times$ 10$^{-20}$ W cm$^{-2}$ to 16 $\times$ 10$^{-20}$ W cm$^{-2}$
for the ortho 2$_{12}$-1$_{01}$ line for different $X_0$-$X_{\rm{d}}$
combinations. The best fit, using both the detection and upper limit,
is found for an outer abundance of 10$^{-7}$ and an inner abundance
between 10$^{-6}$ and 10$^{-5}$. The detected line is reproduced
within 30\%, while the upper limit is only 50\% higher than the
intensity produced by the model. For lower values of $X_{\rm d}$, of
order 10$^{-8}$, the integrated emission in our model is almost an
order of magnitude lower than the observed 2$_{20}$-1$_{11}$
line. Such low outer abundances are found for cold regions in the
envelopes around high-mass YSOs \citep{Boonman03}, but appear
inconsistent with the data for this low-mass YSO, unless the detected
emission originates fully in then outflow. Higher inner abundances
cannot be ruled out. However, the self-absorption in the models
becomes stronger, resulting in a worse fit to the detected line and
upper limit.

Other water lines within the ISO-LWS domain were not detected. The
derived upper limits were found to be at least a factor 5 higher than
fluxes predicted from all possible models. At short ($>$ 80 $\mu$m)
wavelengths, the difference is a few orders of magnitude.  Since all
lines are unresolved, no conclusion can be reached whether the
detected water emission is dominated by outflow contributions or by
thermal excitation within the circumstellar envelope with a high
abundance. This example illustrates the need for spectrally and
spatially resolved data of a number of lines to disentangle the
different physical components.

\begin{table*}[!htb]
\caption{Integrated intensities, $\int T_{\rm{MB}}\Delta V$ [K km
    s$^{-1}$], for H$_2$O lines in the wide parameter grid for an
  abundance combination of $X_0=10^{-4}$ and $X_{\rm d}=10^{-7}$. Column 2 to 4 show the luminosity variation  with a density of 10$^6$ cm$^{-3}$. Column 5 to 7 show the variation of density with a luminosity of 7 $ \rm{L}_{\odot}$ }
\begin{center}
\label{table:h2ogrid_test}
\begin{tabular}{c c c c c c c}
\hline \hline
Transition   &  \multicolumn{6}{c}{Integrated Intensity ( $\int T_{\rm{MB}}\Delta V$ [K km
    s$^{-1}$])} \\ \hline
& \multicolumn{3}{c}{Lum. (w. $n_0$ = 10$^6$ cm$^{-3}$)} & \multicolumn{3}{c}{Dens. (w. $L= 7$ L$_{\odot}$)} \\ 
 & 2 & 7 & 25 & 0.4 & 1 & 5\\ \hline
& \multicolumn{6}{c}{$p=1.5$} \\ \hline
\multicolumn{7}{c}{Ortho-H$_2$O transitions}\\ \hline
1$_{10}$-1$_{01}$ & 0.37 & 0.95 & 0.70 &0.43 & 0.95 & 2.1\\ 
2$_{12}$-1$_{01}$ & 0.35 & 0.36 &-1.2 & 0.30& 0.36  & -4.4\\
2$_{21}$-2$_{12}$ & 0.29 & 0.87 & 1.5 & 0.60& 0.87  & 0.74\\
3$_{12}$-3$_{03}$ & 0.21 & 0.60 & 1.5 & 0.38& 0.60  & 0.89\\
3$_{12}$-2$_{21}$ & 0.21 & 0.67 & 1.6 & 0.40& 0.67  & 0.84\\
3$_{21}$-3$_{12}$ & 0.37 & 1.1  & 2.7 & 0.59& 1.1   & 1.7\\ \hline
\multicolumn{7}{c}{Para-H$_2$O transitions}\\ \hline
1$_{11}$-0$_{00}$ & 0.13 & 1.0& 0.31& 0.43&1.0 & 1.6\\
2$_{02}$-1$_{11}$ & 0.58 & 1.5& 3.0  & 0.72&  1.5&1.1 \\
2$_{11}$-2$_{02}$ & 0.34 & 1.0& 2.0 & 0.41& 1.0& 2.7\\
2$_{20}$-2$_{11}$ & 0.31 & 1.0& 2.3 & 0.47& 1.0& 1.4\\
3$_{31}$-4$_{04}$ & 4.3(-2) &0.15 & 0.58 & 0.18 & 0.15 & 1.3(-3)\\
4$_{22}$-3$_{31}$ & 4.1(-2) & 0.13& 0.48&0.13 &0.13 & 9.6(-2)\\
4$_{22}$-4$_{13}$ & 9.5(-2) & 0.26& 0.74& 0.18& 0.26 &2.3(-3) \\
5$_{24}$-4$_{31}$ &  5.0(-2)& 0.15& 0.57& 0.16& 0.15& 8.7(-2)\\ \hline 
  & \multicolumn{6}{c}{$p=2.0$} \\ \hline
\multicolumn{7}{c}{Ortho-H$_2$O transitions}\\ \hline
1$_{10}$-1$_{01}$ & 0.12 & 0.24& 0.19& 0.10 &  0.24& 0.82\\ 
2$_{12}$-1$_{01}$ & 0.89 &-1.6 &-3.1 & 0.71 & -1.6 & -5.0\\
2$_{21}$-2$_{12}$ & 0.13 & 0.14& 0.15& 0.13 &  0.14& -1.4\\
3$_{12}$-3$_{03}$ & 0.12 & 0.47 & 1.3 & 0.35 &  0.47 & 0.51\\
3$_{12}$-2$_{21}$ & 0.13 & 0.48& 1.2 & 0.29 &  0.48& 0.49\\
3$_{21}$-3$_{12}$ & 0.20 & 0.87& 2.1 & 0.45 &  0.87& 1.2\\ \hline
\multicolumn{7}{c}{Para-H$_2$O transitions}\\ \hline
1$_{11}$-0$_{00}$ & 0.30 & 9.4(-2)& 0.23& 0.52& 9.4(-2) & 9.0(-2)\\
2$_{02}$-1$_{11}$ & 0.65 & 1.4& 3.1& 0.90& 1.4& 1.8\\
2$_{11}$-2$_{02}$ & 0.46 & 0.94& 2.1     & 0.56& 0.94& 2.2\\
2$_{20}$-2$_{11}$ & 0.32 & 0.78& 2.1     & 0.54& 0.78& 1.3\\
3$_{31}$-4$_{04}$ & -3.7(-3) & -6.7(-3) & 0.13 & 7.1(-2)& -6.7(-3) & -1.2(-3) \\
4$_{22}$-3$_{31}$ & 1.0(-2) &5.9(-2) &0.31 & 9.3(-2) & 5.9(-2)& 2.3(-3)\\
4$_{22}$-4$_{13}$ & 4.7(-2) & 0.19 & 0.74 & 0.22& 0.19& 8.1(-2)\\
5$_{24}$-4$_{31}$ & 8.3(-3) & 6.5(-2)& 0.34 & 0.11 & 6.5(-2)& -6.9(-3)\\ \hline \hline
\end{tabular}
\end{center}
\end{table*}

\section{General parameter study}

\begin{figure*}[!htb]
\centering
\includegraphics[width=120pt]{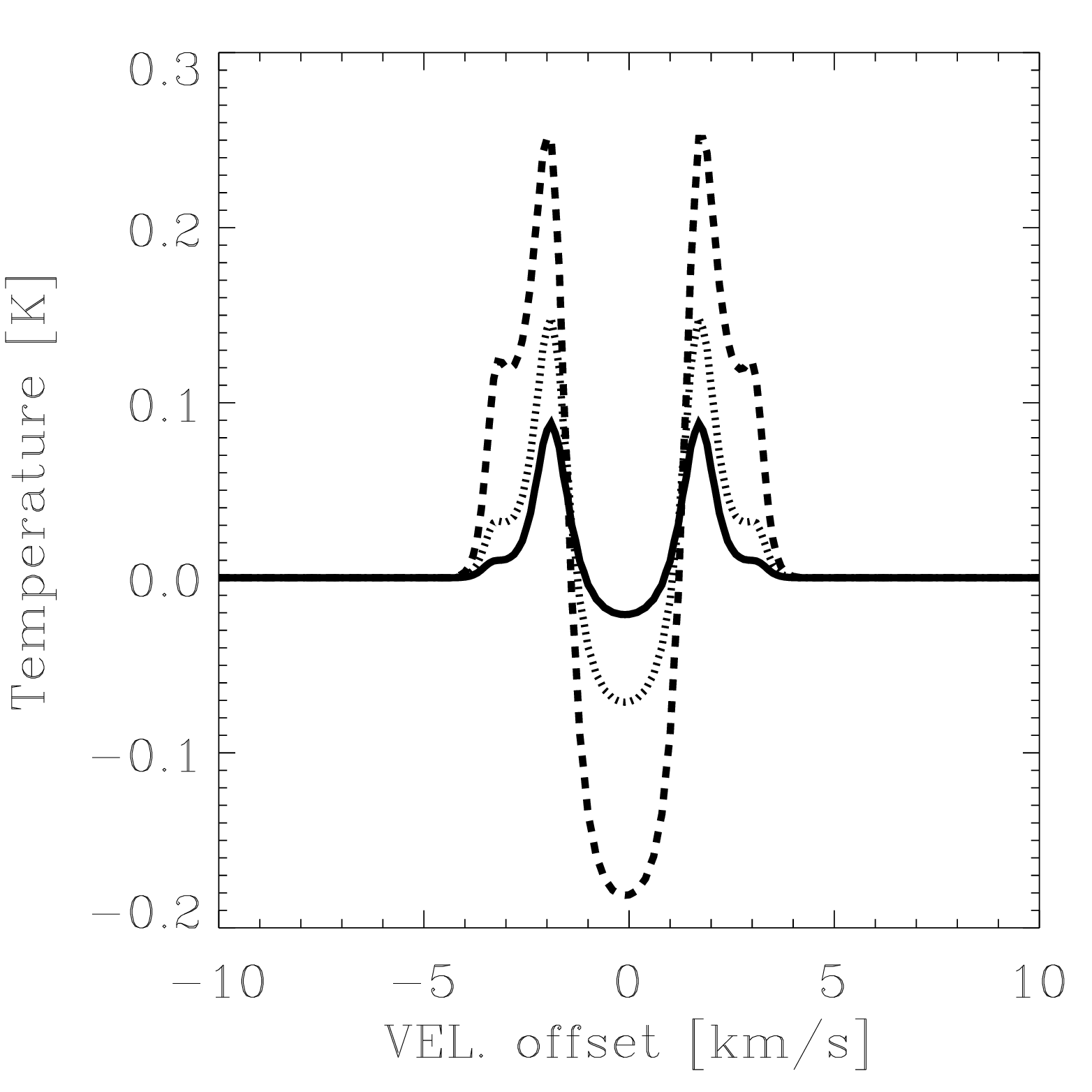}
\includegraphics[width=120pt]{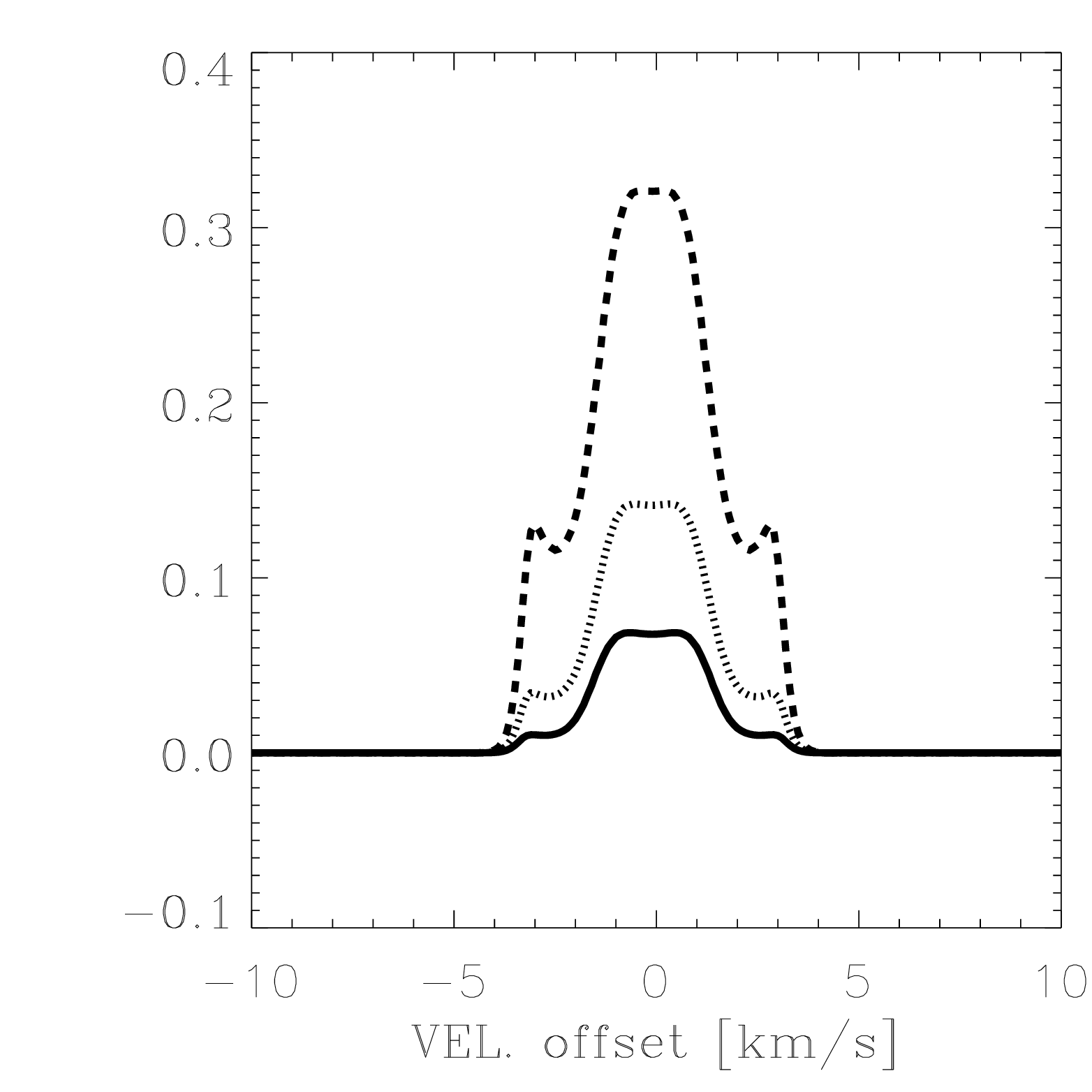}
\includegraphics[width=120pt]{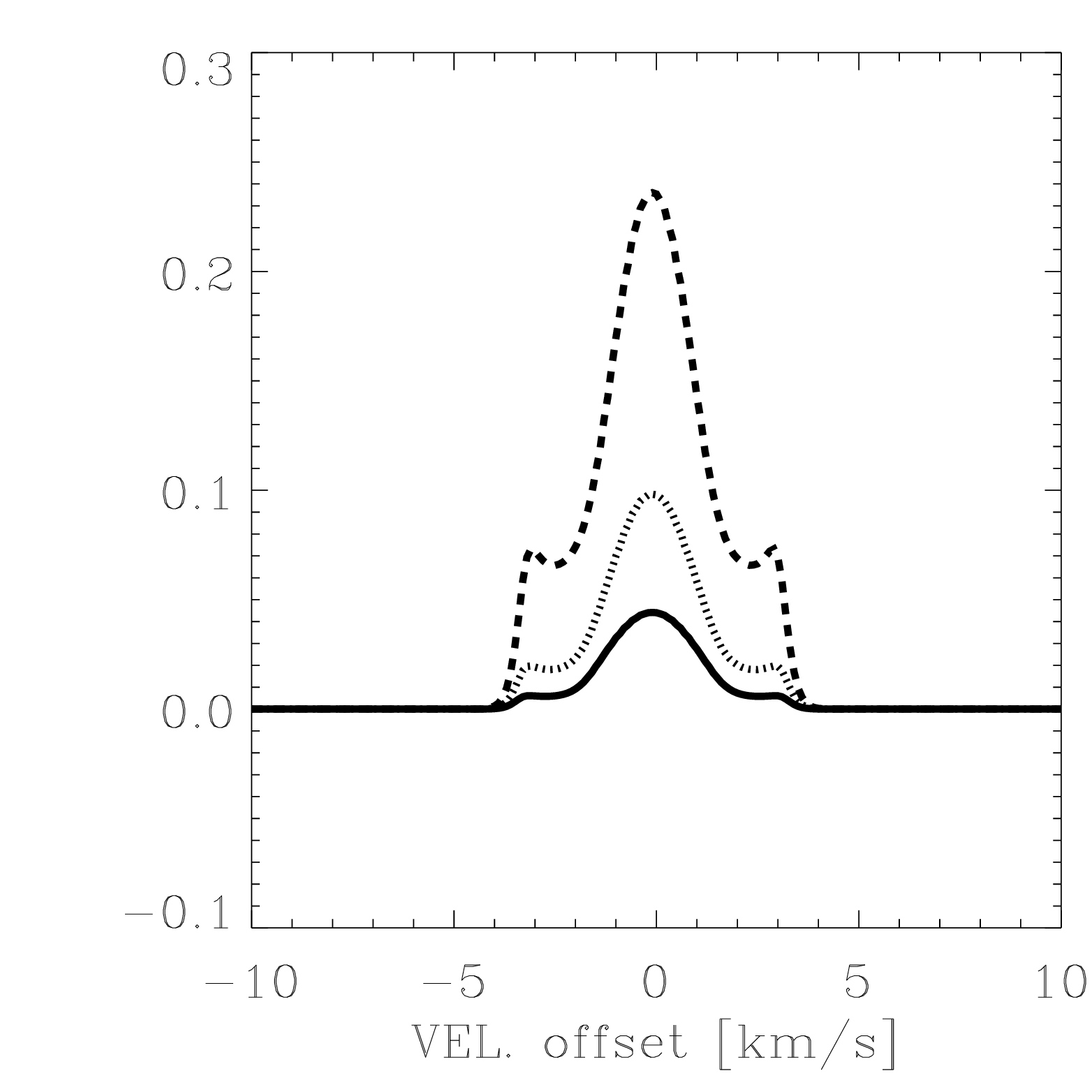}
\includegraphics[width=120pt]{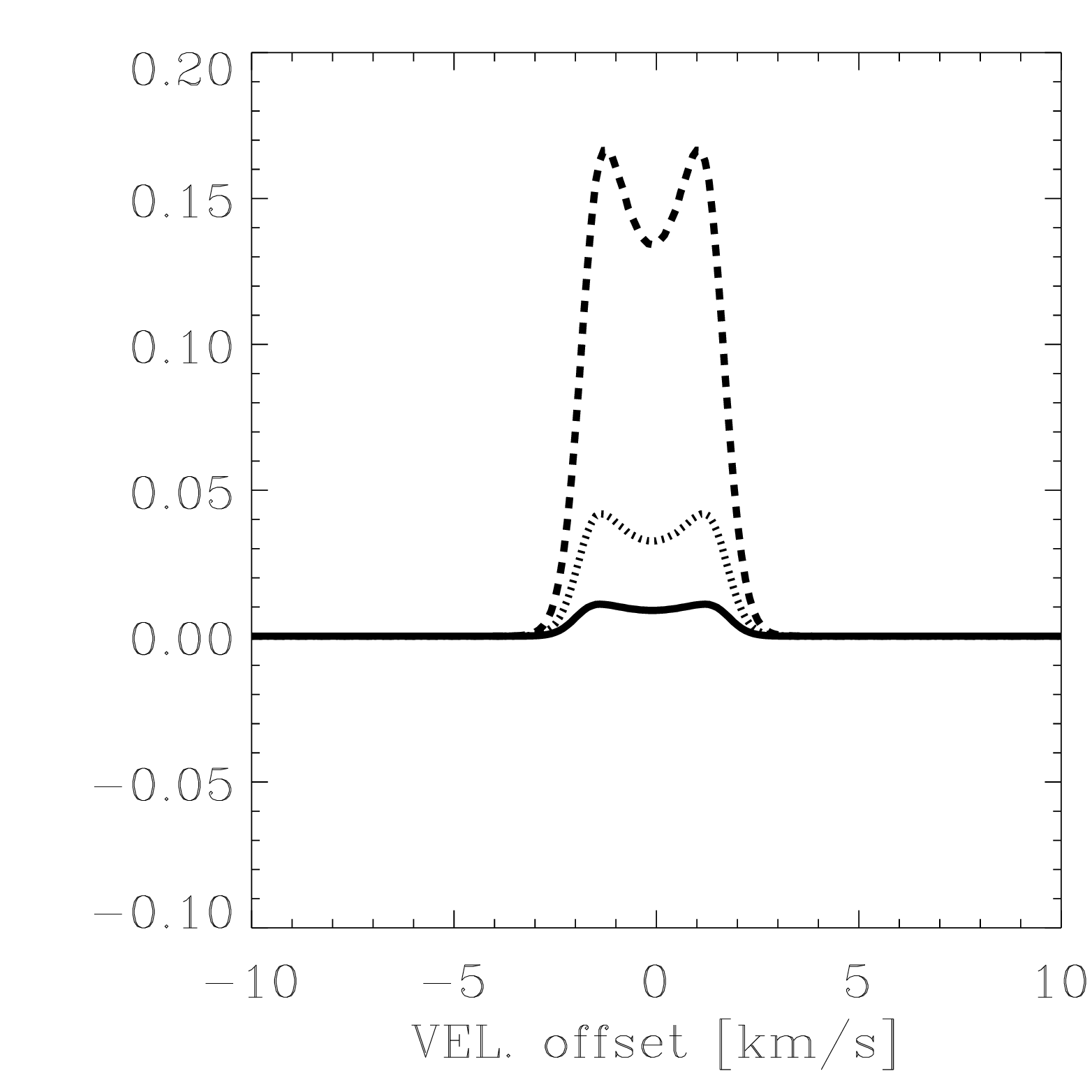}
\caption{Examples of the luminosity dependence of line profiles for a
model with $p$=1.5, $X_0$=10$^{-4}$, $X_{\rm{d}}$=10$^{-8}$,
$n_0({\rm{H}_2})$=10$^6$ cm$^{-3}$. From left to right are shown the
1$_{11}$-0$_{00}$ 1113 GHz, 2$_{02}$-1$_{11}$ 988 GHz,
2$_{11}$-2$_{02}$ 752 GHz and 3$_{31}$-4$_{04}$ 1893 GHz
lines. Luminosities are 2 (solid), 7 (dot) and 25 (dash)
L$_{\odot}$.  These lines reflect the different characteristic line
profiles discussed in $\S$ 3.4.  It can be seen that the line shapes
stay the same with increasing luminosity, while the line strengths and
depths of the absorption are stronger at higher luminosities.  }

\label{Fig:Lumspec}
\end{figure*}

\begin{figure*}[!htb]
\centering
\includegraphics[width=120pt]{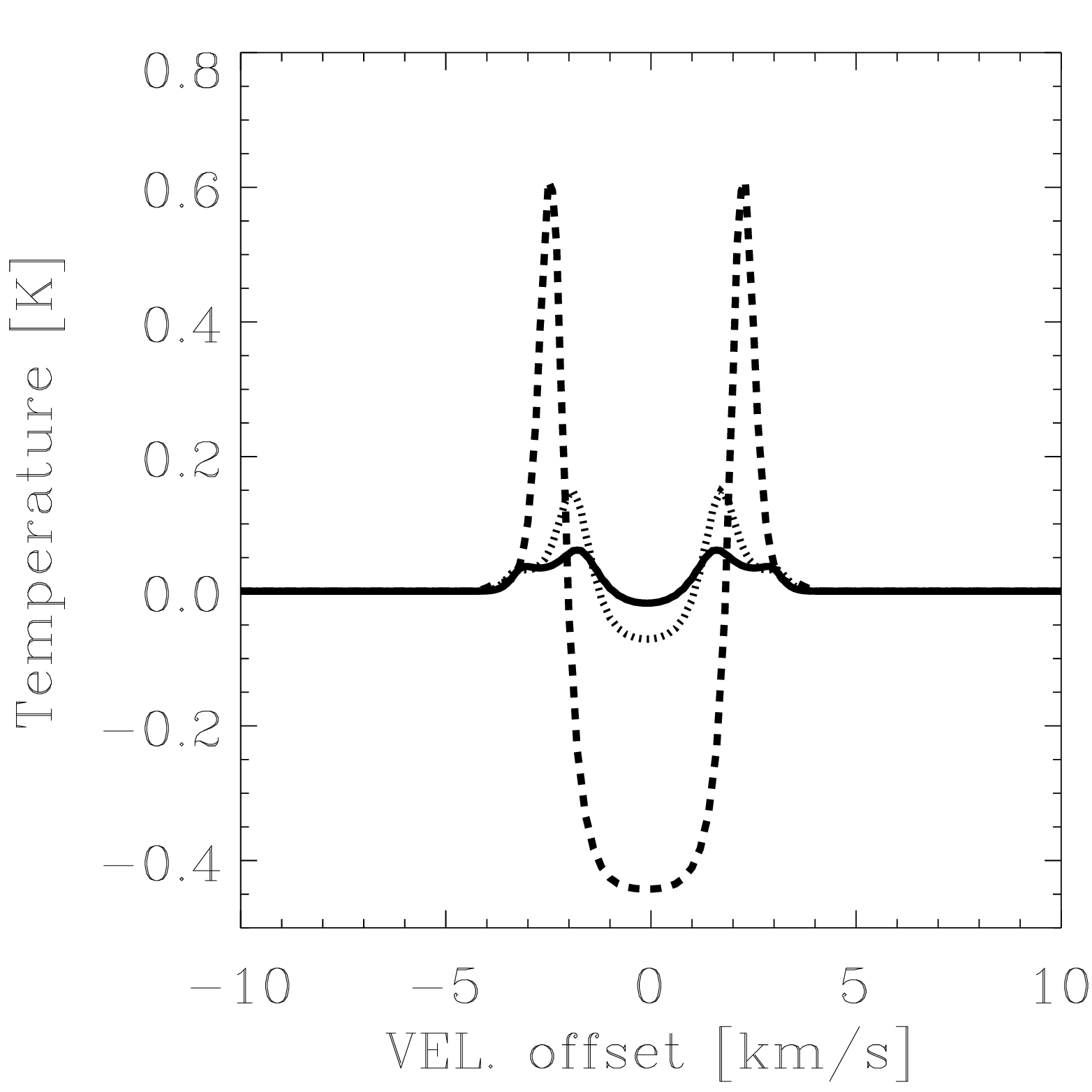}
\includegraphics[width=120pt]{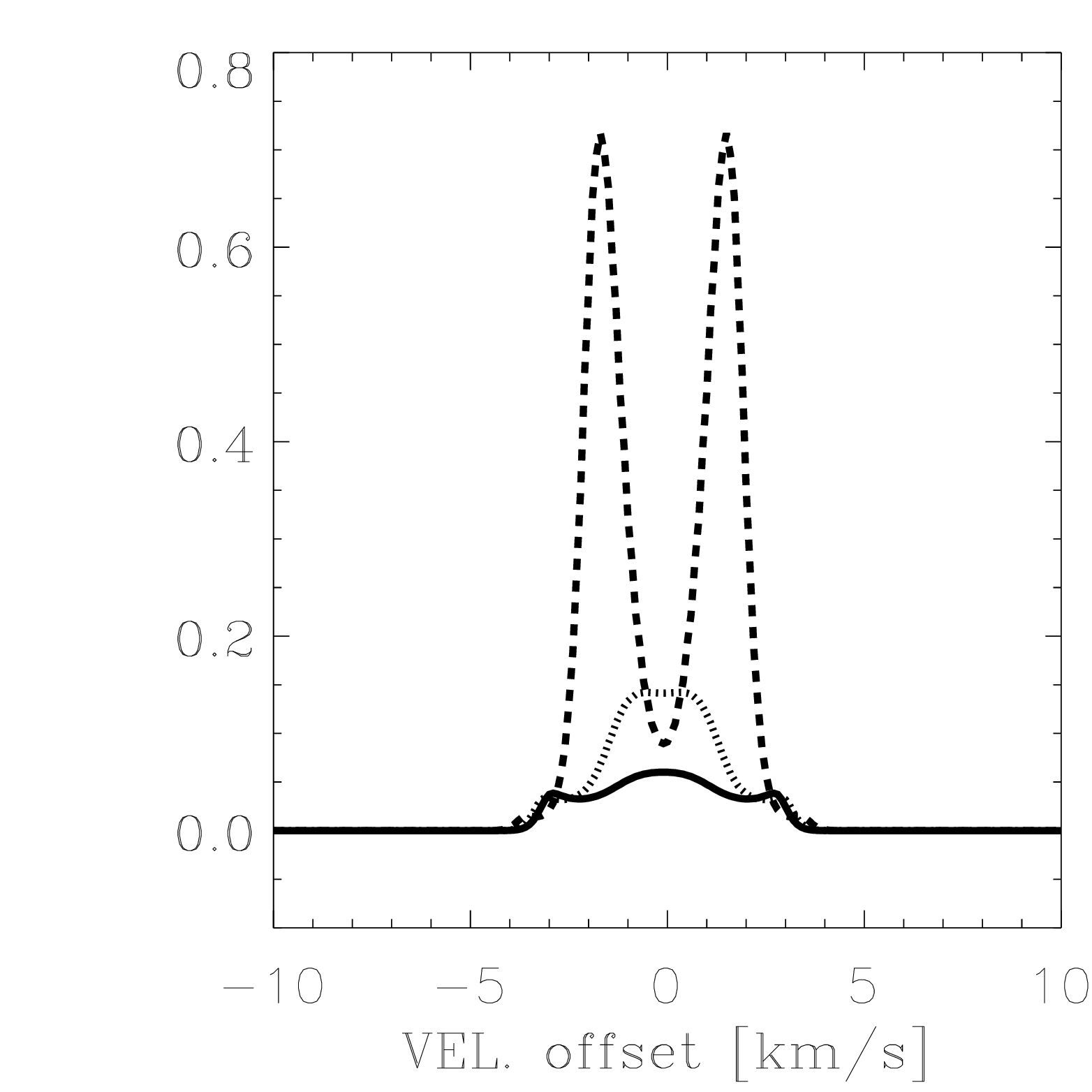}
\includegraphics[width=120pt]{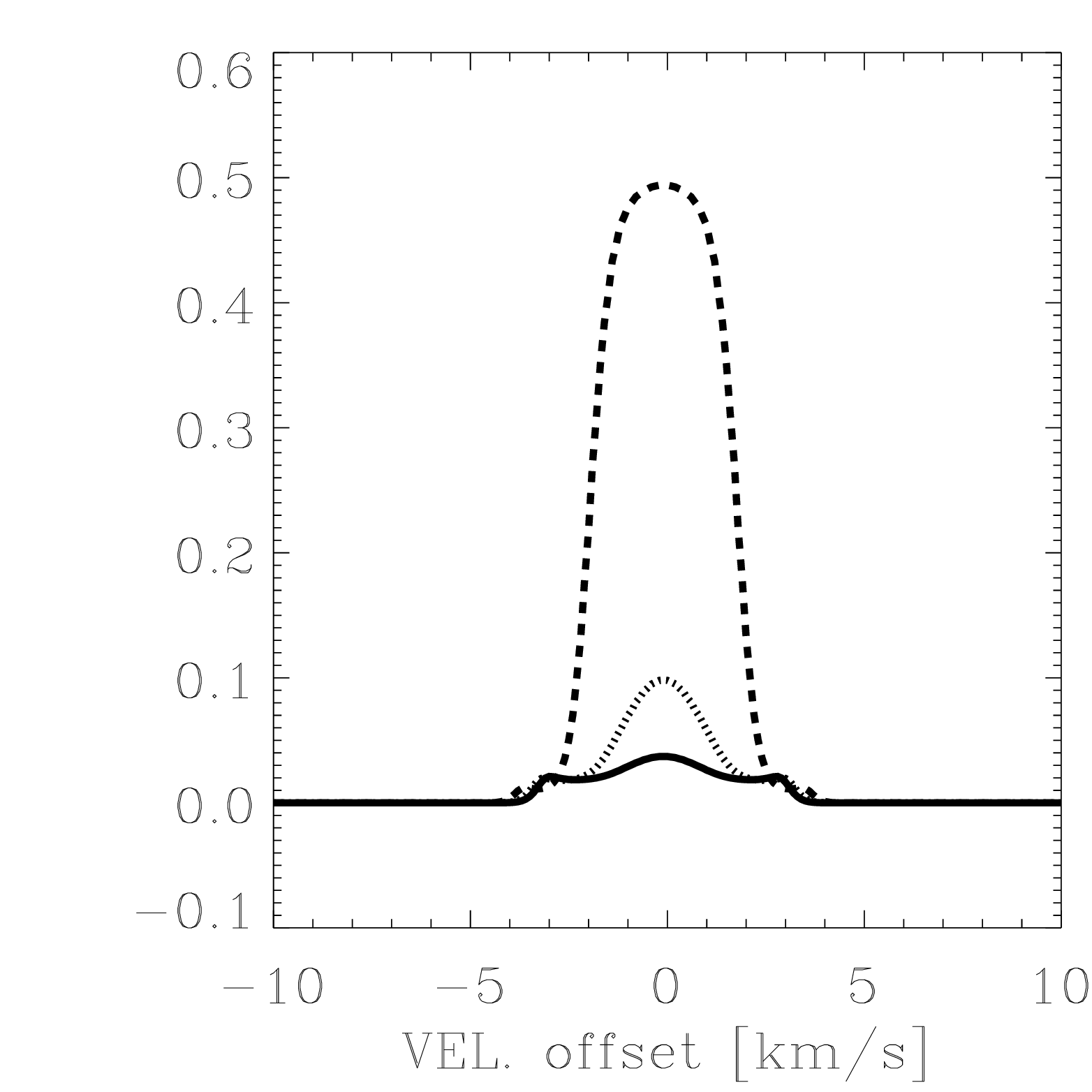}
\includegraphics[width=120pt]{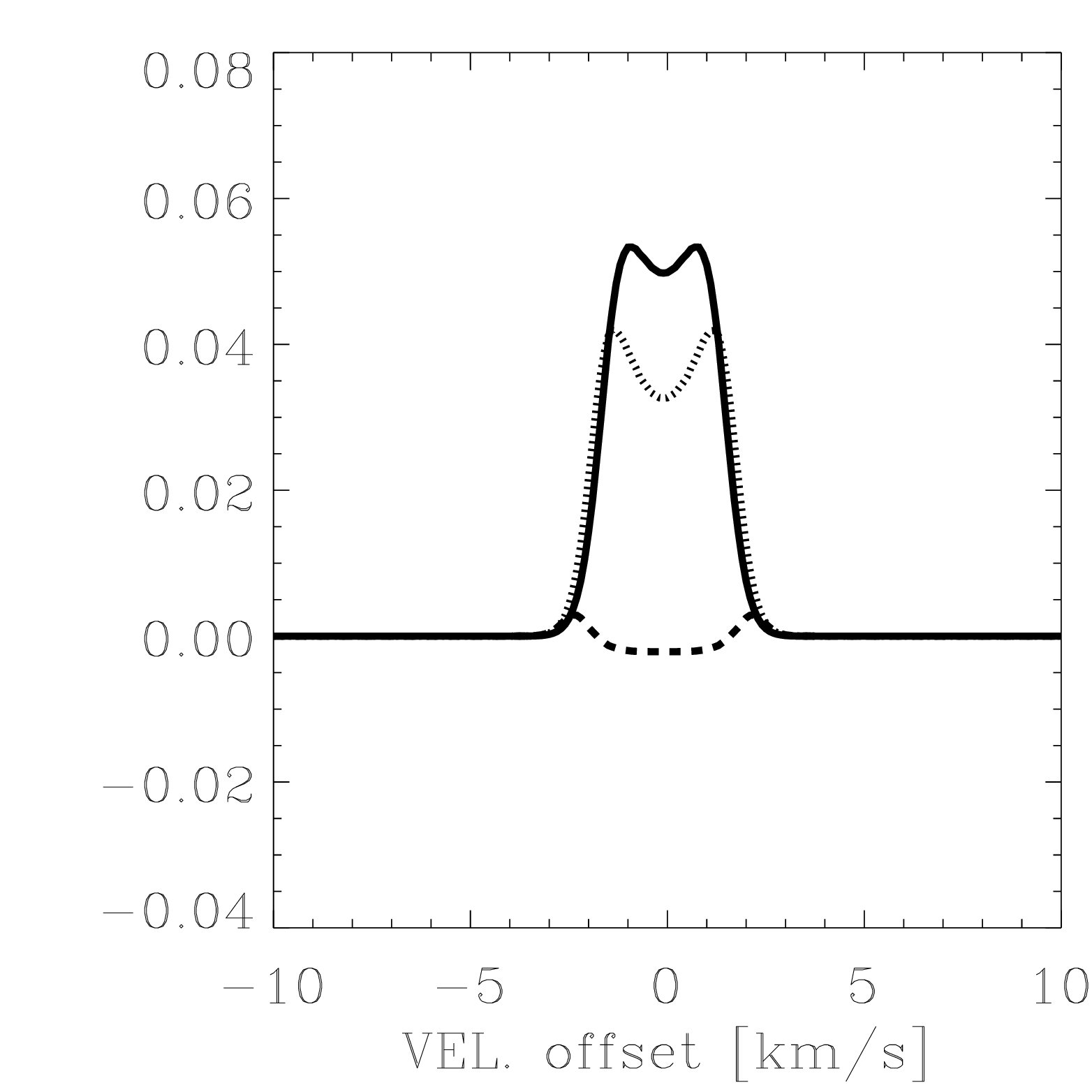}

\caption{Examples of the density dependence of line profiles for a
model with $p$=1.5, $X_0$=10$^{-4}$, $X_{\rm{d}}$=10$^{-8}$, $L$=7
L$_\odot$. From left to right are shown the 1$_{11}$-0$_{00}$ 1113
GHz, 2$_{02}$-1$_{11}$ 988 GHz, 2$_{11}$-2$_{02}$ 752 GHz and
3$_{31}$-4$_{04}$ 1893 GHz lines. Densities are 4$\times10^5$ (solid),
10$^6$ (dot) and 5$\times10^6$ (dash) cm$^{-3}$. For most lines, the
higher density, and thus higher water column, increases the emission,
although absorption features are also enhanced. For the
3$_{31}$-4$_{04}$ line at 1893 GHz, the dust becomes optically thick
at lower temperatures (i.e. larger radii) within the envelope for
higher densities.  }
\label{Fig:densspec}
\end{figure*}

Table \ref{table:h2ogrid_test} and Figures \ref{Fig:Lumspec} and
\ref{Fig:densspec} present the results for a grid probing the
parameters from Table \ref{table:results} for a constant abundance
combination of 10$^{-4}$ for $X_0$ and 10$^{-7}$ for
$X_{\rm{d}}$. Results for the full grid, including all probed
abundances, can be found in the online appendix. The main variables of
the grid - the luminosity, the density and the density profile
steepness - are discussed within this section.  The dependence on
abundance mirrors the behaviour seen for the L~483 model discussed in
\S 3. For most combinations of envelope parameters, the H$_2^{16}$O
transitions studied here are optically thick, sometimes with optically
thin line wings.  For H$_2^{18}$O lines, most transitions are
optically thin, although the three ground state lines are optically
thick at line center.  The only optically thin H$_2^{16}$O lines are
those for which the upper level of the transition has an excitation
temperature above 200--250 K, specifically the lines with $J=3$ and
higher. These lines are only optically thin if the densities are low
enough, $<$10$^7$ cm$^{-3}$, and abundances are low, $X_0<10^{-5}$ and
$X_{\rm{d}}<10^{-7}$.
For typical low-mass envelopes, such densities are only encountered
within the inner region, where enhanced abundances are encountered.
Thus, because many higher excited H$_2^{16}$O lines become optically
thick only in the very dense warm inner region, they are still good
tracers of the inner region (see \S 3.3.1 and
Fig. \ref{Fig:examspec2_L483} right).
For high densities and thus higher envelope masses, the larger dust
column plays a major role by heavily influencing the water line
emission. This is discussed in $\S 5$.

\subsection{Luminosity}
Fig. \ref{Fig:Lumspec} shows four lines for various luminosities. For
all lines, both the line strength as well as the absorption features
are stronger for higher luminosities.  For optically thin lines, a
best fit can be found with a slope that has a power law index of
$\sim$0.8, i.e. $I$ $\propto$ $L^{0.8}$ .  Fig. \ref{Fig:slope} shows
two such power laws for the 2$_{20}$-2$_{11}$ 1228 GHz line that
differentiate between the exact conditions and correspond to $0.45
L^{0.8}$ and $0.15 L^{0.8}$.  The slope of the power law is determined
by the temperature profile, which determines the amount of warm gas
present in the dense inner region, and the filling factor of this warm
gas within the beam. This filling factor is not readily apparent,
since models have different values for $R_{\rm{in}}$ and different
radii where $T$=100 K is reached depending on luminosity and density profile
(see Table~2). 
The peak temperatures of these optically thin lines naturally follow a
similar dependence. Even for lines which display an optically thick
line center (e.g., the 3$_{31}$-4$_{04}$ line in
Fig. \ref{Fig:Lumspec}), the peak temperatures depend on luminosity in
a similar way as the completely optically thin lines.  

For optically thick lines, a more complex dependency is seen in
Fig. \ref{Fig:Lumspec} (e.g., the $1_{11}-0_{00}$ line). For such
self-absorbed lines, the wings (which are optically thin) increase
with $L$, while any absorption features also become more prominent.
Although the peak temperatures do not follow the proposed power law of
the optically thin lines, they will become easier to detect for more
luminous sources.

\subsection{Density}
\begin{figure}[!htb]
\begin{center}
\includegraphics[width=250pt]{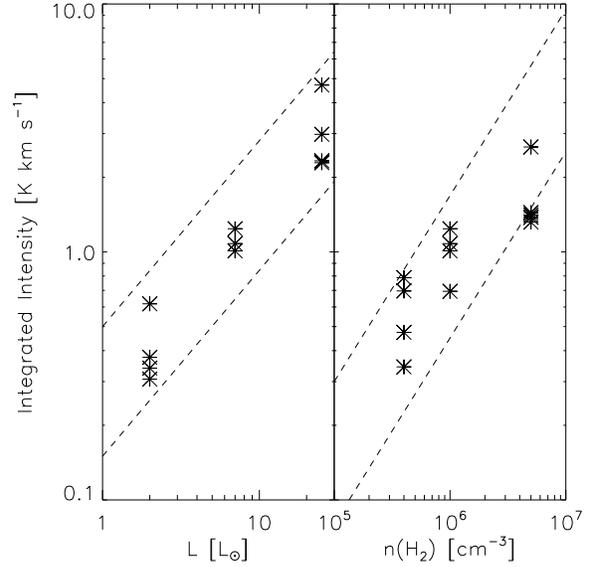}
\end{center}
\caption{The modelled integrated intensity of the optically thin
2$_{20}$-2$_{11}$ 1228 GHz line for $X_0$=$10^{-4}$. Intensities are
shown for the 3 different values of $X_{\rm{d}}$ for a constant
density of 10$^{6}$ cm$^{-3}$ ({\it left}), and for a constant
luminosity of 7 L$_{\odot}$ ({\it right}). Power laws with indices of
0.8 ({\it left}) and 0.75 ({\it right}) are over-plotted. In the right
panel, at 5$\times10^6$ cm$^{-3}$, this line becomes optically thick
for $X_{\rm{d}}$=10$^{-6}$, with its intensity dropping below the
relation. }
\label{Fig:slope}
\end{figure}

Fig. \ref{Fig:densspec} shows four lines for a range of densities,
varying by more than an order of magnitude, corresponding to envelope
masses ranging from 0.15 to 1.93 M$_{\rm{\odot}}$. The density at 1000
AU has a strong influence on the water line emission. For optically
thin lines, models with a constant $p$ and constant luminosity show
that a power law with an index of 0.75 accurately describes the
dependence of line strengths on $n_0(\rm{H}_2)$ (see
Fig. \ref{Fig:slope}). For optically thick lines the line will
decrease in strength at higher densities, since the emission from the
warm inner part is veiled behind a larger, cold water column.

A higher density can also result in the dust becoming optically
thick. This has an effect on the higher excited lines at high
frequencies. For example, the optically thin 3$_{31}$-4$_{04}$ 1893
GHz line (see Fig.\ 6, right) decreases by nearly two orders of
magnitude in strength when the density increases from $0.4\times 10^6$
to $5\times 10^6$ cm$^{-3}$.

The dust is optically thin in the low-density models and optically
thick at high density.  Since the emission of this higher excitation
line originates in the warm inner region, optically thick dust can
obscure over 99\% of the emission of this high frequency line (see
$\S$ 5).

\subsection{Density profile index}

The steepness of the power law used to define the density
distribution, $p$, also influences water line emission. Table
\ref{table:h2ogrid_test} shows that lines differ subtly between
$p$=$2$ and $p$=$1.5$. The line intensities of models with $p$=$2$ are
lower by a factor ranging from 0.5 to 3 than those for models with
$p$=$1.5$. Absorption features also differ in their depths for similar
models with only $p$ varied. This is caused by differing temperature
profiles within the Herschel beams. Assuming equal $n_0({\rm{H}_2})$
at 1000 AU, the warm ($T>$100 K) region is significantly smaller for a
model with $p$=$2$ than for $p$=$1.5$. Although local densities are
higher by almost an order of magnitude, the beam dilution is too
severe to overcome such small regions.

\subsection{H$_2^{18}$O}

The models were also calculated for the isotopologue
H$_2^{18}$O. Similar to the L~483 models, transitions connected with
the ground state lines are optically thick for all models. H$_2^{18}$O
excited lines with $J_{\rm{up}}=2$ and 3 do have optically thick
centers for some combinations of abundance, luminosities and/or
densities, but are optically thin at densities below 10$^6$
cm$^{-3}$. Higher excitation H$_2^{18}$O lines are always optically
thin.  The best-fit power law slopes for the optically thin H$_2^{18}$O
lines are the same as those for H$_2^{16}$O. Of course, these
conclusions are limited by the optical depth of dust at the frequency
of the line, since H$_2^{18}$O lines at high frequencies are affected
by dust. The precise frequency depends on the combination of
luminosity and the density profile (\S 5). Results for H$_2^{18}$O can
be found in the online appendix.

\subsection{Line profiles}

The H$_2$O line profiles shown in Fig.\ \ref{Fig:Lumspec} and
\ref{Fig:densspec} all contain an optically thin and optically thick
part as discussed in $\S 3.4$. Since the line wings of most transitions
are optically thin, a 'bump' is present in these wings at the velocity
for which that particular line becomes optically thick. This velocity
depends solely on the total column of water at this velocity and not
on the luminosity. The height of this bump is determined by both the
luminosity and density profile.  The line center changes according to
the actual optical depth of both the dust and the water itself. Line
centers can (i) be effectively thin and show a Gaussian profile; (ii)
be dominated by a small region and be flat-topped; (iii) show a
self-absorption feature
determined by the optical depth of the line only; (iv) have absorption
into the dust continuum.

\subsection{Model limitations}

Our spherically symmetric envelope models have several limitations,
which can affect the interpretation of observed line strengths.

\begin{itemize}
\item The presence of a large velocity gradient, such as seen in
infall and rotation of envelopes around high-mass stars, can
significantly broaden the water line and simultaneously make it less
optically thick. The velocity field in these models was taken to be
small, with an average infall velocity of 4 km s$^{-1}$ at
$R_{\rm{in}}$. For low-mass sources, such an infall velocity combined
with a turbulent width of 1 km s$^{-1}$ has been shown to accurately
reproduce observed emission line profiles \citep{Jorgensen04}.

\item Outflow contributions. No outflow contribution has been taken
  into account in our models. It is assumed that emission from outflow
  material can be identified with sufficient spectral resolution in
  observational data with Herschel-HIFI. However, material heated by
  outflow shocks in high density regions will not necessarily be
  accelerated to high velocities. This material can add significantly
  to the emission of water lines, but requires a detailed shock model
  to quantify \citep[e.g.,][]{Kaufman96}. In addition, outflow
  cavities allow emission from the high density regions of the inner
  envelope to escape along paths with much lower density than
  encountered in the cold outer envelope.

\item Large extended molecular cloud or fore-ground cloud
  contributions can both emit rotational water lines or absorb
  radiation coming from the protostellar envelope. This is especially
  likely for the three transitions connected to the ground-states of
  ortho- and para-water. For example, for a cloud column density of
  10$^{22}$ cm$^{-2}$ and an assumed water abundance of 10$^{-8}$ with
  a gas density of 10$^5$ cm$^{-3}$, the ground state ortho-H$_2$O 557
  GHz line will have an intensity of 50 mK if the emission fills the
  beam. The optical depth of this line is large, $\tau \approx 20$,
  enough to absorb radiation coming from behind this cloud.
  Results from SWAS and ODIN (See $\S$ 6.2) indeed do not show
    the deep absorption in the 557 GHz line, seen in Fig
    \ref{Fig:examspec_L483}--\ref{Fig:densspec}. It is likely that
    extended cold cloud material dominates close ($<$0.5 km s$^{-1}$)
    to the line center in these data due to the much larger observing
    beams than Herschel. For excited lines, emission and optical
  depth in a foreground cloud are negligible. At high ($>$ 1 THz)
  frequencies the dust within such cloud complexes can absorb the
  emission from excited water, however.

\item A large massive gas disk can contribute to the emission of
water. However, most gas within disks is at low ($T\sim$30~K)
temperatures. Most of the water will be frozen out, except in the
inner few AU and in the warm surface layers. 
In addition, high dust columns can completely obscure frequencies
above 1 THz. Water emission from disks is thus expected to be orders
of magnitude lower for lines at all frequencies, except for the
transitions connected to the ground state below 1 THz. However, the
small angular size of disks causes much of the emission to be diluted
to much lower values (a few mK) within the Herschel beam than commonly
found for envelopes (a few 100 mK to a few K).

\end{itemize}

\section{Effects on H$_2$O excitation and line formation: dust and micro-turbulence}
\subsection{Dust}
The presence of dust in protostellar envelopes influences water
emission lines significantly in two ways, as can be seen in Table
\ref{table:dustspec}. First, the dust can become optically thick,
preventing water emission from deep in the envelope to escape. Second,
the far-infrared radiation can pump the water lines.  Dust becomes
optically thick at specific columns, depending on the value of the
opacity, $\kappa$, at a given frequency (see Fig 5c in
\citet{Ossenkopf94} for the precise relation).
Since $\kappa_\nu$ $\propto$ $\nu^\beta$, with the dust opacity index
$\beta$ typically 1.5, this is especially likely at higher
frequencies.  Table \ref{table:dust}, column 4, shows the total gas and 
dust column densities necessary to reach a dust optical depth equal to
1 for each line.

\begin{figure}
\begin{center}

\end{center}
\caption{FIGURE TO BE DOWNLOADED FROM WEBSITE Cartoon of protostellar envelope models. For two different
models, the $\tau=1$ surfaces at 557 and 1669 GHz for the dust are
shown. A model with low-density ($n_0({\rm{H}_2})$=4$\times10^5$
cm$^{-3}$, $M$=0.15 M$_{\rm{\odot}}$) is shown in the top panel; here
the 557 GHz line contour is absent, since dust is optically thin
throughout the envelope at this frequency. A model with a high density
($n_0({\rm{H}_2})$=5$\times10^6$ cm$^{-3}$, $M$=1.93 M$_{\rm{\odot}}$)
is shown in the bottom panel. The long-dashed line indicates the
$T=100$~K contour. In both models, the 1669 GHz line is unable to
probe the inner warm region, regardless of its own optical depth,
since the dust $\tau=1$ surface lies far outside the $T=100$ K
radius. }
\label{Fig:duststruct}
\end{figure}

Fig. \ref{Fig:duststruct} illustrates this effect in a cartoon for two
different envelope models. Two models, one with a low density (low
$M_{\rm{env}}\approx 0.15$ M$_\odot$) and one with a high density
(high $M_{\rm{env}}\approx 1.93$ M$_\odot$), are compared for two
ground-state lines.  The dust at 557 GHz only becomes optically thick
for the more massive envelope due to very high column densities, while
the low-density model remains optically thin throughout. The $\tau=1$
surfaces correspond to temperatures well over 100 K. Since the 557 GHz
water line is optically thick at lower columns, the dust will have no
effect on this line, except for weakening the line due to dust
absorption. Even the H$_2^{18}$O column is large enough to prevent the
dust opacity from influencing the 548 GHz line. In contrast, the dust
at 1669 GHz becomes optically thick at a much smaller column, which
corresponds to a radius and temperature much further out into the
envelope. The actual radius depends on the model in question.
In the low-density model, the dust only turns optically thick
at radii close to the 100 K water evaporation radius, while dust in
the high-density model already obscures the line in the outer regions
of the envelope. Optically thin lines at this frequency will not be
able to probe the warm inner region, since the optically thick dust
effectively absorbs all water line emission.

\begin{figure*}
\begin{center}
\includegraphics[width=120pt]{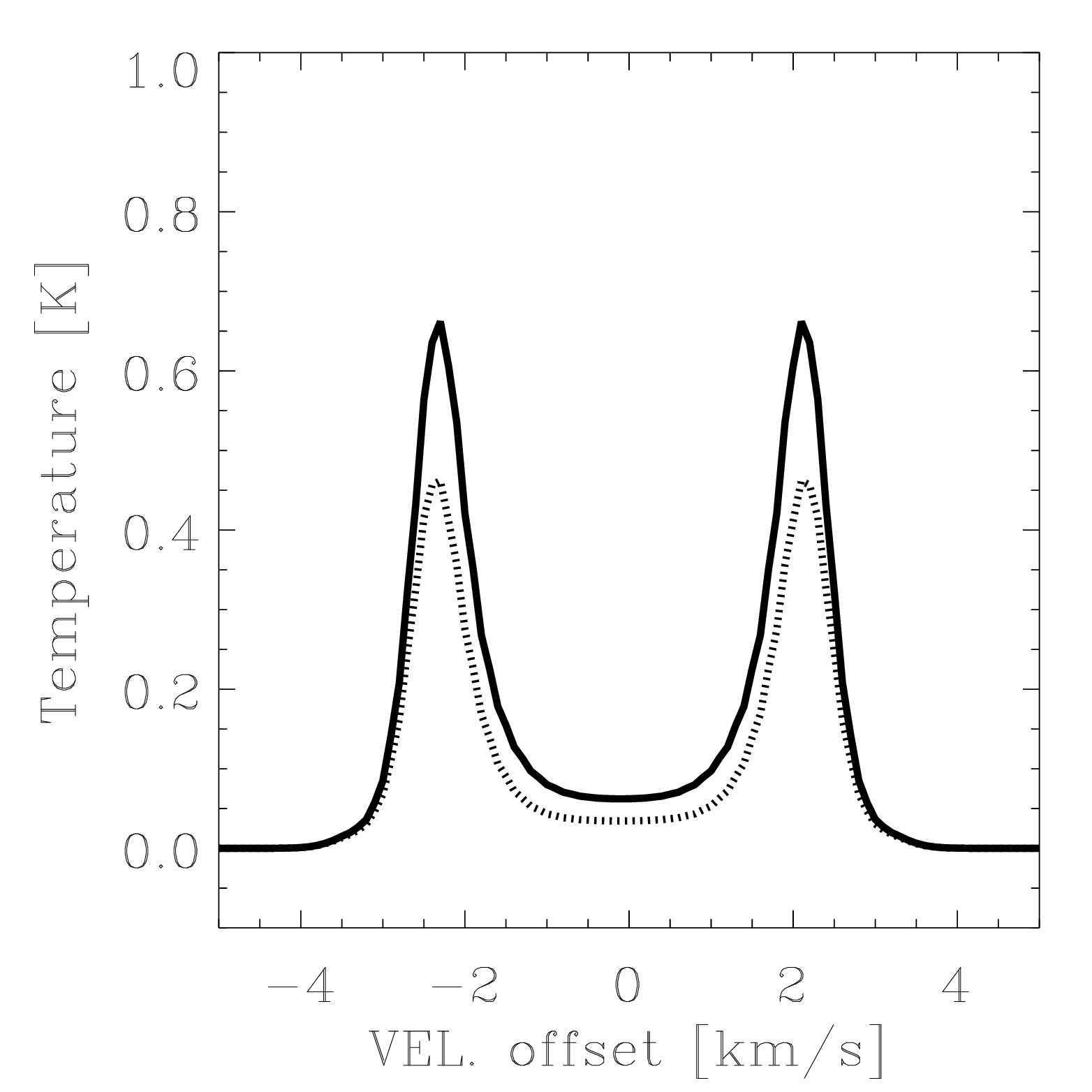}  
\includegraphics[width=120pt]{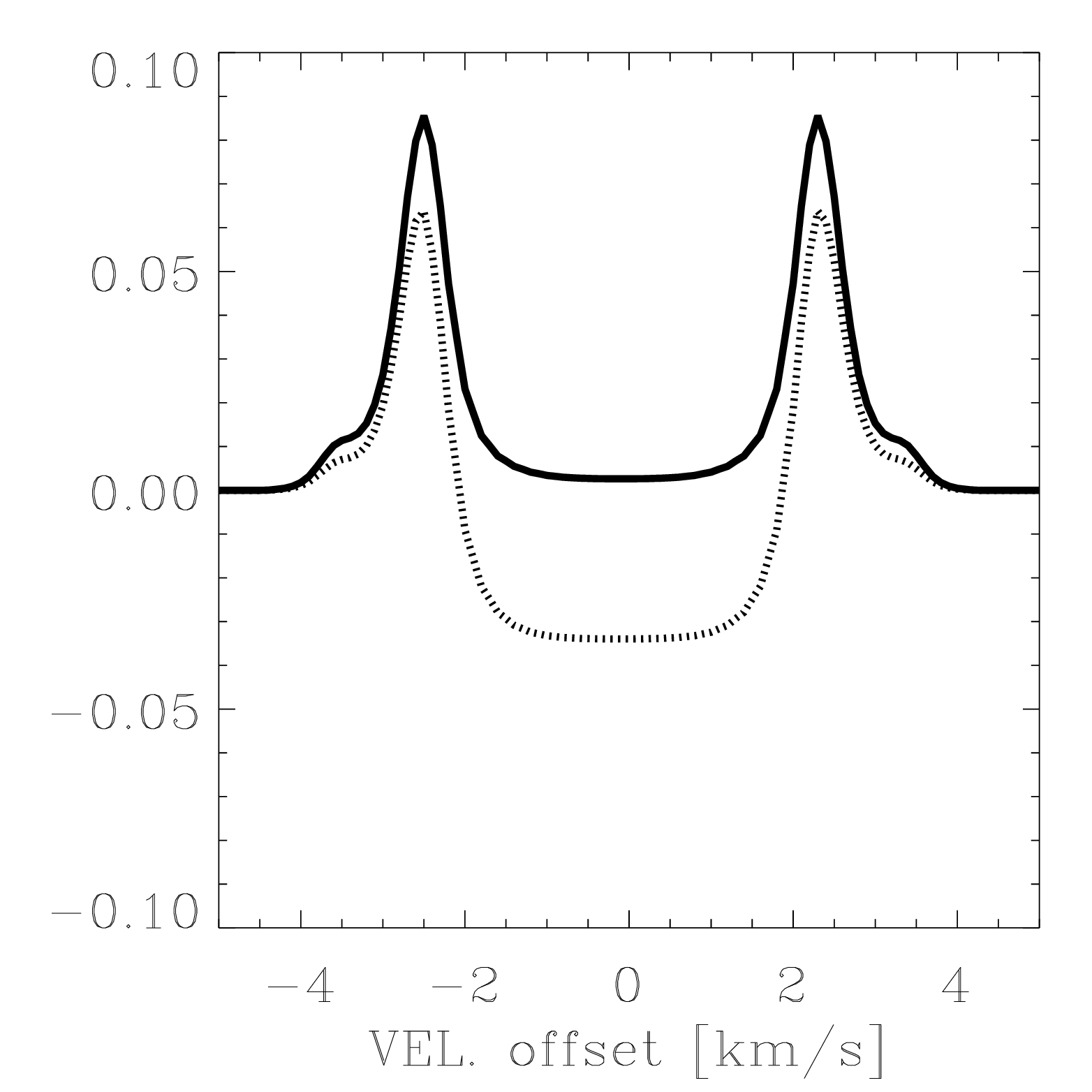} 
\includegraphics[width=120pt]{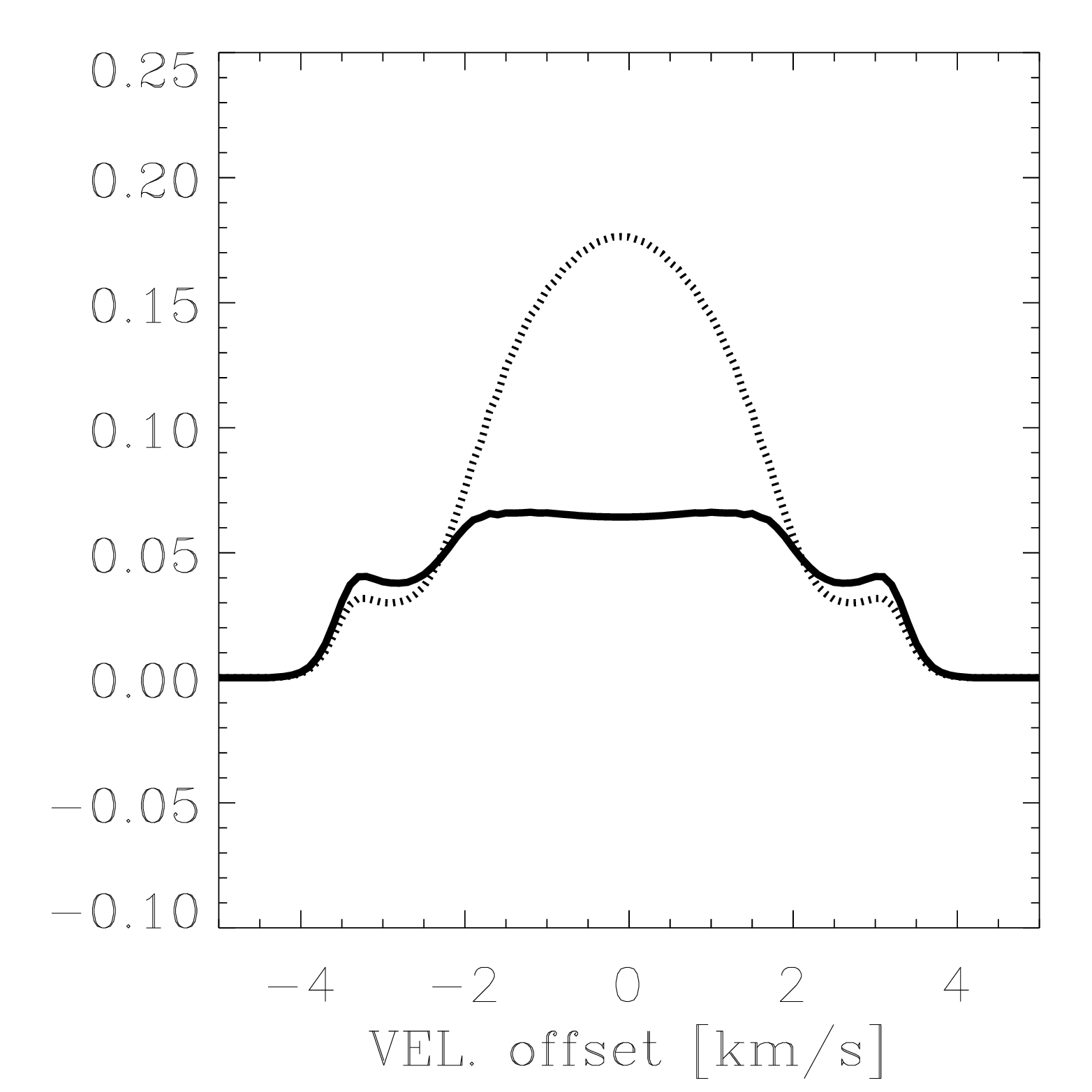}  
\end{center}
\caption{Example spectra of the 1$_{10}$-1$_{01}$ 557 GHz line ({\it
left}), 2$_{12}$-1$_{01}$ 1669 GHz line ({\it middle}) and
3$_{12}$-3$_{03}$ 1097 GHz line ({\it right}) for a model with
$p$=1.5, $L=7$ L$_{\rm{\odot}}$ and $n_{\rm{H}_2}$=1$\times10^6$
cm$^{-3}$.  The solid lines represent the spectra with no dust within
the envelope, while the dotted line shows the same line with dust
included in the model. The influence of dust can be clearly seen. The
peaks of the optically thick lines are reduced in strength, the depth
of the absorption changes, and the optically thin lines are
significantly pumped by far-infrared radiation. }
\label{Fig:dustspec}
\end{figure*}

Fig. \ref{Fig:dustspec} shows example spectra of the 1$_{10}$-1$_{01}$
line at 557 GHz, 2$_{12}$-1$_{01}$ line at 1669 GHz and the
$3_{12}-3_{03}$ line at 1097 GHz for a model with $L$=7 L$_{\odot}$,
$n_{\rm{H}_2}$=1$\times10^6$ cm$^{-3}$, $p$=1.5, $X_0$=10$^{-4}$ and
$X_{\rm{d}}$=10$^{-7}$, with and without dust. The spectra with dust
are significantly weaker by up to 50\%, even for optically thin dust
at 557 GHz, since dust within the line of sight absorbs the water line
photons.
Lines that are optically thin at a frequency where optically thick
dust exists are weaker by factors up to a few or more if dust is
present, due to less water being `visible'. Optically thin lines can
be pumped by far-infrared dust radiation coming from the region at
radii larger than the $\tau=1$ surface of the dust at the frequency of
the line.

\begin{table}
\caption{Integrated intensity and peak brightness of lines with and without dust for a model with $p$=1.5, $L=7$ L$_{\rm{bol}}$ and $n_{\rm{H}_2}$=1$\times10^6$ cm$^{-3}$. }
\label{table:dustspec}
\begin{center}
\begin{tabular}{l l l l l}
\hline \hline
Transition & \multicolumn{2}{c}{Dust} & \multicolumn{2}{c}{No Dust} \\
 & Intensity  & $T_{\rm{peak}}$ &Intensity  & $T_{\rm{peak}}$ \\ 
& [K km s$^{-1}$]& [K] & [K km s$^{-1}$] & [K] \\ \hline
1$_{10}$-1$_{01}$ & 0.95 & 0.47 & 1.41  & 0.65 \\
2$_{12}$-1$_{01}$ & -2.6(-2) & 0.26 & 0.16 & 0.39 \\
3$_{12}$-3$_{03}$ & 0.67 & 0.18 & 0.39 & 0.06 \\ \hline
\end{tabular}
\end{center}
\end{table}
\begin{table*}
\caption{The dust properties at selected water line frequencies for
four typical models. 
The column density
 $N_{\rm{H_2}}$ required to produce $\tau = 1$ 
for the dust at each frequency 
is given with the corresponding radii and 
temperatures.  }
\label{table:dust}
\begin{center}
\begin{tabular}{l l l l l l l l l l l l}
\hline \hline
Transition & Frequency & $\kappa_{OH5}$ & $N_{\rm{H_2}}$ 
& \multicolumn{8}{c}{$r$ and $T$ ($\tau_{\rm{dust}}=1$)} \\ 
  & & & & \multicolumn{2}{c}{Model 1}$^c$ & \multicolumn{2}{c}{Model 2}$^d$ & \multicolumn{2}{c}{Model 3}$^e$ & \multicolumn{2}{c}{Model 4}$^f$\\ 
 & GHz & 10$^{-2}$ cm$^2$g$^{-1}$$^a$ & 10$^{23}$ cm$^{-2}$  & AU & K & AU & K & AU & K & AU & K\\ \hline
\multicolumn{8}{c}{Ortho-H$_2$O transitions}\\ \hline
1$_{10}$-1$_{01}$& 556& 4.56 & 54.8 & -$^b$ & - &- & -& - & - & - & -\\
2$_{12}$-1$_{01}$& 1669& 32.7& 7.6 & 45 & 97 & 62 & 73 & 9 &188 & - & - \\
2$_{21}$-2$_{12}$& 1661& 32.3& 7.2  &48 & 92 & 63 & 72 & 10 &179& - &-\\
3$_{12}$-3$_{03}$& 1097& 17.3& 14.5 & -& - & 32 &111& - & -& - &- \\
3$_{12}$-2$_{12}$& 1153& 18.5& 13.5 & 15 &240 & 34 & 107 & -& - & - & -\\
3$_{21}$-3$_{12}$& 1162& 18.7& 13.3 & 16 &217 & 35 & 105 & -& - & - &-\\ \hline
\multicolumn{8}{c}{Para-H$_2$O transitions} \\ \hline
1$_{11}$-0$_{00}$ & 1113& 17.7& 14.1& - & -& 33& 109 & -& - & - &-\\
2$_{02}$-1$_{11}$ & 987& 14.4& 17.3 & -& -& 27& 128 & -& - & - &-\\
2$_{11}$-2$_{02}$ & 752& 8.45& 29.5 & -& -& 16& 213& -& - & - & -\\
2$_{20}$-2$_{11}$ & 1228& 20.0& 12.4& 18 &185 & 37& 101& -& - & - & -\\
3$_{31}$-4$_{04}$ & 1893& 40.9& 6.1 & 64 &78& 74& 65 & 12 &154 & - &-\\
4$_{22}$-3$_{31}$  & 916& 12.2& 20.4& -& - & 38& 99& -& - & - &-\\
4$_{22}$-4$_{13}$ & 1207& 19.6& 12.7& 17 &193& 23& 144 & -& - & - &-\\
5$_{24}$-4$_{31}$ & 970& 13.9& 17.9 & -& - & 26& 131 & -& - & - & -\\
\hline
\end{tabular}\\
 \end{center}
$^a$ Total gas and dust \\
$^b$ - indicates that the dust is optically thin 
throughout the entire envelope\\
$^c$ See text, $L$=7 L$_{\odot}$, $p$=1.5, $n_0$=5$\times 10^6$ cm$^{-3}$\\
$^d$ See text, $L$=7 L$_{\odot}$, $p$=2, $n_0$=5$\times10^6$ cm$^{-3}$\\
$^e$ See text, $L$=2 L$_{\odot}$, $p$=2, $n_0$=4$\times10^5$ cm$^{-3}$\\
$^f$ See text, $L$=25 L$_{\odot}$, $p$=1.5, $n_0$=1$\times 10^6$ cm$^{-3}$ \\

\end{table*}

For each envelope model the column density of dust depends on a
combination of luminosity $L$ (through the location of $R_{\rm in}$,
Table~2) and envelope structure parameters $n_0$ and $p$. If the
$\tau=1$ surface is approximated as a hard surface where all water
photons are absorbed, a typical temperature can be extracted for each
combination of $L$, $n_0$ and $p$. This temperature is the upper limit
to which water lines can probe for each model before dust becomes
optically thick.  Table \ref{table:dust} presents the $\tau=1$
temperatures for four astronomically distinct models, based on the
results from Table 6 of \citet{Jorgensen02}. Model 1, with $L$=7
L$_{\odot}$, $p$=1.5 and $n_0$=5$\times 10^6$ cm$^{-3}$, corresponds
to a very young Class 0 YSO, which is dominated by a large, cold and
massive protostellar envelope. A good example is the well-studied NGC
1333 IRAS 4B source in the Perseus molecular cloud.  Lines with a
frequency above 1600 GHz are unable to probe the region with $T$$>$100
K. Dust is optically thin throughout the envelope for frequencies
lower than a 1000 GHz. Model 2, with $L$=7 L$_{\odot}$ , $p$=2 and
$n_0$=5$\times10^6$ cm$^{-3}$, has similar parameters but its mass is
more centrally condensed. Due to the very high column densities, dust
becomes optically thick at all frequencies above 700 GHz, making it
even more difficult to probe the central warm region than in Model
1. Highly excited water lines are effectively shielded from observers
within this envelope at all frequencies.  The third model, $L$=2
L$_{\odot}$, $p$=2, $n_0$=4$\times10^5$ cm$^{-3}$, corresponds to a
more evolved source with a low luminosity and a small envelope with
most of the mass concentrated in the center due to a steep density
profile. The relative amount of warm gas in this envelope is
significantly higher. Such a model would correspond to most Class I
sources, e.g., TMC-1A.  The last model included has $L$=25
L$_{\odot}$, $p$=1.5, $n_0$=1$\times 10^6$ cm$^{-3}$ corresponds to a
higher luminosity source, but with a low total mass. The mass of the
envelope is distributed over larger radii and the luminosity is
higher. For these last two models, dust will not hide the inner parts
of the envelope at all. Only the very warm region ($T>$150 K) is
veiled near the inner radius of Model 3.

\begin{figure*}
\begin{center}
\includegraphics[width=120pt]{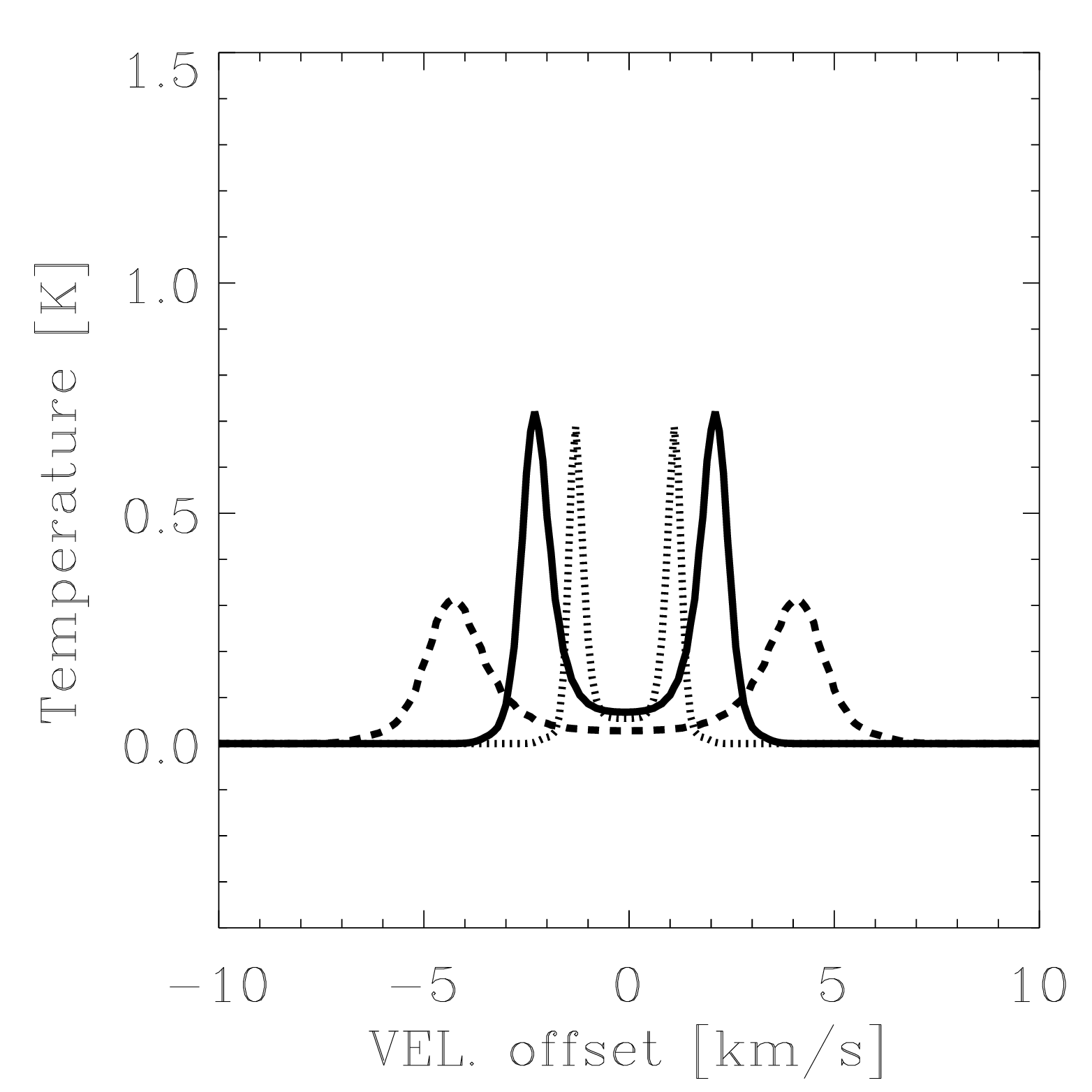}  
\includegraphics[width=120pt]{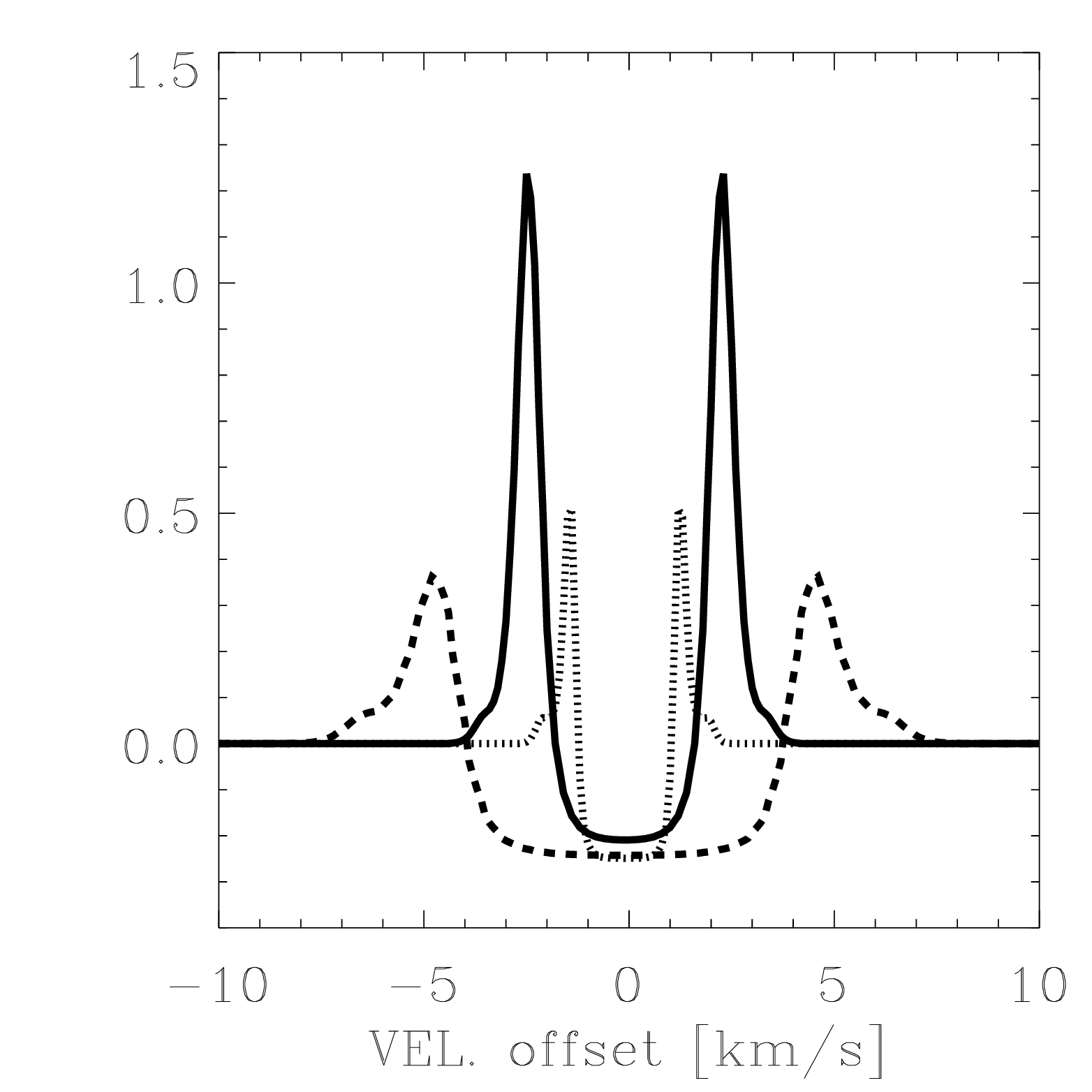} 
\includegraphics[width=120pt]{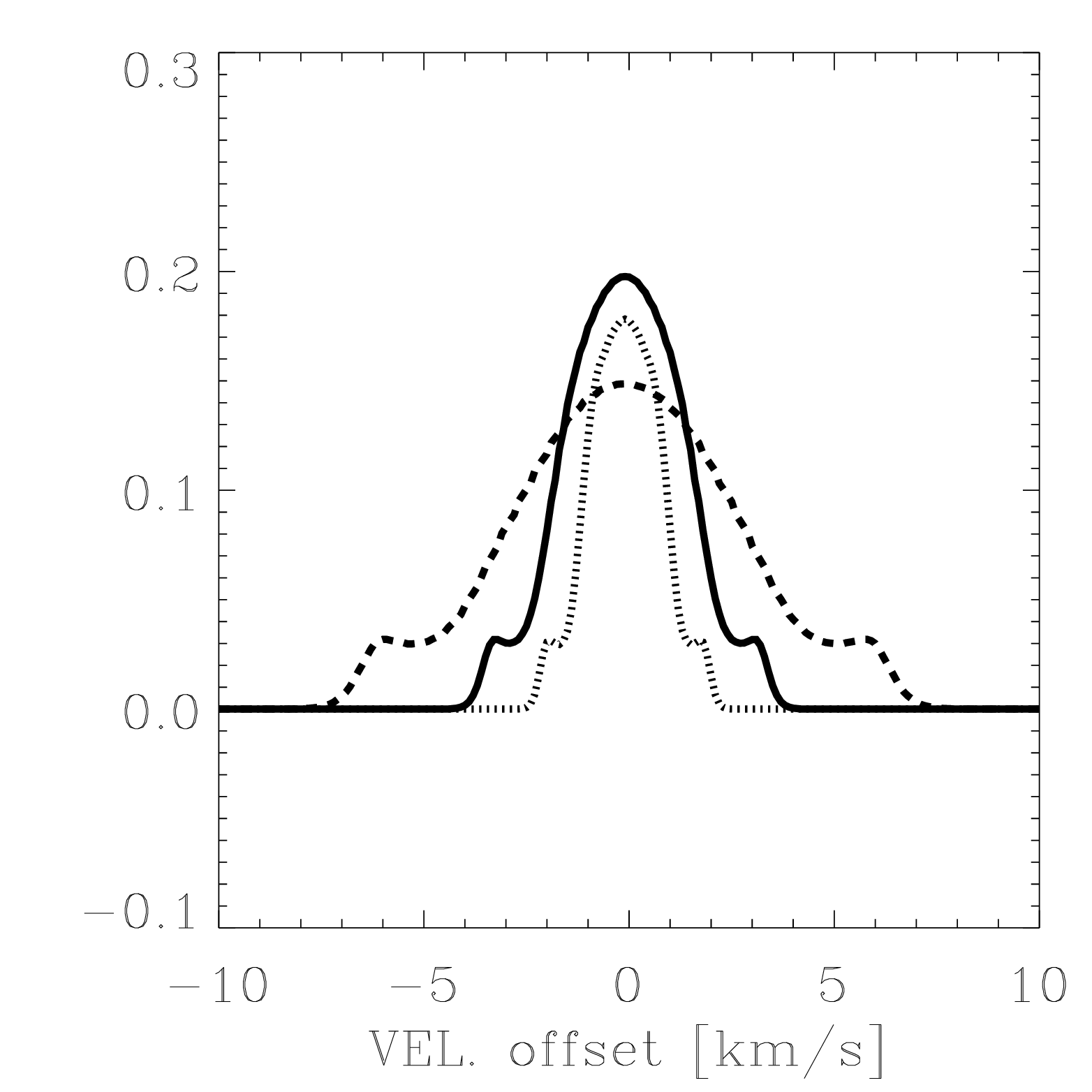}  
\end{center}
\caption{Example spectra of the 1$_{10}$-1$_{01}$ 557 GHz line ({\it
left}), 2$_{12}$-1$_{01}$ 1669 GHz line ({\it middle}) and
3$_{12}$-3$_{03}$ 1097 GHz line ({\it right}) for a model with
$p$=1.5, $L=7$ L$_{\rm{\odot}}$ and $n_{\rm{H}_2}$=1$\times10^6$
cm$^{-3}$. The dotted line is for a model with $\Delta V= 0.83$ km s$^{-1}$. The solid line for $\Delta V= 1.67$ km s$^{-1}$ and the dashed line for $\Delta V= 3.34$ km s$^{-1}$.  The large effect of micro-turbulence for all lines can clearly be seen.}
\label{Fig:turbspec}
\end{figure*}

\subsection{Micro-turbulence}

The structure of micro-turbulence, defined in our models as a
  velocity dispersion, during the embedded phases of star formation is
  not well known and theoretical models do not yet agree on a correct
  treatment. A good review of the theoretical considerations and
  problems is given in \citet[their $\S$
    3.2]{Ward-thompson01}. \citet{Jorgensen04} conclude that the
  inclusion of micro-turbulence as a velocity dispersion gives a more
  accurate reproduction of line intensities and profiles as opposed to
  the inclusion of infall (see also \citet{Jorgensen05}). \citet[$\S$
    4.6]{Ward-thompson01} test different values of micro-turbulence,
  as well as different power law values for the variation of
  micro-turbulence with radius. This results in a large variation in the
  molecular line profiles of CS and HCO$^+$.  Similarly,
  Fig. \ref{Fig:turbspec} shows that the choice of
 the velocity dispersion can have a large
  effect on all water lines, both optically thick and thin. All three
  lines (557, 1669 and 1097 GHz) show large differences in both line
  profile and in total integrated emission up to a
  factor of 3, when the velocity dispersion  is varied between
  0.83 (dotted, $b=$0.5), 1.67 (solid, $b=$1) and 3.34 (dashed, $b=$2)
  km s$^{-1}$.  Both absorption into the continuum and in the line are
  broader for larger values of micro-turbulence, while peak separation
  increases with increased velocity dispersion. Optically thin lines and line
  wings broaden if $b$ is larger. Such effects are similar
  to those found by \citet{Ward-thompson01} \\

For molecular lines in low-mass protostellar envelopes,
  \citet{Ward-thompson01} and \citet{Jorgensen05} both adopt a
  constant value for the micro-turbulence with radius. In our case, a
  velocity dispersion, $\Delta V$ $\sim$1.67 km s$^{-1}$ ($b$=1 km
  s$^{-1}$) has been chosen, which best reproduces line emission for a
  large variety of molecules in sources such as L483, including CO, CS,
  but also complex molecules such as H$_2$CO and CH$_3$OH tracing both
  the warm and cold region. We recommend that for the modelling of
  water lines from specific sources, the velocity dispersion for the
  cloud in question as derived from other molecular lines is adopted.

\section{Observational studies}

The conclusions presented in previous paragraphs can be compared to
current  observational studies. However, due to the
approximations used within the models and the unique combinations of
parameters for individual sources, it is recommended that these models
are used only as a starting point for interpreting water emission of
protostellar envelopes from specific sources. As discussed in \S 4.6,
there are several factors which will complicate any analysis of
observed water lines.  To compare the envelope models presented above
with any individual observation other than those from Herschel, the
line intensities have to be corrected for the actual beam size of the
telescope.

\subsection{ISO-LWS}

Water has been detected in several low-mass YSOs (both Class 0 and
Class I) with ISO-LWS such as NGC 1333 IRAS 2/4, IRAS 16293-2422,
Elias 29, L 1448, L 483 and SSV 13
\citep[e.g.,][]{Liseau96,Ceccarelli99,Nisini99,Giannini01,Maret02}.  Up
to 14 water lines were detected, ranging from 180.5 $\mu$m to 67.3
$\mu$m. In most sources the total number of lines is significantly
lower, but sufficient to prove the existence of water. For example,
only a single line was detected for the Class I source Elias 29
\citep{Ceccarelli99}.

As an illustration, we consider the case of NGC 1333 IRAS4A/B, for
which \citet{Maret02} reported the emission of 14 water lines together
with high-J CO, OH, [O~I] and [C~II]. A model was put forward to explain the
origin of these FIR lines using a collapsing envelope model without
an outflow. Based on the model developed by \citet{Ceccarelli96}, the
envelope and stellar parameters were derived from the best fit model
to the 14 water lines. The water abundances in the envelope were found to
include a jump at 100~K with a best fit of $X_0$=5$\times10^{-6}$ and
$X_{\rm{d}}$=5$\times10^{-7}$. 
A lower inner abundance is excluded by their
model results,
but a higher $X_0$ abundance of
the order of $5\times10^{-5}$ cannot be excluded.  

The physical parameters derived by \citet{Jorgensen02} and abundances
proposed by \citet{Maret02} correspond best with our model with
$n_0=$10$^6$ cm$^{-3}$, $L$=7 L$_\odot$ and $p=1.5$ 
with abundances 10$^{-6}$ for
$X_0$ and 10$^{-6}$/10$^{-7}$ for $X_{\rm{d}}$ . 
However, our predicted line strengths differ from the observed and
predicted line strengths in \citet{Maret02}, especially for low
excitation lines. Our 2$_{21}$-2$_{12}$ and 2$_{12}$-1$_{01}$ 
predictions are 0.5 $10^{-20}$ and 1.1 $10^{-20}$ W cm$^{-2}$ for the
10${^{-6}}$/10$^{-7}$ abundance combination within the ISO-LWS beam,
significantly lower than the observed values of 11 and 27 $10^{-20}$ W
cm$^{-2}$. Although a inner abundance of 10$^{-4}$  produces brighter lines for the presented models, these lines are still an order of magnitude lower than the observed fluxes in \citet{Maret02}.

The most likely explanation is the different treatment of dust between
the two models, which lowers the fluxes of the high-frequency lines in
our case.  However, this also suggests that our models can only
explain the ISO-LWS fluxes if there are cavities in the envelopes
through which the photons can escape. Recent results with the Spitzer
Space Telescope show numerous spectrally unresolved water lines from
20 to 30 $\mu$m, which suggest that NGC 1333 IRAS 4B is indeed viewed
almost face-on, allowing the ro-vibrational lines to escape the
protostellar system unhindered along the outflow axis
\citep{Watson07}.  This can also help the longer wavelength lines
coming from the inner envelope. Another possibility is that
significant contributions come from large-scale outflows. This is
confirmed by SWAS (see $\S 6.2$).

\subsection{SWAS}
SWAS\footnote{See
  http://cfa-www.harvard.edu/swas/ for more info} observed only the
ground-state transition of o-H$_2$O at 557 GHz
and o-H$_2^{18}$O at 548 GHz.  Although the beam of SWAS
(3.3$'$x4.5$'$) is significantly larger than that of ISO-LWS, the
increased spectral resolution of better than 1 km s$^{-1}$ spectrally
resolves most lines in star-forming regions.

SWAS observed the NGC 1333 IRAS4 region \citep{Neufeld00, Bergin03}
where, contrary to the analysis of \citet{Maret02}, they observed a
wide (24.4 km s$^{-1}$) and bright ($\int T^*_A dV = 0.5-1.9$ K km
s$^{-1}$) line. It was concluded that these lines originate within the
outflow of one or more sources inside the NGC 1333 region. The model
proposed by \citet{Maret02} would have produced a narrow ($<$4 km
s$^{-1}$) and weak (0.02-0.04 K km s$^{-1}$) 557 GHz line for each
source contained within the beam. Our grid gives a typical integrated
intensity within a SWAS beam of  $\int T_{\rm MB} dV =
$ 0.5 K km s$^{-1}$ for a $X_0$/$X_{\rm{d}}$ combination of
10$^{-4}$/10$^{-6}$ with a typical line width of 6 km s$^{-1}$.

\citet{Ashby00} report on SWAS data for B335, an isolated Class 0
source with a 3$\sigma$ upper limit of $<$0.06 K km
s$^{-1}$. \citet{Evans05} used a wide variety of molecular emission
lines to model the chemistry and physical structure of this source.
The low luminosity (4.5 L$_{\odot}$) and low density
(4$\times$10$^5$ cm$^{-3}$) found by \citet{Evans05} indicate that
this class 0 source would be represented best by a model
with $p$=1.5, $L$=2 L$_{\odot}$ and $n_0(\rm{H}_2)$=4$\times$
10$^5$ cm$^{-3}$. 
Our predicted 557 GHz line strength corrected for the much larger SWAS
beam is 0.02-0.06 K km s$^{-1}$ for outer abundances from 10$^{-8}$ to
10$^{-6}$, just below the upper limits from SWAS. The only model that
would have been detectable with SWAS has a constant abundance of
$10^{-6}$. The high jump abundance model with $X_0$=10$^{-4}$ and
$X_{\rm{d}}$=10$^{-6}$ cannot be ruled out, since most of the line is
self-absorbed.  Similar conclusion holds for most other Class 0
sources detected with SWAS.

Using the SWAS archive\footnote{See
http://lambda.gsfc.nasa.gov/product/swas/ for more information}, upper
limits were found on several Class I sources, 
including RCrA IRS 5 and 7. 
Only upper limits of 20--50 mK in peak temperature could be derived. These
observed limits are higher by up to a order of magnitude than the
predictions presented in $\S$4 for models that would fit the physical
parameters of Class I sources. 
The cold envelope material is constrained to 10$''$-30$''$, a small
region within the SWAS beam. Thus, our models are consistent with the
scarce observational data on Class I sources.

In summary, both the ISO and SWAS data 
stress the need for Herschel Space Observatory data to properly constrain
the origin of the observed H$_2$O emission and disentangle large-scale
and outflow contributions from the emission originating in the
quiescent gas of the protostellar envelope.

\section{Summary and conclusions}

This paper presents several models examining the excitation of water
and its oxygen-18 isotopologue in envelopes around low-mass
protostars, with parameters covering the entire range of luminosity
and envelope characteristics of observed Class 0 and I sources. A
model for the Class 0 source L~483 was used to illustrate the
dependence of rotational water lines on the water abundances in the
inner ($T>100$ K) region $X_0$ and outer ($T<100$ K) region
$X_{\rm{d}}$. In turn, a larger grid of models was explored for the
full range of parameters and likely abundances. The main conclusions
are

- The strengths and line shapes of all rotational water lines depend
sensitively on the physical parameters of the envelope. Thus, an
accurate physical model, including a well-constrained density profile,
is a necessary pre-requisite to model water emission and derive
abundances.

- Ground state lines from both H$_2^{16}$O and H$_2^{18}$O become
optically thick at low water columns. Even though water abundances are
expected to be very low in the outer envelope (10$^{-8}$ with respect
to H$_2$), the ground-state lines are still dominated by emission from
the cold ($T<30$~K) outer ($r>1000$ AU) envelope. 
Ground state line profiles are characterized by a double peak feature
with an absorption around narrow, optically thin emission line
wings. The total integrated intensity can not be used to trace the
water abundance.  Spectrally resolving the ground state lines is
essential if these lines are to be used as probes for the water
abundance.

- Higher excitation lines ($E_{\rm{up}}>150$ K) are usually dominated
by emission coming from the warm inner region. At very high excitation
($E_{\rm{up}}>200-250 K$), the lines are optically thin. Although
these lines are weak ($<$0.4 K km s$^{-1}$), they provide valuable
information about the characteristics of the warm gas present in the
inner regions of the protostellar envelopes.  Power laws with an index
of 0.8 and 0.75 were found to accurately represent the dependence of
optically thin lines on luminosity and density.

- H$_2^{18}$O is a reliable alternative tracer of the abundances in
both the warm ($T>100$ K) and cold ($T<100$ K) regions of the
envelope. However, such lines are much weaker with predicted peak
temperatures of only a few tens of mK, requiring very long integration
times.

- The presence of dust greatly influences the emission of most water
lines. The optical depth of dust, which depends on the luminosity (due
to varying $R_{\rm{in}}$ in our models) and density profile
parameters, can prevent optically thin water lines from escaping the
inner envelope, resulting in orders of magnitude reduction of the
water fluxes.  In addition, even optically thin dust absorbs up to
50\% of the total line emission. Furthermore, the far-infrared
continuum can pump optically thin water lines.

- The value of the micro-turbulent velocity dispersion and the
  structure of any systematic velocity field can significantly
  influence the line profiles and intensities of all water lines. The
  adoption of a micro-turbulent width that is constant with radius, as
  inferred from line profile fits of other molecules such as CS and
  HCO$^+$, is expected to produce the best-fitting H$_2$O line
  profiles, but the actual values need to be determined on a
  source-by-source basis.

To conclude, although water is a complex molecule to model, total 
intenstities and line profiles of rotational water transitions can be used to
constrain essential information about the protostellar
envelope. 

The increased sensitivity and resolution, both spectrally and
spatially, available with HIFI and PACS on Herschel compared with
previous missions will provide the technology to observe water lines
covering a wide range of excitations for the water molecule. The
increased spectral resolution will be essential to locate the origin
of warm gaseous water and understanding the structure of low-mass
protostellar envelopes.  Abundances in both the cold outer region and
the warm inner region are best probed by a combination of lines and
must include detailed line profiles.

\begin{acknowledgements}
T.v.K. would like to thank Dieter Poelman, Floris van der Tak and
Antonio Crapsi for valuable discussion concerning YSO envelope
modelling in general and the use of RATRAN . T.v.K. and astrochemistry
at Leiden Observatory are supported by a Spinoza grant and by grant
614.041.004 from The Netherlands Organization for Scientific Research
(NWO).
\end{acknowledgements}

\bibliographystyle{aa}
\bibliography{biblio}

\Online
Images too large for astro-ph. Will be published by A\&A. See website http://www.strw.leidenuniv.nl/~kempen/water.php
\begin{landscape} 
\begin{table}[!htb]
\label{table:h2ogrid_large4}
\caption{Integrated intensities, $\int T_{\rm{MB}}\Delta V$ [K km s$^{-1}$], for  H$_2$O lines in the wide parameter grid for all abundances. Ortho lines for $p$=1.5}
\begin{center}
\tiny
\begin{tabular}{c@{} c   c   c c  c   c c  c   c c  c  c c  c  c c  c  c}
\hline \hline
 Density (cm$^{-3}$) & \multicolumn{18}{c}{Integrated Intensity $\int T_{\rm{MB}}\Delta V$ [K km s$^{-1}$] } \\ \hline
 & \multicolumn{6}{c}{$L=2$ (L$_{\odot}$)} & \multicolumn{6}{c}{$L=25$ (L$_{\odot}$)} & \multicolumn{6}{c}{$L=7$ (L$_{\odot}$)} \\ 
& \multicolumn{6}{c}{Abundance$^a$} &  \multicolumn{6}{c}{Abundance$^a$} & \multicolumn{6}{c}{Abundance$^a$} \\ 
& 1a & 1b & 1c & 2a & 2b & 2c & 1a & 1b & 1c & 2a & 2b & 2c & 1a & 1b & 1c & 2a & 2b & 2c \\ \hline
\tiny & \multicolumn{18}{c}{ 1$_{10}-1_{01}$ } \\ \hline
 4$\times 10^5$ &  0.68     &  0.27     &  4.7(-2) &   2.1     &  0.18     &  6.2(-2) &   1.2     &  0.55     &  0.27     &   1.3     &  0.23     &  0.14     &   4.1     &  0.43     &  0.12     &  0.59     &  0.19     &  9.1(-2) \\ 
 1$\times 10^6$ &   2.7     &   2.7     &  0.37     &  0.24     &   2.9     &  0.64     &  0.16     &   6.2     &  0.70     &  0.48     &   2.2     &  0.54     &  0.29     &   2.5     &  0.95     &  0.17     &   1.8     &  0.56     \\ 
5$\times 10^6$ &   3.6     &   2.3     &  0.22     &   3.9     &   2.1     &   1.7     &   11.2     &   1.9     &  0.33     &   15.0     &   2.0     &  0.16     &   8.6    &   2.1     &  0.22     &   3.6     &   3.9     &  0.43     \\ \hline
 & \multicolumn{18}{c}{$2_{12}$-1$_{01}$ } \\ \hline
 4$\times 10^5$ &  2.9(-2) &  7.2(-2) & -2.5(-2) &   2.1     & -1.3(-2) &  4.2(-2) &  0.26     &  0.82     &  0.92     &  0.11     & -0.37     &  0.28     &   5.0     &  0.30     &  0.19     & -0.24     & -0.21     &  2.48(-3) \\ 
 1$\times 10^6$ &   3.7     & -0.35     &  3.6(-2) &   1.2     & -0.21     & -0.17     &   1.0     &  -1.2     & -0.14     &  -1.7     &  -1.7     & -0.75     & -0.43     & -0.36     & -0.54     & -0.83     & -0.80     &  0.88     \\ 
 5$\times 10^6$ &  -1.3     &  -1.1     &  -1.7     &  -1.4     &  -1.6     & -3.6(-2) &  -9.1     &  -9.1     &  -7.5     &  -5.5     &  -9.1     &  -7.7     &  -3.9     &  -4.4     &  -3.8     &  -3.9     &  -3.6     &  -3.7     \\ \hline
 & \multicolumn{18}{c}{2$_{21}$-2$_{12}$} \\ \hline
 4$\times 10^5$ &  0.20     &  0.51     &  0.16     &  0.40     &  0.17     &  7.4(-2) &   1.4     &   1.2     &   1.2    &  0.73     &  0.67     &  0.65     &   1.2     &  0.60     &  0.45     &  0.41     &  0.32     &  0.17     \\ 
 1$\times 10^6$ &   4.8     &  0.29     &  0.25     &  0.13     &  0.42     &  0.14     &   6.3     &   1.5     &   1.6     &  0.42     &   1.2     &  0.67     &  2.3(-2) &  0.87     &  0.46     &  0.20     &  0.55     &  0.38     \\ 
 5$\times 10^6$ & -0.29     &  0.16     &  0.10     & -0.59     & -0.30     &  0.48     &  -5.1     &  -2.0     &   1.0     &   2.0     &  -2.0     &  0.42     &  -1.3     & -0.74     &  0.50     & -0.54     & -0.69     &  0.29     \\ \hline
 & \multicolumn{18}{c}{3$_{12}$-3$_{03}$} \\ \hline
 4$\times 10^5$ &  0.26     &  0.18     &  8.3(-2) &  0.32     &  0.11     &  5.8(-2) &   2.0     &   1.1     &  0.91     &   1.7     &  0.75     &  0.77     &  0.95     &  0.38     &  0.25     &  0.65     &  0.26     &  0.18     \\ 
 1$\times 10^6$ &  0.82     &  0.21     &  9.1(-2) &  0.37     &  0.22     &  6.9(-2) &   8.3     &   1.5     &  0.92     &   3.1     &   1.3     &  0.69     &   1.1     &  0.60     &  0.25     &   1.3     &  0.47     &  0.23     \\ 
 5$\times 10^6$ &  0.57     &  0.32     &  6.9(-2) &  0.69     &  0.28     &  8.8(-2) &   4.3     &   2.7     &  0.92     &   4.1     &   2.5     &  0.77     &   1.9     &  0.89     &  0.23     &  0.98     &  0.95     &  0.23     \\ \hline
 & \multicolumn{18}{c}{3$_{12}$-2$_{21}$} \\ \hline
 4$\times 10^5$ &  0.19     &  0.19     &  8.3(-2) &   0.31     &  0.10     &  5.2(-2) &   1.5     &   1.1     &  0.70     &   1.4     &  0.69     &  0.44     &  0.83     &  0.40     &  0.24     &  0.47     &  0.26     &  0.14     \\ 
 1$\times 10^6$ &  0.93     &  0.21     &  0.12     &  0.33     &  0.23     &  9.2(-2) &   5.8     &   1.6     &   1.1     &   2.1     &   1.4     &  0.69     &  0.78     &  0.67     &  0.29     &  0.71     &  0.50     &  0.30     \\ 
 5$\times 10^6$ &  0.37     &  0.37     &  0.12     &  0.33     &  0.30     &  0.18     &   2.4     &   2.9     &   1.5     &   3.3     &   2.5     &   1.3     &   1.1     &  0.84     &  0.42     &   1.0     &  0.99     &  0.42     \\ \hline
 & \multicolumn{18}{c}{3$_{21}$-3$_{12}$} \\ \hline
 4$\times 10^5$ &  0.328     &  0.302     &  0.103     &  0.575     &  0.155     &  5.604(-2) &   2.84     &   1.4     &  0.753     &   2.67     &  0.929     &  0.672     &   1.52     &  0.592     &  0.308     &  0.867     &  0.372     &  0.152     \\ 
 1$\times 10^6$ &   1.44     &  0.317     &  0.16     &  0.633     &  0.372     &  0.118     &   12.1     &   2.69     &   1.28     &   4.49     &   2.29     &  0.832     &   1.53     &   1.09     &  0.374     &   1.59     &  0.824     &  0.398     \\ 
5$\times 10^6$ &  0.785      &  0.609      &  0.157      &  0.809      &  0.547      &  0.236      &   7.34      &   5.44      &   2.28      &   9.35      &   5.17      &   2.05      &   2.8      &   1.64      &  0.60      &   1.96      &   1.85      &  0.612      \\ \hline
\end{tabular}\\
\end{center}
\end{table}
$^a$ The abundance combinations are as follows: {\bf 1: $X_0$ = 10$^{-4}$, 2: $X_0$ = 10$^{-6}$ and a:$X_d$ = 10$^{-6}$, b: $X_d$= 10$^{-7}$, c:$X_d$ = 10$^{-8}$}\\
Note: A(B) = A $\times 10^{\rm{B}}$
\end{landscape}

\begin{landscape} 
\begin{table}[!htb]
\label{table:h2ogrid_large3}
\caption{Integrated intensities, $\int T_{\rm{MB}}\Delta V$ [K km s$^{-1}$], for  H$_2$O lines in the wide parameter grid for all abundances, Ortho lines for $p$=2.0}
\begin{center}
\tiny
\begin{tabular}{c@{} c   c   c c  c   c c  c   c c  c  c c  c  c c  c  c}
\hline \hline
 Density (cm$^{-3}$) & \multicolumn{18}{c}{Integrated Intensity $\int T_{\rm{MB}}\Delta V$ [K km s$^{-1}$] } \\ \hline
 & \multicolumn{6}{c}{$L=2$ (L$_{\odot}$)} & \multicolumn{6}{c}{$L=25$ (L$_{\odot}$)} & \multicolumn{6}{c}{$L=7$ (L$_{\odot}$)} \\ 
& \multicolumn{6}{c}{Abundance$^a$} &  \multicolumn{6}{c}{Abundance$^a$} & \multicolumn{6}{c}{Abundance$^a$} \\ 
& 1a & 1b & 1c & 2a & 2b & 2c & 1a & 1b & 1c & 2a & 2b & 2c & 1a & 1b & 1c & 2a & 2b & 2c \\ \hline
\tiny & \multicolumn{18}{c}{ 1$_{10}-1_{01}$ } \\ \hline
 4$\times 10^5$ &  0.37     &  5.1(-2) &  0.23     &   1.2     &  5.1(-2) &  0.11     &  0.45     &  0.28     &  0.25     &  0.63     &  0.25     &  0.19     &   1.5     &  0.10     &  0.20     &   1.6     &  0.69     &  0.11     \\ 
 1$\times 10^6$ &   2.4     &   2.4     &  0.12     &  0.19     &   1.1     &  0.27     &  0.11     &   2.8     &  0.19     &  0.41     &  0.85     &  0.19     &  0.28     &   2.8     &  0.24     &  0.11     &   3.3     &  0.16     \\ 
 5$\times 10^6$ &   4.0     &   2.2     &  0.36     &   5.0     &   1.7     &  0.59     &   14.0     &   1.6     & -6.2(-2) &   4.5     &   2.4     &   1.2     &   9.0     &  0.82     &  0.78     &   7.3     &  0.89     &  0.11     \\ \hline
 & \multicolumn{18}{c}{$2_{12}$-1$_{01}$ } \\ \hline
  4$\times 10^5$ & -0.27     & -0.43     &  0.43     &  0.35     & -0.43     & -0.10     & -0.98     & -0.73     &  9.6(-2) & -0.68     & -0.79     & -0.11     &  0.27     & -0.71     &  0.17     &  5.7(-2) &  0.42     & -0.31     \\ 
 1$\times 10^6$ & -0.16     & -0.89     & -0.47     & -0.53     & -0.80     & -0.62     &  -1.3     &  -3.08     &  -1.4     &  -3.2     &  -3.1     &  -1.8     & -0.88     &  -1.6     &  -1.3     &  6.3(-3) &  -1.7     & -0.88     \\ 
5$\times 10^6$ &  -1.1     &  -1.5     &  -1.9     & -0.62     &  -2.1     &  -1.7     &  -7.2     &  -9.3     &  -8.4     &  -10.4     &  -8.7     &  -6.5     &  -3.2     &  -5.0     &  -3.6     &  -3.6     &  -4.9     &  -4.2     \\ \hline
 & \multicolumn{18}{c}{2$_{21}$-2$_{12}$} \\ \hline
 4$\times 10^5$ & -0.11     & -6.6(-3) &  0.11     &  3.6(-2) & -1.4(-2) &  6.3(-2) &  0.47     &  0.94     &  0.91     &  0.22     &  0.77     &  0.46     &  0.57     &  0.13     &  0.20     &   1.2     &  0.55     &  0.17     \\ 
 1$\times 10^6$ & -0.23     & -0.13     &  6.2(-2) & -0.35     &  2.6(-2) &  0.11     &  0.14     &  0.15     &  0.53     & -0.95     &  1.9(-2) &  0.41     &  0.28     &  0.14     &  0.20     & -0.23     & -0.13     &  0.18     \\ 
 5$\times 10^6$ & -0.58     & -0.58     &  8.5(-4) & -0.93     & -0.7     &  5.6(-3) &  -4.8     &  -3.07     & -0.36     &  -6.3     &  -3.2     &  -1.1     &  -2.1     &  -1.4     & -0.18     &  -1.0     &  -1.5     & -0.11     \\ \hline
 & \multicolumn{18}{c}{3$_{12}$-3$_{03}$} \\ \hline
 4$\times 10^5$ &  0.23     &  0.13     &  6.3(-2) &  0.25     &  0.13     &  5.2(-2) &   2.0     &   1.1     &  0.70     &   1.6     &  0.98     &  0.54     &  0.88     &  0.35     &  0.20     &   1.1     &  0.43     &  0.16     \\ 
 1$\times 10^6$ &  0.31     &  0.12     &  4.6(-2) &  0.29     &  0.16     &  4.7(-2) &   2.9     &   1.3     &  0.63     &   2.6     &   1.2     &  0.53     &   1.2     &  0.47     &  0.15     &  0.85     &  0.39     &  0.15     \\ 
5$\times 10^6$ &  0.19     &  0.15     &  2.9(-2) &  0.18     &  0.16     &  3.2(-2) &   2.4     &   1.9     &  0.44     &   3.1     &   1.9     &  0.56     &  0.78     &  0.51     &  0.14     &   1.06     &  0.47     &  0.13     \\ \hline
 & \multicolumn{18}{c}{3$_{12}$-2$_{21}$} \\ \hline
 4$\times 10^5$ &  0.18     &  9.3(-2) &  8.9(-2) &  0.19     &  8.1(-2) &  7.1(-2) &   1.18     &  0.99     &  0.73     &   1.2     &  0.97     &  0.53     &  0.69     &  0.29     &  0.23     &  0.83     &  0.46     &  0.19     \\ 
 1$\times 10^6$ &  0.20     &  0.13     &  7.3(-2) &  0.21     &  0.16     &  6.8(-2) &   2.1     &   1.2     &  0.86     &   1.4     &   1.1     &  0.73     &  0.80     &  0.48     &  0.20     &  0.73     &  0.40     &  0.23     \\ 
 5$\times 10^6$ &  0.20     &  0.14     &  6.7(-2) &  0.15     &  0.11     &  7.3(-2) &   2.7    &   1.6     &  0.80     &  0.89     &   1.8     &  0.98     &  0.48     &  0.49     &  0.28     &  0.76     &  0.55     &  0.24     \\ \hline
 & \multicolumn{18}{c}{3$_{21}$-3$_{12}$} \\ \hline
 4$\times 10^5$ &  0.29     &  0.14     &  0.13     &  0.35     &  0.13     &  0.10     &   2.3     &   1.6     &  0.93     &   2.4     &   1.6     &  0.65     &   1.3     &  0.45     &  0.33     &   1.6     &  0.77     &  0.26     \\ 
 1$\times 10^6$ &  0.45     &  0.20     &  0.11     &  0.41     &  0.27     &  0.10     &   4.4     &   2.1     &   1.2     &   3.4     &   2.0     &   1.1     &   1.6     &  0.87     &  0.30     &   1.5     &  0.73     &  0.35     \\ 
5$\times 10^6$ &  0.49      &  0.35      &  0.11      &  0.48      &  0.33      &  0.1      &   6.3      &   4.0      &   1.5      &   4.9      &   4.3      &   1.8      &   1.7      &   1.2      &  0.49      &   1.9      &   1.3      &  0.43      \\ \hline
\end{tabular}\\
\tiny
\end{center}
\end{table}
$^a$ The abundance combinations are as follows: {\bf 1: $X_0$ = 10$^{-4}$, 2: $X_0$ = 10$^{-6}$ and a:$X_d$ = 10$^{-6}$, b: $X_d$= 10$^{-7}$, c:$X_d$ = 10$^{-8}$}\\
Note: A(B) = A $\times 10^{\rm{B}}$
\end{landscape}

\begin{landscape} 
\begin{table}[!htb]
\label{table:h2ogrid_large2}
\caption{Integrated intensities, $\int T_{\rm{MB}}\Delta V$ [K km
s$^{-1}$], for H$_2$O lines in the wide parameter grid for all
abundances. Para lines for $p$=1.5}
\begin{center}
\tiny
\begin{tabular}{c@{} c c c c  c  c c  c  c c  c  c c  c  c c  c  c}
\hline \hline
 Density (cm$^{-3}$) & \multicolumn{18}{c}{Integrated Intensity $\int T_{\rm{MB}}\Delta V$ [K km s$^{-1}$] } \\ \hline
 & \multicolumn{6}{c}{$L=2$ (L$_{\odot}$)} & \multicolumn{6}{c}{$L=25$ (L$_{\odot}$)} & \multicolumn{6}{c}{$L=7$ (L$_{\odot}$)} \\ 
& \multicolumn{6}{c}{Abundance$^a$} &  \multicolumn{6}{c}{Abundance$^a$} & \multicolumn{6}{c}{Abundance$^a$} \\
& 1a & 1b & 1c & 2a & 2b & 2c & 1a & 1b & 1c & 2a & 2b & 2c & 1a & 1b & 1c & 2a & 2b & 2c\\ \hline
\tiny & \multicolumn{18}{c}{1$_{11}$-0$_{00}$}\\ \hline
 4$\times 10^5$ &  0.50     &  0.23     &  9.0(-2) &   3.0     &  0.22     &  6.9(-2) &   1.4     &  0.52     &  0.48     &   2.8     &  0.20     &  0.14     &  0.64     &  0.43     &  0.18     &   1.0     &  0.26     &  9.4(-2)  \\
 1$\times 10^6$ &  1.0     &  1.0     &  0.13     &  0.15     &  0.62     &  0.38     &  0.13     &   1.5     &  0.31     &  0.28     &   1.2     &   1.2     &  4.7(-2) &   1.5     &   1.0     &  0.19     &   2.9     &  5.4(-3)  \\
 5$\times 10^6$ &   1.5     & -1.6(-2) &  4.0(-2) &   1.9     &  0.19     &  0.22     & -0.45     &  -1.8     & -0.70     &   6.2     &  -1.1     &  -2.3     &  0.88     &   1.6     & -0.66     &   3.1     &  0.9     & -0.55      \\ \hline
 & \multicolumn{18}{c}{ 2$_{02}$-1$_{11}$}\\ \hline
 4$\times 10^5$ &  0.65     &  0.32     &  0.12     &   3.8     &  0.31     &  9.0(-2) &   3.2     &   1.6     &  0.87     &   2.8     &   1.2     &  0.43     &   1.2     &  0.72     &  0.28     &   1.3     &  0.62     &  0.19      \\
  1$\times 10^6$ &   1.3     &  0.58     &  0.24     &   1.0     &  0.64     &  0.22     &   6.7     &   3.0     &   1.3     &   3.9     &   3.1     &  0.99     &   2.0     &   1.5     &  0.53     &   1.7     &   1.2     &  0.46      \\
5$\times 10^6$ &   1.7     &  0.76     &  0.60     &   1.6     &  0.82     &  0.6     &   6.8     &   5.2     &   4.3     &   15.1     &   4.6     &   4.2     &   5.1     &   1.1     &   1.9     &   3.4     &   1.1     &   1.9      \\\hline
 & \multicolumn{18}{c}{ 2$_{11}$-2$_{02}$}\\\hline
 4$\times 10^5$ &  0.37     &  0.17     &  6.2(-2) &   1.4     &  0.17     &  4.8(-2) &   2.3     &  1.0     &  0.51     &   2.4     &  0.78     &  0.26     &  0.78     &  0.41     &  0.16     &  0.87     &  0.36     &  1.0(-1)  \\
 1$\times 10^6$ &  0.75     &  0.34     &  0.13     &  0.60     &  0.39     &  0.12     &   4.9     &   2.0     &  0.79     &   3.5     &   2.4     &  0.62     &   1.5     &   1.0     &  0.29     &   1.4     &  0.72     &  0.26      \\
 5$\times 10^6$ &   1.3     &  0.82     &  0.47     &   1.4     &  0.85     &  0.48     &   10.4     &   6.5     &   5.0     &   25.7     &   6.8     &   4.4     &   4.9     &   2.7     &   1.8     &   3.9     &   2.3     &   1.8  \\    \hline
 & \multicolumn{18}{c}{2$_{20}$-2$_{11}$}\\\hline
 4$\times 10^5$ &  0.31     &  0.18     &  8.8(-2) &   1.1     &  0.15     &  4.7(-2) &   2.3     &   1.1     &  0.69     &   2.2     &  0.74     &  0.22     &  0.70     &  0.47     &  0.18     &  0.79     &  0.34     &  9.4(-2)  \\
 1$\times 10^6$ &  0.62     &  0.31     &  0.13     &  0.38     &  0.34     &  0.11     &   4.7     &   2.3     &  0.94    &   3.0     &   2.3     &  0.59     &   1.2     &   1.0     &  0.36     &   1.1     &  0.69     &  0.25   \\   
 5$\times 10^6$ &  0.37     &  0.45     &  0.33     &  0.44     &  0.47     &  0.33     &   3.9     &   4.1     &   4.3     &   16.8     &   4.5     &   3.8     &   1.4     &   1.4     &   1.4     &   2.7     &   1.3     &   1.4    \\  \hline
 & \multicolumn{18}{c}{3$_{31}$-4$_{04}$}\\\hline
 4$\times 10^5$ &  5.9(-2) &  5.8(-2) &  5.7(-2) &  4.6(-3) &  1.8(-3) &  1.7(-3) &  0.57     &  0.56     &  0.56     &  1.4(-2) &  8.0(-3) &  7.6(-3) &  0.18     &  0.18     &  0.18     &  7.4(-3) &  4.0(-3) &  3.8(-3)  \\
 1$\times 10^6$ &  4.7(-2) &  4.2(-2) &  4.1(-2) &  7.4(-3) &  3.8(-3) &  3.4(-3) &  0.60     &  0.58     &  0.58     &  4.7(-2) &  2.6(-2) &  2.6(-2) &  0.16     &  0.15     &  0.15     &  1.8(-2) &  1.0(-2) &  9.8(-3)  \\
 5$\times 10^6$ &  1.1(-3) & -7.8(-4) & -9.6(-4) &  2.6(-3) &  6.8(-4) &  5.0(-4) &  0.20     &  8.1(-2) &  6.9(-2) &  0.2     &  6.9(-2) &  5.5(-2) &  2.9(-2) &  1.3(-3) & -1.5(-3) &  3.6(-2) &  9.6(-3) &  7.6(-3)  \\\hline
 & \multicolumn{18}{c}{ 4$_{22}$-3$_{31}$}\\\hline
 4$\times 10^5$ &  4.8(-2) &  4.2(-2) &  4.2(-2) &  2.2(-2) &  1.0(-2) &  9.8(-3) &  0.56     &  0.56     &  0.56     &  9.7(-2) &  6.2(-2) &  6.0(-2) &  0.15     &  0.13     &  0.13     &  4.4(-2) &  2.8(-2) &  2.7(-2)  \\
 1$\times 10^6$ &  4.9(-2) &  4.1(-2) &  4.3(-2) &  3.1(-2) &  1.7(-2) &  1.6(-2) &  0.59     &  0.48     &  0.47     &  0.27     &  0.18     &  0.17     &  0.16     &  0.13     &  0.12     &  9.0(-2) &  5.6(-2) &  5.3(-2)  \\
 5$\times 10^6$ &  4.0(-2) &  1.9(-2) &  1.5(-2) &  3.5(-2) &  1.4(-2) &  1.1(-2) &  0.87     &  0.39     &  0.31     &  0.78     &  0.31     &  0.23     &  0.22     &  9.6(-2) &  6.8(-2) &  0.19     &  8.0(-2) &  5.2(-2)  \\\hline
 & \multicolumn{18}{c}{4$_{22}$-4$_{13}$}\\\hline
 4$\times 10^5$ &  0.15     &  6.6(-2) &  5.1(-2) &  0.18     &  5.7(-2) &  4.5(-2) &   1.0     &  0.64     &  0.63     &   1.1     &  0.54     &  0.51     &  0.37     &  0.18     &  0.15     &  0.39     &  0.17     &  0.15      \\
 1$\times 10^6$ &  0.20     &  9.5(-2) &  5.1(-2) &  0.19     &  8.4(-2) &  4.3(-2) &   2.2     &  0.74     &  0.55     &   2.0     &  0.75     &  0.60     &  0.58     &  0.26     &  0.15     &  0.53     &  0.24     &  0.15      \\
 5$\times 10^6$ &  0.14     &  6.6(-2) &  2.3(-2) &  0.13     &  6.4(-2) &  2.1(-2) &   3.4     &   1.8     &  0.66     &   3.4     &   1.8     &  0.60     &  0.86     &  0.41     &  0.15     &  0.87     &  0.41     &  0.15      \\\hline
 & \multicolumn{18}{c}{5$_{24}$-4$_{31}$}\\\hline
4$\times 10^5$ &  5.0(-2) &  5.0(-2) &  5.0(-2) &  6.7(-3) &  6.6(-3) &  0.59     &  0.61     &  0.62     &  3.6(-2) &  2.6(-2) &  2.7(-2) &  0.16     &  0.16     &  0.16     &  2.0(-2) &  1.6(-2) &  1.5(-2) &  4.3(-2)  \\
1$\times 10^6$ &  5.0(-2) &  5.0(-2) &  5.1(-2) &  1.9(-2) &  1.4(-2) &  1.4(-2) &  0.58     &  0.57     &  0.57     &  0.14     &  0.12     &  0.12     &  0.17     &  0.15     &  0.15     &  5.2(-2) &  4.3(-2) &  4.3(-2)  \\
5$\times 10^6$ &  2.6(-2)  &  1.8(-2)  &  2.0(-2)  &  1.2(-2)  &  1.1(-2)  &  0.54      &  0.39      &  0.37      &  0.43      &  0.27      &  0.25      &  0.14      &  8.8(-2)  &  7.9(-2)  &  0.12      &  6.3(-2)  &  5.6(-2)  &  4.3(-2)   \\ \hline
\end{tabular}\\
\tiny
\end{center}
\end{table}
$^a$ The abundance combinations are as follows: {\bf 1: $X_0$ = 10$^{-4}$, 2: $X_0$ = 10$^{-6}$ and a:$X_d$ = 10$^{-6}$, b: $X_d$= 10$^{-7}$, c:$X_d$ = 10$^{-8}$}\\
Note: A(B) = A $\times 10^{\rm{B}}$
\end{landscape}

\begin{landscape} 
\begin{table}[!htb]
\label{table:h2ogrid_large1}
\caption{Integrated intensities, $\int T_{\rm{MB}}\Delta V$ [K km s$^{-1}$], for  H$_2$O lines in the wide parameter grid for all abundances for para p =2.0}
\begin{center}
\tiny
\begin{tabular}{c@{}c   c   c c  c   c c  c   c c  c  c c  c  c c  c  c}
\hline \hline
 Density (cm$^{-3}$) & \multicolumn{18}{c}{Integrated Intensity $\int T_{\rm{MB}}\Delta V$ [K km s$^{-1}$] } \\ \hline
 & \multicolumn{6}{c}{$L=2$ (L$_{\odot}$)} & \multicolumn{6}{c}{$L=25$ (L$_{\odot}$)} & \multicolumn{6}{c}{$L=7$ (L$_{\odot}$)} \\ 
& \multicolumn{6}{c}{Abundance$^a$} &  \multicolumn{6}{c}{Abundance$^a$} & \multicolumn{6}{c}{Abundance$^a$} \\ 
& 1a & 1b & 1c & 2a & 2b & 2c & 1a & 1b & 1c & 2a & 2b & 2c & 1a & 1b & 1c & 2a & 2b & 2c \\ \hline
\tiny & \multicolumn{18}{c}{1$_{11}$-0$_{00}$}\\ \hline
  4$\times 10^5$ &   3.9     &  0.11     &  9.3(-2) &  0.51     &  0.26     &  8.4(-2) &  0.7     &  0.43     &  0.38     &   2.8     &  0.24     &  0.22     &  0.29     &  0.52     &  0.15     &   1.4     &  0.17     &  0.12     \\ 
1$\times 10^6$ &  0.34     &  0.34     &  0.30     &  0.10     &  0.71     &  0.30     &  0.12     &  0.44     &  0.23     &  8.1(-2) &  0.21     & -0.11     &  3.6(-3) &  0.66     &  9.4(-2) &  8.3(-2) &   1.6     &   2.7     \\ 
 5$\times 10^6$ &   1.7     &  0.98     & -0.26     &  0.91     &  0.44     & -0.10     &  -1.9     &   2.0     &  -1.8     &  -3.5     &  0.16     &  -2.1     & -0.2     & -9.0(-2) & -0.18     &   6.1     & -0.49     & -0.57     \\ \hline
 & \multicolumn{18}{c}{ 2$_{02}$-1$_{11}$}\\ \hline
 4$\times 10^5$ &   2.9     &  0.40     &  0.18     &  0.76     &  0.4     &  0.18     &   2.8     &   1.9     &  0.95     &   2.9     &   1.7     &  0.74     &   1.2     &  0.90     &  0.40     &   1.9     &  0.8     &  0.35     \\ 
 1$\times 10^6$ &  0.90     &  0.66     &  0.31     &  0.86     &  0.65     &  0.29     &   4.1     &   3.1     &   1.6     &   3.6     &   2.8     &   1.5     &   1.9     &   1.4     &  0.65     &   1.8     &   1.6     &  0.62     \\ 
 5$\times 10^6$ &   1.4     &  0.94     &  0.46     &   1.36     &  0.68     &  0.47     &   2.2     &   3.8     &   2.7     &   7.6     &   3.57     &   2.6     &  0.77     &   1.8     &   1.1     &   5.5     &   1.6     &   1.2     \\ \hline
 & \multicolumn{18}{c}{ 2$_{11}$-2$_{02}$}\\\hline
 4$\times 10^5$ &   1.3     &  0.24     &  9.8(-2) &  0.46     &  0.26     &  9.4(-2) &   1.9     &   1.1     &  0.54     &   2.2     &   1.0     &  0.43     &  0.79     &  0.56     &  0.21     &   1.16     &  0.49     &  0.19     \\ 
 1$\times 10^6$ &  0.66     &  0.46     &  0.19     &  0.66     &  0.45     &  0.19     &   3.0     &   2.1     &  0.89     &   2.8     &   1.9     &  0.83     &   1.4     &  0.94     &  0.40     &   1.5     &   1.2     &  0.38     \\ 
 5$\times 10^6$ &   1.4     &  0.99     &  0.45     &   1.3     &  0.88     &  0.47     &   4.7     &   5.9     &   2.5     &   6.1     &   5.2     &   2.4     &   1.9     &   2.2     &   1.1     &   4.7     &   2.1     &   1.1     \\ \hline
 & \multicolumn{18}{c}{2$_{20}$-2$_{11}$}\\\hline
 4$\times 10^5$ &   1.1     &  0.19     &  8.6(-2) &  0.31     &  0.21     &  8.1(-2) &   1.8     &   1.3     &  0.75     &   2.1     &   1.1     &  0.4     &  0.58     &  0.54     &  0.24     &   1.0     &  0.47     &  0.18     \\ 
 1$\times 10^6$ &  0.30     &  0.32     &  0.13     &  0.34     &  0.32     &  0.13     &   2.5     &   2.1     &  0.96     &   1.9     &   1.8     &  0.81     &  0.93     &  0.78     &  0.33     &   1.1     &   1.0     &  0.31     \\ 
 5$\times 10^6$ &  0.50     &  0.50     &  0.25     &  0.26     &  0.43     &  0.26     &   1.2     &   4.1     &   1.9     &   1.8     &   3.5     &   1.9     &  0.18     &   1.3     &  0.75     &   2.6     &   1.2     &  0.74     \\ \hline
 & \multicolumn{18}{c}{3$_{31}$-4$_{04}$}\\\hline
 4$\times 10^5$ &  8.4(-3) &  5.5(-3) &  5.0(-3) &  5.8(-3) &  2.6(-3) &  2.3(-3) &  0.46     &  0.45     &  0.42     &  5.7(-2) &  4.0(-2) &  3.8(-2) &  7.9(-2) &  7.1(-2) &  7.0(-2) &  2.2(-2) &  1.3(-2) &  1.2(-2) \\ 
 1$\times 10^6$ & -1.8(-3) & -3.7(-3) & -3.9(-3) &  1.6(-3) &  1.1(-5) & -1.6(-4) &  0.16     &  0.13     &  0.13     &  6.4(-2) &  3.9(-2) &  3.6(-2) &  1.6(-3) & -6.7(-3) & -7.3(-3) &  1.3(-2) &  4.5(-3) &  3.5(-3) \\ 
 5$\times 10^6$ & -2.5(-4) & -3.2(-5) & -1.1(-5) & -2.9(-4) & -2.5(-5) & -2.6(-6) & -1.7(-2) & -3.0(-2) & -3.1(-2) &     8.6(-3) & -2.3(-3) & -3.3(-3) & -1.5(-3) & -1.2(-3) & -1.2(-3) &  2.4(-4) & -1.9(-4) & -1.7(-4) \\ \hline
 & \multicolumn{18}{c}{ 4$_{22}$-3$_{31}$}\\\hline
 4$\times 10^5$ &  3.7(-2) &  2.2(-2) &  2.0(-2) &  2.9(-2) &  1.5(-2) &  1.3(-2) &  0.44     &  0.40     &  0.43     &  0.26     &  0.20     &  0.19     &  0.12     &  9.3(-2) &  9.2(-2) &  9.1(-2) &  5.7(-2) &  5.4(-2) \\ 
 1$\times 10^6$ &  2.7(-2) &  1.0(-2) &  7.2(-3) &  2.5(-2) &  9.1(-3) &  6.0(-3) &  0.41     &  0.31     &  0.29     &  0.30     &  0.20     &  0.19     &  0.11     &  5.9(-2) &  5.2(-2) &  9.2(-2) &  4.6(-2) &  3.8(-2) \\ 
 5$\times 10^6$ &  5.7(-3) &  3.0(-4) & -8.6(-4) &  6.0(-3) &  7.3(-4) & -3.9(-4) &  0.24     &  5.2(-2) &  1.9(-2) &  0.25     &  6.0(-2) &  2.8(-2) &  3.3(-2) &  2.3(-3) & -5.5(-3) &  3.8(-2) &  6.1(-3) & -1.5(-3) \\ \hline
  & \multicolumn{18}{c}{4$_{22}$-4$_{13}$}\\\hline
 4$\times 10^5$ &  0.14     &  6.8(-2) &  2.8(-2) &  0.13     &  6.3(-2) &  2.4(-2) &   1.4     &  0.61     &  0.42     &   1.4     &  0.60     &  0.49     &  0.40     &  0.22     &  0.12     &  0.40     &  0.21     &  0.11     \\ 
 1$\times 10^6$ &  0.10     &  4.7(-2) &  1.2(-2) &  0.10     &  4.7(-2) &  1.2(-2) &   1.8     &  0.74     &  0.35     &   1.7     &  0.69     &  0.34     &  0.42     &  0.19     &  7.0(-2) &  0.43     &  0.19     &  6.4(-2) \\ 
 5$\times 10^6$ &  4.2(-2) &  8.5(-3) &  1.1(-3) &  4.16(-2) &  8.6(-3) &  1.2(-3) &   1.5     &  0.55     &  7.1(-2) &   1.4     &  0.56     &  8.6(-2) &  0.32     &  8.0(-2) &  7.8(-3) &  0.31     &  8.3(-2) &  1.0(-2) \\ \hline
& \multicolumn{18}{c}{5$_{24}$-4$_{31}$}\\\hline
4$\times 10^5$ &  3.1(-2) &  2.6(-2) &  2.6(-2) &  2.0(-2) &  1.5(-2) &  1.4(-2) &  0.50     &  0.49     &  0.51     &  0.19     &  0.17     &  0.16     &  0.13     &  0.12     &  0.12     &  6.7(-2) &  5.5(-2) &  5.4(-2) \\ 
1$\times 10^6$ &  1.5(-2) &  8.3(-3) &  7.6(-3) &  1.3(-2) &  7.0(-3) &  6.2(-3) &  0.40     &  0.34     &  0.35     &  0.23     &  0.19     &  0.19     &  8.2(-2) &  6.5(-2) &  6.2(-2) &  6.3(-2) &  4.3(-2) &  4.1(-2) \\ 
5$\times 10^6$ & -9.9(-5)  & -9.4(-4)  & -1.1(-3)  &  3.3(-4)  & -3.7(-4)  & -4.9(-4)  &  6.0(-2)  &  1.3(-2)  &  7.5(-3)  &  7.7(-2)  &  3.0(-2)  &  2.4(-2)  &  2.3(-3)  & -6.9(-3)  & -8.7(-3)  &  7.8(-3)  & -1.0(-3)  & -2.6(-3)  \\ \hline
\end{tabular}\\
\tiny
\end{center}
\end{table}
$^a$ The abundance combinations are as follows: {\bf 1: $X_0$ = 10$^{-4}$, 2: $X_0$ = 10$^{-6}$ and a:$X_d$ = 10$^{-6}$, b: $X_d$= 10$^{-7}$, c:$X_d$ = 10$^{-8}$}\\
Note: $A(B)$ denotes $A\times 10^{{B}}$
\end{landscape}
\end{document}